\documentclass{article}
\usepackage{graphicx} 

\usepackage{mathtools}

\usepackage{color,xcolor,ucs}
\usepackage{mathtools}   \usepackage{tikz} 
\usepackage{ amssymb }
\usepackage{extarrows} 
\usepackage{pgf,tikz}
\usepackage{float}
\usetikzlibrary{positioning}
\usetikzlibrary{shapes.geometric}
\usetikzlibrary{shapes.misc}
\usetikzlibrary{arrows}
\usepackage{caption}
\usepackage{mathrsfs}
\usetikzlibrary{arrows,shapes,automata,backgrounds,petri,positioning}
\usetikzlibrary{decorations.pathmorphing}
\usetikzlibrary{decorations.shapes}
\usetikzlibrary{decorations.text}
\usetikzlibrary{decorations.fractals}
\usetikzlibrary{decorations.footprints}
\usetikzlibrary{shadows}
\usetikzlibrary{calc}
\usetikzlibrary{spy}
\usepackage{amsmath}
\usepackage{array}
\usepackage{ amssymb }
\usepackage{braket}
\usepackage{qcircuit}
\usepackage{soul}
\usepackage{braket} 
\usepackage{relsize}

\usepackage{amsmath}
\usepackage{ amssymb }
\usepackage{braket}
\usepackage{qcircuit}
\usepackage{soul}
\usepackage{braket}

\title{{\LARGE Malleability of transformations on the ciphertext in noisy Quantum public key encryption              }}
\author{Pete Rigas \footnote{Newport Beach, CA 92625, pbr43@cornell.edu, ORCiD: 0009-0003-1053-9720}}
\date{}

\begin{document}

\maketitle

\abstract{We characterize a noisy variant of a Quantum public encryption protocol recently introduced by Malavolta and Walter which demonstrated that the notion of everlasting security can be rigorously formulated for Quantum key distribution after two rounds of interaction between Alice and Bob. To address one possible direction of research that is related to injecting noise in the cryptographic protocol related to Quantum key distribution we formulate arguments for further examining the notion of everlasting security through malleability assumptions on transformations of the ciphertext. Assumptions surrounding malleability were introduced by Maurer and Tackmann for the purposes of comparing how authenticate then encrypt, and encrypt then authenticate, protocols behave through a variety of expressions for the forwarding error, deleting error, and reconstruction probabilities. Such probabilities are put to further use for obtaining connections between the indistinguishability and security threshold for a cryptographic protocol of interests. To further build upon such associations we demonstrate, through an adaptation of the Gentle Measurement Lemma from Quantum information theory, how upper bounds on the trace distance can be used to generalize the negligibility function obtained by Malavolta and Walter in the noiseless setting. Besides the fact that the negligibility function in the noisy setting is related to a higher security threshold it continues to remain of interest to determine whether computations provided in this work for upper bounding the trace distance can be related to other settings that are centered more on game-theoretic approaches.  \textit{{Keywords}: Quantum games, non-locality, Quantum computation, amplification} \footnote{\textbf{MSC Class}: 81P02; 81Q02}}

\maketitle

\section{Introduction}

\subsection{Overview}

\noindent Quantum Key Distribution (QKD) has attracted significant research attention through theoretical and experimental investigations alike, related to security through Information-theoretic, or Computational, Cryptography, [1]; the connection between generalizations of indistinguishability, through the notion of indifferentiability in Abstract Cryptography, along with the possible sets of resources used by an adversary Eve, or by Alice and Bob, [2]; whether any amount of Quantum information imparts Cryptographic security, [4]; Cryptographic security in general [11,12]; asymmetry between the security of Alice and Bob [13]; amplification results, [5]; confidentiality and integrity, [6]; amongst several closely related topics, [7,8,9,10]. Recently, an information-theoretic formalism presented in [3] presents many intriguing connections between Public Key Encryption (PKE) and QKD, specifically through: distinguishing between Information-theoretic, and Computational, Cryptography, particularly through whether an adversary is capable of receiving one copy, or arbitrarily many copies, of a Quantum secret key; computations with respect to the trace distance; collision probabilities from hashing functions; pseudo random functions (PRFs); secureness of the Quantum one-way function which essentially makes computing the preimage of a function under a Quantum mapping, through a suitable encoding, practically infeasible for an adversary; one time signature (OTS) schemes (OTSs); the design of hybrid experiments.

Separate from security notions of the aforementioned results everlasting security defined in [3] is related to the Information-theoretic notion of security. Such a notion is especially desirable to quantify, along with its connections to PKE. Straightforwardly while one may expect that the notion of everlasting security is considerably strong than that of ordinary security the theoretical reasons are deeply rooted in the construction of secret keys, along with the keys that Alice and Bob construct for verifying output that could be shared between them. Under such circumstances Alice and Bob must not only prevent the adversary from compromising the security of their protocol with one copy of the Quantum secret key, but also take into account constructions of their public, verification and signature, keys intrinsic to OTSs. Namely Alice and Bob must: (1) construct candidate secret Quantum keys from verification keys; (2) generate signature keys to ensure that the secret keys returned by OTSs are legitimate; (3) construct the projection operator $\Pi$ (defined formally at the end of the Introduction) which is used for checking whether a state belongs to the subspace spanned by the \textit{unforgeable} states; (4) decoding of the ciphertext through the generation of a Quantum secret key given the choice of a fixed security threshold and PRF.

Ultimately the security of PKE, from the security of QKD, relies upon the same limitations underlying Quantum computation that have been identified in other studies of QKD security [10]. However, everlasting security of QKD in [3] is shown to depend upon only two rounds of interaction. This feature of the cryptographic protocol is reflected through the Two-Message (ie, two rounds of interaction) QKD introduced in [3], in which Alice and Bob make use of a noninteractive protocol for agreeing on a shared secret key after two rounds of interaction. Generally speaking the noninteractive protocol within Quantum Information theory is of interest to examine, as demonstrated through previous work of the author [15]; such protocols allow for Alice and Bob to agree on a shared secret, transmitted through either a Quantum key or other cryptographic objects, which could impart prospective advantages of Quantum information that are not present in Classical information which relate to entanglement. Whether the exact setting is in QKD or not, one enforces constraints of noninteractive protocols between Alice and Bob, or between countably many players, through the stipulation that the participants not share any additional information amongst themselves beyond a single message.

More specifically, the assumptions surrounding noninteractive protocols greatly differ from those enforced in other cryptographic protocols in QKD, in particular the one presented in [10]. Namely in the QKD cryptographic protocol presented by Ostrev in [10], Alice and Bob interact several times through: (1) the choice of a random matrix, as well as the transpose of its inverse, for computing the intermediate quantities $v_A, v_B, w_A, w_B$ that Alice and Bob use for determining their shared secret key; (2) the use of isometries (ie, length preserving transformations) for performing measurements with respect to the computational basis; (3) the computation of functions $g$ of the parity check matrices; (4) generalizations of isometries, as well as functions of the parity check matrices, examined by the author in the context of the security threshold [14]. Whether one could incorporate assumptions between interactive, and noninteractive, cryptographic protocols for QKD is of great interest to further explore.

However, while interactive protocols between Alice and Bob for QKD are capable of modeling conditions underlying noisy Quantum computation in near-term experiments, approximating functions of the parity check matrices can be computationally expensive; given the fact that the parity check matrices that Alice and Bob use for agreeing on a shared secret key are themselves random matrices, registers in Quantum circuits must not only store information about each entry but also must incorporate sources of noise. Interactive protocols, in comparison to their noninteractive counterparts, do not rely upon such computations which could be of interest towards: rigorously comparing the computational complexity, through the expected algorithmic runtime of an interactive protocol versus that of a noninteractive protocol; simplifying the number of preprocessing steps that Alice and Bob use for generating candidate secret keys in QKD; determining how the noninteractive assumption on the cryptographic protocol is dependent upon two, or more, rounds of encoding, decoding and hashing of the ciphertext; generating verification keys from evaluations of a Hash function.

There remain several fascinating directions of research from the interaction between PKE and QKD developed in [3]. One such direction that immediately comes to mind relates to making use of the observations provided in [9], in which Maurer, Ruedlinger and Tackmann discuss: (1) transformations on the ciphertext, formally through assumptions related to malleability; (2) game-based security notions which quantify the probability of Alice and Bob hiding information that Eve wishes to compromise; (3) forgery through CCA and CPA (Chosen Ciphertext Attack and Chosen Plaintext Attack, respectively) protocols, which can serve the purpose of allowing the adversary to modify ciphertext and plaintext candidates used by Alice and Bob; (4) encoding, decoding, authentication and transmission mechanisms on inner and outer interfaces; (5) requirements on real and ideal systems, along with the choice of corresponding simulators; (6) indistinguishability, through the difference of two probabilities corresponding to two unique random experiments. By and large the analysis related to such notions bridges cryptographic security that Alice and Bob use for across inner and outer interfaces and discrete probability.

Moreover, while the aforementioned notions provided above are related to cryptographic protocols that are either dependent, or independent, of QKD, the connection between QKD and PKE could provide quantitative observations surrounding: (1) connections between the transformations that Eve could use when Alice and Bob are transmitting information over a noisy Quantum channel; (2) quantification of the trace distance between real and idealized states, similar to those provided for the indistinguishability and indifferentiability metrics; (3) limitations of noisy Quantum computation, and corresponding tensor network representations. To make use of the infrastructure surrounding OTSs introduced in [3], we: (1) demonstrate how security threshold for idealized QKD can be obtained; (2) quantitative connections between the number of measurements in the isometries used by Alice and Bob and the accompanying security threshold; (3) measurements associated with two parity checks, which are defined through a marginalization of random matrices; (4) connections with Positive Operator Valued Measurements (POVMs) and Completely Positive Trace Preserving (CPTP) maps (CPTPs), particularly through the fact that requirements on POVMs and CPTPs together imply several properties of unconditional security; (5) potential adaptations related to: uncertainty propagation, limits of information transmission over noisy Quantum channels, security of Quantum one-way functions, and limitations of various related cryptographic protocols. Indeed, while the requirements underlying unconditional, and as considered in [3], everlasting, security could be difficult to experimentally realize such protocols are nevertheless an invaluable starting point. Along the lines of previous efforts of the author which included terms of the form,

{\small

\begin{align*}
   {\tiny \frac{1}{\sqrt{2}}}  \bigg\{  {\tiny  \ket{F \big( \alpha \big) , F \big( \alpha + \beta^{\prime} \big)}_{S(S^{\prime} + T^{\prime})} +    \ket{F \big( \beta  \big) , F \big( \alpha + \beta^{\prime} \big)}_{T(S^{\prime} + T^{\prime})}  }   \bigg\}     , \\ \\                         {\tiny \frac{1}{\sqrt{2}} }  \bigg\{   {\tiny  \ket{F ( \alpha^{\prime} + \beta^{\prime\prime} ) , F \big( \alpha^{\prime}  \big) }_{(S+T) S^{\prime}} +   \ket{F ( \alpha^{\prime} + \beta^{\prime\prime} ) , F \big( \beta^{\prime\prime} \big) }_{(S+T) T^{\prime}}  }               \bigg\}        , \\ 
\end{align*}

}

\noindent where,

{\small

\begin{align*}
       {\tiny F \equiv } \textit{Idealized function}   , \end{align*}

       \begin{align*} {\tiny \alpha \neq \alpha^{\prime} \neq \beta  \neq \beta^{\prime} \equiv}  \textit{Constants taken over the base field}     ,  \\ \\  {\tiny S , T \equiv } \textit{Resources used by Alice and Bob for parameters $\alpha, \beta $}   , \\ \\  {\tiny S^{\prime} , T^{\prime} \equiv } \textit{Resources used by Alice and Bob for parameters $\alpha^{\prime}, \beta^{\prime} $}    , \\ 
\end{align*}

}

\noindent in non-idealized isometries used by Alice and Bob in QKD. For hash functions that are two-universal, one takes the base field to be $\textbf{F}_2$, ie the field with two elements.

\bigskip

\noindent As with any other Quantum algorithm (or, protocol), noise emerges as a fundamental limitation to not only the quality of readout but also to potential advantages over Classical computation. Albeit the fact that several comparisons have already been established, as described in [14,15,16], the fields of Quantum computation and Quantum Information theory alike stand to benefit from additional comparisons and differences between established methods. For QKD protocols dependent upon QPKE, the authors of [3] raise several intriguing directions of research related to such limitations. To further develop how the idealized QPKE-QKD protocol can be adapted to noisy Quantum computation, we draw the attention of the reader surrounding: QKD protocols, and their security thresholds, that are independent of QPKE which have previously been examined by the author, [14]; the introduction of QPKE, [3], along with its connections to QKD for distinguishing computational, versus information-theoretic, notions of security; notions of malleable functions on the plaintext, which can be used to discuss how ordinary, and strong, notions of unforgeability expressed for the QPKE-QKD protocol can be generalized from the notion of unforgeability for the Chosen Message Attack (CMA) provided in [9]. Comparing the notions of unforgeability for OTSs (UF-OTS) with the notion of strong unforgeability from CMAs (SUF-CMA) could be of use for quantifying how attacks on the plaintext, through the notion of malleability, interact with the notions of computational and information-theoretic security.

Through conditions enforced on such functions of the plaintext, one is also capable of introducing more noise into the QPKE-QKD protocols, as suggested in [3]. Generally speaking the presence of noise in QKD-dependent protocols could serve to corrupt information amongst blocks of the secret key which Alice and Bob are planning to share at the end of their protocol. This security vulnerability can be expressed in terms of properties of the plaintexts and ciphertexts which Alice and Bob construct throughout a protocol. First, Alice manipulates a plaintext object at the beginning of the protocol that she wishes to share with Bob. She not only is interested in determining properties of the initial plaintext encoding that would prevent Eve from compromising the security of the ciphertext produced in the middle of the protocol, but also for the transformation on the ciphertext to output another plaintext for Bob. Although the QKD-QPKE protocol, along with its information-theoretic and computational security notions provided in [3] is expressed in terms of the action of a suitably defined projection operation,

{\small

\begin{align*}
 {\tiny \Pi \equiv        \underset{\sigma_0 \in \Sigma_0}{\sum}}  \big\{  {\tiny \ket{0} \otimes \ket{\sigma_0} \bra{\sigma_0} \otimes  \bra{0} }  \big\} + {\tiny \underset{\sigma_1 \in \Sigma_1}{\sum} }  \big\{  {\tiny \ket{1} \otimes   \ket{\sigma_1} \bra{\sigma_1}  \otimes \bra{1} }  \big\}  ,  \\ 
\end{align*}

}

\noindent which is decomposed in terms of two inner products over the possible spaces of signatures for Alice, and Bob, respectively, incorporating additional terms (and hence, measurements) into generalized representations of the projection would allow one to artificially inject noise into one step of the protocol, from which any downstream impacts on the security threshold can be quantified. Under noise one would expect that the projection operator associated with the QPKE-QKD protocol would take the form,

{\small

\begin{align*}
 {\tiny \widetilde{\Pi} \equiv        \underset{\sigma_0 \in \Sigma_0}{\sum} } \big\{ \tiny{ \ket{0} \otimes \ket{\sigma_0 + \textit{noise} } \bra{\sigma_0 + \textit{noise} } \otimes  \bra{0}}   \big\}  + {\tiny \underset{\sigma_1 \in \Sigma_1}{\sum}}  \big\{  \tiny{ \ket{1} \otimes   \ket{\sigma_1 + \textit{noise} } \bra{\sigma_1 + \textit{noise} }  \otimes \bra{1} } \big\}    , \\ 
\end{align*}

}

\noindent To this end, additional measurements introduced through the projection operator in the QPKE-QKD setting would serve the purpose of not only a higher security threshold (ie, some constant bounded by a larger constant away from zero) than the protocol developed in [3] corresponding to no noise, but also could provide intriguing comparisons to the presence of noise for QKD-dependent protocols which are independent of QPKE, as analyzed by the author in [14]. From the two-universal hash functions analyzed for QKD protocols, [10,14], hash functions that do not satisfy any universality conditions can nevertheless be put to use for a stronger notion of security, in the computational sense.

While various formulations of security are dependent upon the number of copies of the secret key that Eve can receive, the number of copies of the secret key are also of fundamental importance in the encoding and decoding phases of the protocol. If Eve is to receive only one copy of the secret key for the information-theoretic notion of security, then the encoding and decoding protocols that Alice and Bob use from a candidate secret key are dependent upon pseudo random functions. On the other hand, if Eve is to receive countably many copies of the secret key for the computational notion of security, then the encoding and decoding are expressed in terms of a direct summand of the message with the ciphertext. Such a decomposition of the secret key after decoding has been widely used in [6,7,8,9], for characterizing notions of security similar to those in the computational sense. Relatedly, albeit the fact that everlasting security from information-theoretic security is defined independently of hashing functions that are two-universal, weakening the universality assumption on the hashing function continues to be sufficient for a notion of security that is quite strong. 

Separate from the computational framework provided in [14] (which primarily relied upon incorporating additional measurements into isometries used for QKD) we devote attention towards encoding and decoding protocols as described in security of the information-theoretic sense. As previously alluded to, even if everlasting security in the information-theoretic sense has the advantage of requiring only one copy of the secret key for Eve, each step associated with encoding and decoding given an instance of the secret key are undoubtedly more complicated than the steps associated with encoding and decoding for security of the computational-sense. To quantitatively comment on how noise would impact the information-theoretic sense of security, in the following we: draw the attention of the reader to expansions for outer products taken with respect to the entangled basis; collision probabilities for hashing functions that are dependent, and independent, of universality assumptions; comparisons between negiligibility and indistinguishability, which is useful for comparing notions of ordinary and strong unforgeability.

\subsection{This paper's contributions}

\noindent This paper formalizes how up to constants lower bounds for the trace can be related to upper bounds on the trace distance, hence partially generalizing the notion of everlasting security with noise. Specifically, central to the following computations with respect to the trace distance is the existence of a constant $\mathcal{C}^{\prime\prime}$ taken sufficiently small for which,

{\small

\begin{align*}
 \mathrm{Tr} \bigg\{                \sqrt{\frac{{\tiny                  \big( \widetilde{\rho - \tau} \big)^{\dagger} \times \bigg(   \frac{\big( \widetilde{\rho - \tau} \big) }{ \rho - \tau }    \bigg)^{\dagger}     }}{{\tiny        \frac{\big( \rho - \tau \big)}{\big( \widetilde{\rho - \tau} \big)^2 }   }}}    \bigg/  \sqrt{  {\tiny \big( \rho - \tau \big)^{\dagger }  \times  \big( \rho - \tau \big)   }  }     \bigg\}    \geq   \frac{1}{\mathcal{C}^{\prime\prime}}  \cdot    \mathrm{Tr} \bigg\{              \bigg(    {\tiny              \frac{  \big( \widetilde{\rho - \tau } \big)^{\dagger} }{\big( \rho - \tau \big)^{\dagger}}  }     \bigg/    {\tiny               \frac{ \rho - \tau }{ \widetilde{\rho - \tau}   }                     }    \bigg)                                \bigg\}  ,  \\ 
\end{align*}

}

\noindent given noiseless quantum states $\rho$ and $\tau$, along with the noisy quantum states $\widetilde{\rho}$ and $\widetilde{\tau}$. The above estimate is used to provide a lower bound for,

{\small

\begin{align*}
        \mathrm{Tr} \bigg\{   \sqrt{ \frac{ {\tiny \big( \widetilde{\rho - \tau} \big)^{\dagger}}}{{\tiny \big( \rho - \tau \big)^{\dagger } }} \frac{{\tiny \big( \widetilde{\rho - \tau } \big)}}{{\tiny \big( \rho - \tau \big) } }}          \bigg\} , \\ 
        \end{align*}
        }

        \noindent and also for,

       {\small

\begin{align*}
  \mathrm{Tr} \bigg\{  \frac{    \sqrt{ {\tiny    \big( \widetilde{\rho - \tau} \big)^{\dagger} \times  \big( \widetilde{\rho - \tau } \big)  } }   }{     \sqrt{  {\tiny \big( \rho - \tau \big)^{\dagger }  \times  \big( \rho - \tau \big)   }  }  }      \bigg\}    . \\ 
\end{align*}
}

        \noindent Comparing the two above trace distances, either under a single square root function,

{\small

\begin{align*}
 \sqrt{ \frac{ {\tiny \big( \widetilde{\rho - \tau} \big)^{\dagger}}}{{\tiny \big( \rho - \tau \big)^{\dagger } }} \frac{{\tiny \big( \widetilde{\rho - \tau } \big)}}{{\tiny \big( \rho - \tau \big) } }} 
   , \\
\end{align*}

}

        \noindent as opposed to two square root functions,

{\small

\begin{align*}
  \frac{    \sqrt{ {\tiny    \big( \widetilde{\rho - \tau} \big)^{\dagger} \times  \big( \widetilde{\rho - \tau } \big)  } }   }{     \sqrt{  {\tiny \big( \rho - \tau \big)^{\dagger }  \times  \big( \rho - \tau \big)   }  }  }    , \\ 
\end{align*}

}

\noindent serves the purpose of quantifying the negligibility function,

{\small

\begin{align*}
   {\tiny \big\{ \mathrm{NEGL} \big(      \lambda^{\prime}  - \lambda     \big)  \big\}^{-1}  }   , 
\end{align*}

}

\noindent given the choice of noiseless and noisy security parameters for the Quantum public key encryption protocol, $\lambda$ and $\lambda^{\prime}$, respectively. The above negligibility function under specific assumptions on the noise in the public key encryption protocol is related to the neglibility function defined in \textbf{Definition} \textit{12}.

Under the correctness assumption of the noisy quantum public key encryption protocol, for the probability measure $\mathcal{P}$ over the noisy Quantum public key encryption protocol,

{\small

\begin{align*}
  \mathcal{P} \bigg[ {\tiny  1 = \widetilde{\mathrm{Ver}} \big( \widetilde{\mathrm{vk}} , \widetilde{m^{*}} , \widetilde{\sigma^{*}} \big)}  \textit{ with } {\tiny \big( \widetilde{m} , \widetilde{\sigma} \big) \neq \big( \widetilde{m^{*}} , \widetilde{\sigma^{*}} \big) :                     \big\{ \big( \widetilde{\mathrm{vk}} , \widetilde{\mathrm{sk}} \big) \leftarrow \widetilde{\mathrm{SGen}} \big( 1^{\lambda^{\prime}} \big) \big\} , \big\{ \sigma \leftarrow \widetilde{\mathrm{Sign}} \big( \widetilde{\mathrm{sk}} , \widetilde{m} \big) \big\} ,}  \\ {\tiny \big\{        \big( \widetilde{m^{*}} , \widetilde{\sigma^{*}} \big) \leftarrow A_{\lambda^{\prime}} \big( \mathrm{vk} , \widetilde{m} , \widetilde{\sigma} \big)       \big\}  }     \bigg] = {\tiny \mathrm{NEGL} \big( \lambda^{\prime} \big) }  , \\ 
\end{align*}

}

\noindent introduced in \textbf{Definition} \textit{40} is an alternative formulation of the neglibility function $\mathrm{NEGL} {\tiny \big( \cdot \big)}$ and,

{\small

\begin{align*}
 \mathcal{P} \big[  {\tiny  1 = \widetilde{\mathrm{Ver}} \big( \widetilde{\mathrm{vk}} , m^{\prime} ,  \widetilde{\mathrm{Sign}} \big(       \widetilde{\mathrm{sk}} , m^{\prime}    \big)  \big)  :                \big( \widetilde{\mathrm{vk}} , \widetilde{\mathrm{sk}} \big) \leftarrow  \widetilde{\mathrm{SGen}} \big( 1^{\lambda^{\prime}} \big)    }        \big] = 1    , \\ 
\end{align*}

}

\noindent given the noisy verification, key generation, and signature, protocols (defined in \textbf{Definition} \textit{34}, \textbf{Definition} \textit{35}, \textbf{Definition} \textit{36},  respectively) are used to conclude,

{\small

\begin{align*}
         1 -  \bigg(  \frac{1}{2 \mathcal{C} \mathcal{C}^{\prime}  \big( \mathcal{C}^{\prime\prime} \big)^3  \big( C C^{\prime} C^{\prime\prime} \big)^2 } \bigg)^2  \cdot         \underset{\rho,\tau,\widetilde{\rho},\widetilde{\tau}}{\mathrm{inf}} \bigg\{         {\mathrm{Tr} \bigg\{   {\tiny \frac{\sqrt{\widetilde{\rho}^{\dagger}}}{\sqrt{\big( \rho - \tau \big)^{\dagger} }}  }  }  \bigg\}    \times  \bigg[ \mathrm{Tr} \bigg\{               {\tiny              \frac{  \big( \widetilde{\rho - \tau } \big)^{\dagger} }{\big( \rho - \tau \big)^{\dagger}}  }     \bigg\} \bigg/  \mathrm{Tr} \bigg\{       {\tiny               \frac{ \rho - \tau }{ \widetilde{\rho - \tau}   }                     }                    \bigg\}      \bigg]              ,        {\mathrm{Tr} \bigg\{  {\tiny   \frac{\sqrt{\widetilde{\rho}}}{\sqrt{\big( \rho - \tau \big)}}      }   } \bigg\}   \\   \\  \\  \times  \bigg[ \mathrm{Tr} \bigg\{               {\tiny              \frac{  \big( \widetilde{\rho - \tau } \big)^{\dagger} }{\big( \rho - \tau \big)^{\dagger}}  }     \bigg\} \bigg/  \mathrm{Tr} \bigg\{       {\tiny               \frac{ \rho - \tau }{ \widetilde{\rho - \tau}   }                     }                    \bigg\}      \bigg]   ,  {\mathrm{Tr}          \bigg\{       {\tiny \frac{\sqrt{\widetilde{\rho}^{\dagger}}}{\sqrt{\big( \rho - \tau \big)^{\dagger} }} }         }  \bigg\}    \times  \bigg[ \mathrm{Tr} \bigg\{               {\tiny              \frac{  \big( \widetilde{\rho - \tau } \big)^{\dagger} }{\big( \rho - \tau \big)^{\dagger}}  }     \bigg\} \bigg/  \mathrm{Tr} \bigg\{       {\tiny               \frac{ \rho - \tau }{ \widetilde{\rho - \tau}   }                     }                    \bigg\}      \bigg]  ,  \end{align*}

         \begin{align*}  \\       {\mathrm{Tr} \bigg\{   {\tiny \frac{\sqrt{\widetilde{\tau}}}{\sqrt{\big( \rho - \tau \big) } }}   } \bigg\}    \times  \bigg[ \mathrm{Tr} \bigg\{               {\tiny              \frac{  \big( \widetilde{\rho - \tau } \big)^{\dagger} }{\big( \rho - \tau \big)^{\dagger}}  }     \bigg\} \bigg/  \mathrm{Tr} \bigg\{       {\tiny               \frac{ \rho - \tau }{ \widetilde{\rho - \tau}   }                     }                    \bigg\}      \bigg]   ,     {\mathrm{Tr}              \bigg\{  {\tiny   \frac{\sqrt{\widetilde{\tau}^{\dagger}}}{\sqrt{\big( \rho - \tau \big)^{\dagger}} }      }              }  \bigg\}    \\ \\ \\      \times  \bigg[ \mathrm{Tr} \bigg\{               {\tiny              \frac{  \big( \widetilde{\rho - \tau } \big)^{\dagger} }{\big( \rho - \tau \big)^{\dagger}}  }     \bigg\} \bigg/  \mathrm{Tr} \bigg\{       {\tiny               \frac{ \rho - \tau }{ \widetilde{\rho - \tau}   }                     }                    \bigg\}      \bigg]    ,    {\mathrm{Tr}         \bigg\{   {\tiny    \frac{\sqrt{\widetilde{\rho}}}{\sqrt{\big( \rho - \tau \big) }}     }      } \bigg\} \\ \\ \\     \times  \bigg[ \mathrm{Tr} \bigg\{               {\tiny              \frac{  \big( \widetilde{\rho - \tau } \big)^{\dagger} }{\big( \rho - \tau \big)^{\dagger}}  }     \bigg\} \bigg/  \mathrm{Tr} \bigg\{       {\tiny               \frac{ \rho - \tau }{ \widetilde{\rho - \tau}   }                     }                    \bigg\}      \bigg]    ,   {\mathrm{Tr}     \bigg\{  {\tiny   \frac{\sqrt{\widetilde{\tau}^{\dagger}}}{\sqrt{\big( \rho - \tau \big)^{\dagger}}}      }           } \bigg\}  \\ \\ \\   \times  \bigg[ \mathrm{Tr} \bigg\{               {\tiny              \frac{  \big( \widetilde{\rho - \tau } \big)^{\dagger} }{\big( \rho - \tau \big)^{\dagger}}  }     \bigg\} \bigg/  \mathrm{Tr} \bigg\{       {\tiny               \frac{ \rho - \tau }{ \widetilde{\rho - \tau}   }                     }                    \bigg\}      \bigg]   ,    {\mathrm{Tr}     \bigg\{  {\tiny   \frac{\sqrt{\widetilde{\tau}}}{\sqrt{\big( \rho - \tau \big)}}      }         }   \bigg\}  \\ \\ \\   \times  \bigg[ \mathrm{Tr} \bigg\{               {\tiny              \frac{  \big( \widetilde{\rho - \tau } \big)^{\dagger} }{\big( \rho - \tau \big)^{\dagger}}  }     \bigg\} \bigg/  \mathrm{Tr} \bigg\{       {\tiny               \frac{ \rho - \tau }{ \widetilde{\rho - \tau}   }                     }                    \bigg\}      \bigg]             \bigg\}^8      ,    \\ 
\end{align*}

}

\noindent for constants ${\tiny \mathcal{C} \neq \mathcal{C}^{\prime} \neq \mathcal{C}^{\prime\prime} \neq C \neq C^{\prime} \neq C^{\prime\prime}} $ taken sufficiently small, noiseless quantum states ${\tiny \rho,\tau}$ and noisy quantum states ${\tiny \widetilde{\rho}, \widetilde{\tau}}$, as demonstrated in the arguments for \textbf{Lemma} in \textit{4.1}. The upper bound on the trace distance from an application of the Gentle Measurement Lemma (abbreviated moving forwards as GML), [18], is applied to demonstrate that the main result,

{\small

\begin{align*}
  \mathrm{Td} \big(    {\tiny \widetilde{\mathrm{Exp}}^{A_{\lambda^{\prime}}}  \big( 1^{\lambda^{\prime}} , 1 \big)     }      ,   {\tiny \widetilde{\mathrm{Exp}}^{A_{\lambda^{\prime}}} \big( 1^{\lambda^{\prime}} , 0 \big)     }    \big)  \lesssim   {\tiny \mathrm{NEGL} \big( \lambda^{\prime} \big) }   , \\ 
\end{align*}

}

\noindent as stated in \textbf{Theorem}, holds. The notion of everlasting security under noise is formally introduced in \textbf{Definition} \textit{37}.

\section{QKD objects}

\noindent We introduce several objects associated with quantum key distribution.

\subsection{Security thresholds from idealized measurement}

\noindent We introduce additional objects discussed in {[10]}, beyond the projection operator and the $n$-qubit Pauli group in the previous section. One assumes that the final state that is produced by the QKD protocol, after implementing \textit{two-universal} hashing, takes the form,

{\small \begin{align*}
     \widetilde{\rho}_{W_A W_B C E} \equiv \ket{\bot\bot}    \bra{\bot\bot}_{W_A W_B} \otimes \widetilde{\rho}_{CE} \big( \bot \big) + \underset{w_A, w_B}{\sum} \big\{ \ket{w_A w_B} \bra{w_A w_B}_{W_A W_B}  \otimes \widetilde{\rho}_{CE} \big( w_A , w_B \big) \big\}    , 
\end{align*} } 

\noindent where the subscript of the above quantum state $\rho$ denotes the registers that Alice and Bob would use for outputting the desired secret key, which has a corresponding transcript of classical communication, $C$, and the symbol '$\bot$' denotes the abort message, after which the QKD hashing protocol would terminate.

\bigskip

\noindent \textbf{Definition} \textit{1}, {[10]} ($\epsilon$ \textit{-security of the QKD protocol}). A QKD protocol is said to be $\epsilon$-secure if the trace distance between  $\widetilde{\rho}_{W_A W_B C E}$, and,

{\small \begin{align*}
    \ket{\bot\bot} \bra{\bot \bot}_{W_A W_B} \otimes \widetilde{\rho}_{CE} \big( \bot \big)       +   \underset{w}{\sum} \frac{1}{\big| W \big|} \big\{ \ket{ww} \bra{ww}_{W_A W_B}    \big\}   \otimes \big( \widetilde{\rho}_{CE} - \widetilde{\rho}_{CE} \big( \bot \big) \big)      ,
\end{align*} }

\noindent is equal to $\epsilon$, where $\big| W \big|$ denotes the size of the secret key space.

\bigskip

\noindent \textbf{Definition} \textit{2}, {[10]} ($\epsilon$-\textit{correctness of the QKD protocol}). A QKD protocol is said to be $\epsilon$-correct if for all input states $\rho_{ABE}$ the probability,

{\small \begin{align*}
    \textbf{P} \big[ W_A \neq W_B \big] =     \underset{w_A \neq w_B}{\sum}     \mathrm{Tr} \big( \widetilde{\rho}_{CE}   \big( w_A , w_B \big) \big)     , 
\end{align*} }

\noindent that the keys outputted by Alice and Bob are not equal, is bounded by $\epsilon$.

\bigskip

\noindent \textbf{Definition} \textit{3}, {[10]} ($\epsilon$-\textit{closeness with respect to the trace distance}). Alice's key is $\epsilon$ \textit{secret} if for all input states $\rho_{ABE}$, the reduced output state $\widetilde{\rho}_{W_A C E}$ is $\epsilon$ close to the corresponding ideal state,

{\small \begin{align*}
   \ket{\bot} \bra{\bot}_{W_A} \otimes \widetilde{\rho}_{CE} \big( \bot \big) + \underset{w}{\sum} \frac{1}{\big| W \big| } \big\{   \ket{w} \bra{w}_{W_A} \big\}  \otimes \big( \widetilde{\rho}_{CE} - \widetilde{\rho}_{CE} \big( \bot \big) \big)    .
\end{align*} } 

\noindent In comparison to the outer product decomposition provided in the previous section, specifically with \textbf{Lemma}, maximally entangled states have previously been expressed through the following decompositions. Instead of obtaining an expansion for the maximally entangled state with elements of the Bell basis with \textit{both} elements of the base field, expansions with one element from the base field are given by:

\bigskip

\noindent \textbf{Lemma} \textit{7}, {[10]} (\textit{outer products of maximally entangled states}). For all $n$, $\alpha, \beta \in \textbf{F}^n_2$,

{\small \begin{align*}
 \underset{\beta^{\prime} \in \textbf{F}^n_2}{\sum} \ket{\psi_{\alpha \beta^{\prime}}} \bra{\psi_{\alpha \beta^{\prime}}} = \underset{z_A \in \textbf{F}^n_2}{\sum} \ket{z_A, z_A + \alpha} \bra{z_A , z_A + \alpha }   , \\ \\   \underset{\alpha^{\prime} \in \textbf{F}^n_2}{\sum} \ket{\psi_{\alpha^{\prime} \beta}} \bra{\psi_{\alpha^{\prime} \beta}} = \underset{x_A \in \textbf{F}^n_2}{\sum} H^{\otimes 2n} \ket{x_A , x_A + \beta} \bra{x_A , x_A + \beta}  H^{\otimes 2n}   .
\end{align*} }

\noindent As alluded to in previous remarks throughout the introduction, one must determine whether Quantum states manipulated in the hashing protocol exhibit "typical" errors. As a function of a subset $S$ of the field with two elements supported over $n$ qubits, denote the image of,

\begin{align*}
   f_S : \textbf{F}^n_2 \longrightarrow S \cup \big\{ \bot \big\}  ,
\end{align*}

\noindent with, 

{\small \[ \left\{\!\begin{array}{ll@{}l} 
\alpha \Longleftrightarrow \alpha \in  S , \\ \bot \text{ otherwise} , 
\end{array}\right. 
\]  }

\bigskip

\noindent corresponding to the set of errors over $S$. Simply put, $S$ denotes the set of possible acceptable errors, namely those corresponding to bit flip and phase flip errors. Besides quantifying the occurrence of such errors, the \textit{collision bound} is defined with the following probabilistic quantity:

\bigskip

\noindent \textbf{Definition} \textit{4}, {[10]} (\textit{collision bounds from hashing functions}). Denote a family of functions $\textbf{H}$, which can be expressed as a mapping with codomain $\textbf{X}$ and image $\textbf{Y}$. The \textit{two-universal} collision probability from the family of hashing functions $\textbf{H}$, $\epsilon$, satisfies,

{\small \begin{align*}
   \underset{h \in \textbf{H}}{\textbf{P}} \big[ {\tiny h \big( x \big) = h \big( x^{\prime} \big) }  \big]  \leq \epsilon . 
\end{align*} }

\noindent In the above inequality, the mass that the probability measure over each hashing function $h$ is distributed uniformly over the two-universal family $\textbf{H}$. Otherwise, the probability measure. over hashing functions assigns an equal mass of $\big| \textbf{Y} \big|^{-1}$ for each such $h$.

\bigskip

\noindent Below, we introduce the collision bound from POVMs for $\mathscr{U}$ and $\mathscr{V}$ isometries defined in \textbf{Definition} \textit{1}.

\bigskip

\noindent \textbf{Definition} \textit{2} (\textit{suitable collision probability bounds from POVMs over isometries}). Denote a family of hashing functions, $\mathscr{H}$, where $\mathscr{H} : \mathscr{X} \longrightarrow \mathscr{Y}$, for two finite sets $\mathscr{X}$ and $\mathscr{Y}$. The \textit{two-universal} collision probability bound, $\Theta > 0 $, satisfies,

\begin{align*}
  \mathscr{P}\mathscr{O}\mathcal{V}\mathscr{M}_{\mathscr{U}} \big[ H \in \mathscr{H} : H \big( x \big) = H \big( x^{\prime} \big)  \big] = \mathscr{P}\mathscr{O}\mathcal{V}\mathscr{M}_{\mathscr{V}} \big[ H \in \mathscr{H} : H \big( x \big) = H \big( x^{\prime} \big)  \big] \\ \textbf{P}_L \big[ H \in \mathscr{H} : H \big( x \big) = H \big( x^{\prime} \big)  \big]  \leq \Theta . 
\end{align*}

\noindent Given the \textit{collision bound} probability above, we conclude our overview of the results from {[10]} with the following statement of the security result. Below, denote the \textit{two-universal} QKD hashing protocol considered in {[10]} with $\pi \big( n , k , r \big)$, where $n$ denotes the total number of qubits that Alice or Bob initially transmit given a suitable encoding, $k$ denotes the size of the syndrome measurements, and $r$ denotes the number of bit flip, and phase flip, errors.

\bigskip

\noindent \textbf{Theorem} \textit{2}, {[10]} (\textit{security of the two-universal QKD hashing protocol}). Fix $n,k,r \in \textbf{N}$ such that $2n h \big( \frac{r}{n} \big) < 2k < n $. Then, the \textit{two-universal} QKD hashing protocol $\pi \big( n , k , r \big)$ is,

\begin{align*}
  2^{-\frac{k}{2} + n h ( \frac{r}{n} ) + \frac{5}{2}}  , 
\end{align*}

\noindent secure.

\bigskip

\noindent Determining how the security of the QKD hashing protocol depends upon the decomposition of outer products of the maximally entangled states is of interest to explore with the following objects. Intuitively, the security level of the hashing protocol will be shown to depend upon braket states,

\begin{align*}
  \bra{\mathcal{L}}  \big[ \otimes \textit{Product representations for} (\textit{Simulated})^{\dagger}, (\textit{Ideal})^{\dagger}, \textit{and Real isometries}    \big]             \ket{\mathcal{L}} \\ \equiv    \bra{\mathcal{L}}  \big[ (\textit{Simulated Isometry})^{\dagger} \big(\textit{Ideal Isometry}\big)^{\dagger} \textit{Real Isometry}  \big]             \ket{\mathcal{L}}  , 
\end{align*}

\noindent of the purified states of random matrices.

\bigskip

\noindent Such cryptographic notions were subsequently considered by the author in [14], specifically from the perspective of isometries that generate additional measurements besides those contained in idealizations.

\bigskip

\noindent \textbf{Definition} \textit{7} (\textit{parity check matrices from the marginals of probability distributions supported over random matrices}). The parity check matrices are defined as the following two marginal distributions,

{\small \begin{align*}
     \textbf{P}_L \big[ \cdot \big] \bigg|_{\textit{first column}} \equiv \mathcal{P}_1   ,  \\ \end{align*}

     \begin{align*} \textbf{P}_{(L^{-1} )^{\mathrm{T}}} \big[ \cdot \big] \bigg|_{\textit{second column}} \equiv \mathcal{P}_2  , 
\end{align*} }

\noindent with support given by the random matrices $L,\big( L^{-1} \big)^{\mathrm{T}}$ over the field with two elements are given as $\mathcal{P}_1$ and $\mathcal{P}_2$, respectively.

\bigskip

\noindent \textbf{Definition} \textit{8} (\textit{Pauli basis}). Denote,

{\small \begin{align*}
   \sigma_1 =    \begin{bmatrix}
   0 & 1   \\ 1 & 0 
    \end{bmatrix}  ,  \\  \\   \sigma_2 =    \begin{bmatrix}
   0 & i   \\ - i  & 0 
    \end{bmatrix}  , \\  \\   \sigma_3 = \begin{bmatrix}
    1 & 0  \\ 0 & -1 
    \end{bmatrix} , 
\end{align*} }

\noindent corresponding to the Pauli operations $\sigma_1$, $\sigma_2$ and $\sigma_3$, respectively.

\bigskip

\noindent The following QKD hashing protocol is closely related to the role of the two-universal functions, which in [10] allow for an identification of asymmetry in composable security.

\bigskip

\noindent \textbf{Definition} \textit{9} (\textit{hashing protocols for Quantum key distribution}). Denote,

{\small \begin{align*}
  x_A \equiv \textit{Alice's measured state that she obtains in the $\ket{+}$, $\ket{-}$ computational basis from Eve's state}  , \\ \\ x_B \equiv \textit{Bob's measured state that she obtains in the $\ket{+}$, $\ket{-}$ computational basis from Eve's state} , \\ \\ z_A \equiv    \textit{Alice's measured state that is initially distributed to her at the beginning of the hashing pro-} \\ \textit{tocol by Eve}        , \\ \\ z_B \equiv    \textit{Bob's measured state that is initially distributed to her at the beginning of the hashing pro-} \\ \textit{tocol by Eve}         . \\ 
\end{align*} }

\noindent For a QKD protocol $\pi^{\prime} \equiv \pi^{\prime} \big( n , k ,r \big)$, where $n$ denotes the total number of qubits in the state which Eve initially distributes to Alice and Bob, $k$ is the number of ancilla qubits in the state which Eve initially distributes to Alice and Bob, and $r$ is half of the number of phase flip, and bit flip, errors, the hashing protocol,



{\small \[ \left\{\!\begin{array}{ll@{}l} 
\textbf{Initialization}, \textit{(0)}: \textit{Alice and Bob initialize a system, } AB, \textit{ that they share between themselves,} \\ \textit{and  which they can compute the trace of with the operator } \mathrm{Tr}_{AB} \big[ \cdot \big], \textit{from n qubit states that} \\   \textit{ they receive from Eve}.  \\ \\  \textbf{Random matrix state purification}, \textit{(1)}: \textit{Alice and Bob pick a random matrix}, L, \textit{which they purify}
\\  \textit{to } \ket{\mathcal{L}}_{\textbf{L} \textbf{L}^{\prime}} \textit{ and which they can compute the trace of with the operator }  \mathrm{Tr}_{\textbf{L}^{\prime}} \big[ \cdot \big]. \\ \\  \textbf{Isometry}, \textit{(2)}: \textit{Alice and Bob apply the isometry } L_1 \textit{ to } z_A, \textit{ and to } z_B, \textit{respectively, which they can } \\  \textit{compute the trace of with the operator } \mathrm{Tr}_{U^{\prime}_A U^{\prime}_B} \big[ \cdot \big]  .  \\ \\   \textbf{Transpose of random matrix inverse}, \textit{(3)}:     \textit{Alice and Bob apply the second, and third, columns of } 
\\  \big( L^{-1} \big)^{\mathrm{T}} \textit{ to the results obtained from the previous step above, which they can compute the trace of} 
\\   \textit{with the operator } \mathrm{Tr}_{V^{\prime}_A V^{\prime}_B} \big[ \cdot \big]     . 
\\  \textbf{Parity check matrices computations}, \textit{(4)}:        \textit{Alice and Bob compute functions } g_1 \textit{ and } g_2 \textit{ of } \\  \textit{ the parity check matrices, which they can compute the trace of with the operator } \mathrm{Tr}_{S^{\prime} T^{\prime}} \big[ \cdot \big] ,  \\ \\    \textbf{Secret key output}, \textit{(5)}: \textit{If the two-universal hashing protocol is not aborted, Alice and Bob output} \\  \textit{ the secret key, which she first outputs as } w_B, \textit{and after which Bob accepts } w_B \\ + \big[ \textit{third column of } \big( L^{-1} \big)^{\mathrm{T}} \big] t \textit{ as a copy of, which they can compute the trace of} \\ \textit{ with the operator } \mathrm{Tr}_{W^{\prime}_A W^{\prime}_B} \big[ \cdot \big] .     
\end{array}\right. 
\]  }


\bigskip 

\noindent associated with $\pi^{\prime}$ is dependent upon the computation of the states,

\begin{align*}
 \ket{\mathcal{P}_1  z_A,  \mathcal{P}_1 z_A , \mathcal{P}_1  z_B,  \mathcal{P}_1 z_B ,  g_1 \big( \mathcal{P}_1 , \mathcal{P}_1 \big( z_A + z_B \big)  \big)    , \mathcal{P}_1 z_B     }_{U_A U^{\prime}_A U_B U^{\prime}_B S U^{\prime\prime}_B}      , \\ \\    \ket{\mathcal{P}_1  z_A,  \mathcal{P}_1 z_A , \mathcal{P}_1  z_B,  \mathcal{P}_1 z_B , \mathcal{P}_1 z_B   ,  g_1 \big( \mathcal{P}_1 , \mathcal{P}_1 \big( z_A + z_B \big) \big)              }_{U_A U^{\prime}_A U_B U^{\prime}_B  U_B U^{\prime\prime}_B  S }       , \end{align*}
 
 \begin{align*}     \ket{\mathcal{P}_1  z_A,  \mathcal{P}_1 z_A , \mathcal{P}_1  z_B,  \mathcal{P}_1 z_B ,  \mathcal{P}_1  z_A     , \mathcal{P}_1 z_B     }_{U_A U^{\prime}_A U_B U^{\prime}_B U^{\prime\prime}_A U^{\prime\prime}_B}          ,  \end{align*}
 
 \begin{align*} \ket{\mathcal{P}_1  z_A,  \mathcal{P}_1 z_A , \mathcal{P}_1  z_B,  \mathcal{P}_1 z_B , \mathcal{P}_1 z_B   ,  \mathcal{P}_1  z_A }_{U_A U^{\prime}_A U_B U^{\prime}_B U^{\prime\prime}_B U^{\prime\prime}_A  }        , \end{align*}
 
 \begin{align*}   \ket{\mathcal{P}_2  x_A,  \mathcal{P}_2 x_A , \mathcal{P}_2  x_B,  \mathcal{P}_2 x_B ,  g_2 \big( \mathcal{P}_1 , \mathcal{P}_2 \big( x_A + x_B \big) \big)  ,  \mathcal{P}_2  x_A  }_{U_A U^{\prime}_A U_B U^{\prime}_B T U^{\prime\prime}_A }       , \\ \\   \ket{\mathcal{P}_2  x_A,  \mathcal{P}_2 x_A , \mathcal{P}_2  x_B,  \mathcal{P}_2 x_B , \mathcal{P}_2 x_B   ,  g_2 \big( \mathcal{P}_1 , \mathcal{P}_2 \big( x_A + x_B \big) \big)         }_{U_A U^{\prime}_A U_B     U^{\prime}_B U^{\prime\prime}_A T     }    , \end{align*}
 
 \begin{align*}    \ket{\mathcal{P}_2  x_A,  \mathcal{P}_2 x_A , \mathcal{P}_2  x_B,  \mathcal{P}_2 x_B , \mathcal{P}_2  x_A  ,  \mathcal{P}_2  x_B  }_{U_A U^{\prime}_A U_B U^{\prime}_B U^{\prime\prime}_A U^{\prime\prime}_B }  , \\ \\   \ket{\mathcal{P}_2  x_A,  \mathcal{P}_2 x_A , \mathcal{P}_2  x_B,  \mathcal{P}_2 x_B , \mathcal{P}_2 x_B   ,  \mathcal{P}_2 x_A }_{U_A U^{\prime}_A U_B U^{\prime}_B U^{\prime\prime}_B U^{\prime\prime\prime}_B U^{\prime\prime}_A }       , 
\end{align*} 

\noindent where,

{\small \begin{align*}
  g_1 \big[ \cdot \big] : = \textit{Function of $\mathcal{P}_1$}  , \\ \\ g_2 \big[ \cdot \big] : = \textit{Function of $\mathcal{P}_2$} . 
\end{align*} }

\bigskip

\noindent \textbf{Definition} \textit{10} (\textit{POVMs corresponding to the two parity check matrices in the QKD hashing protocol}). Given a state $\ket{\psi} \equiv   \ket{\psi_{\alpha \beta}} \equiv \big[ \textbf{I} \otimes   \sigma^{\alpha^{\mathrm{T}}}_1 \sigma^{\beta^{\mathrm{T}}}_3       \big] \ket{\psi}$ supported over the field with two elements, in addition to a function $F$ of the set of anticipated errors in the QKD hashing protocol from bit flips and phase flips the isometries,

{\small \begin{align*}
  {\tiny  \mathscr{U} \equiv     \underset{\alpha \neq \beta^{\prime} \in \textbf{F}^n_2}{\sum}      \ket{\psi_{\alpha \beta^{\prime} }} \bra{\psi_{\alpha \beta^{\prime} }}  \otimes    \frac{1}{\sqrt{2}} } \bigg\{ {\tiny  \ket{F \big( \alpha \big) , F \big( \alpha + \beta^{\prime} \big)}_{S(S^{\prime} + T^{\prime})} +    \ket{F \big( \beta  \big) , F \big( \alpha + \beta^{\prime} \big)}_{T(S^{\prime} + T^{\prime})}  }   \bigg\}      , \\ \\    {\tiny   \mathscr{V} \equiv        \underset{\alpha^{\prime} \neq \beta^{\prime\prime} \in \textbf{F}^n_2}{\sum}  \ket{\psi_{\alpha^{\prime}  \beta^{\prime\prime} }} \bra{\psi_{\alpha^{\prime} \beta^{\prime\prime} }} \otimes   \frac{1}{\sqrt{2}}}  \bigg\{   {\tiny \ket{F ( \alpha^{\prime} + \beta^{\prime\prime} ) , F \big( \alpha^{\prime}  \big) }_{(S+T) S^{\prime}} +   \ket{F ( \alpha^{\prime} + \beta^{\prime\prime} ) , F \big( \beta^{\prime\prime} \big) }_{(S+T) T^{\prime}}              }   \bigg\}   ,  
\end{align*} }

\bigskip

\noindent the Positive Operator Valued Measurements (POVMs) corresponding to $\mathcal{P}_1$ and $\mathcal{P}_2$ above are denoted with,

{\small \begin{align*}
     \textbf{P}_L \big[  \textit{There exists a CSS code such that the first parity check matrix, from the finite dimensional} \\ \textit{representation of L, which can be used for error correction}       \big]    \equiv  \mathscr{P} \mathscr{O} \mathcal{V}  \mathscr{M}_{\mathscr{U}}      , 
\end{align*} } 

\noindent where,

\begin{align*}
 \underset{\mathscr{U}}{\sum } \mathscr{P} \mathscr{O} \mathcal{V}  \mathscr{M}_{\mathscr{U}} = \textbf{I}   , 
\end{align*}

\noindent and,

{\small \begin{align*}
       \textbf{P}_{ ( L^{-1} )^{\mathrm{T}}} \big[  \textit{There exists a CSS code such that the second parity check matrix, from the finite dimensional} \\ \textit{representation of $\big( L^{-1} \big)^{\mathrm{T}}$, which can be used for error correction}          \big]  \equiv          \mathscr{P} \mathscr{O} \mathcal{V}  \mathscr{M}_{\mathscr{V}}         , 
\end{align*} }

\noindent where,

{\small \begin{align*}
   \underset{\mathscr{V}}{\sum } \mathscr{P} \mathscr{O} \mathcal{V}  \mathscr{M}_{\mathscr{V}} = \textbf{I}  . 
\end{align*} }

\bigskip

\noindent \textbf{Theorem} (\textbf{Theorem} \textit{2}, [14]). Denote the hashing protocol with $\pi^{\prime} \big( n,k,r\big)$, along with some $C>0$. For the number of bit flip, and phase flip, errors $2r$, and total number of qubits $n$ that are encoded in states that Alice and Bob initially input to the QKD hashing protocol, the protocol is,

\begin{align*}
  2^{-\frac{k}{2} + n \frac{h}{2} ( \frac{r}{n} ) + \frac{5}{2} ( 5 - \frac{3}{2} ) + \mathrm{log}_2 \sqrt{C}}  ,
\end{align*}

\noindent secure.

\subsection{OTSs and the computational versus information-theoretic security framework}

\noindent We define objects associated with information-theoretic and computational security provided in [3].

\bigskip

\noindent \textbf{Definition} \textit{11} (\textit{Quantum machines with polynomial runtime}, [3]). Fix a security parameter $\lambda > 0$. Define a non-uniform Quantum polynomial-time (QPT) machine $\big\{ A_{\lambda} \big\}_{\lambda \in \textbf{N}}$ as a family of polynomial-size machines $A_{\lambda}$, where each such $A_{\lambda}$ is initialized with a polynomial-time advice state $\ket{\alpha_{\lambda}}$. Under a realization as a Completely Positive Trace Preserving (CPTP) map, the QPT is a concatenation of Quantum channels with corresponding input, output and work registers. Before presenting these definitions, as a matter of notation, we write,

{\small

\begin{align*}
\big\{ x \leftarrow \mathcal{X}   \big\}  , \\ \\ \big\{ y \leftarrow \mathcal{Y}   \big\} , \end{align*}

\begin{align*}    \big\{  1 \leftarrow A_{\lambda} \big( x \big)  \big\}      , \\ \\   \big\{  1 \leftarrow A_{\lambda} \big( y \big)  \big\}       , \\ 
\end{align*}

}

\noindent which respectively correspond to the outputs of the Quantum polynomial-time machines (QPTs), given choice of inputs distributed iid according to $\mathcal{X}$ and $\mathcal{Y}$, respectively. Furthermore, denote the vector of all $1$'s of length $\lambda$ as $\textbf{1}_{\lambda}$.

\bigskip

\noindent \textbf{Definition} \textit{12} (\textit{computational indistinguishability of QPTs}, [3]). We say that two QPTs are computationally indistinguishable if,

{\small

\begin{align*}
    \big|      \textbf{P} \big[   {\tiny   1 \leftarrow A_{\lambda} \big( x \big) :   x \leftarrow \mathcal{X}     }  \big]          -      \textbf{P} \big[   {\tiny  1 \leftarrow A_{\lambda} \big( y \big) :   y \leftarrow \mathcal{Y}  }       \big]          \big| = {\tiny \mathrm{negl} \big( \lambda \big)  }       , \\ 
\end{align*}

}

\noindent given the probability distributions $\mathcal{X}$, $\mathcal{Y}$ are computationally indistinguishable, for ${\tiny x \in \mathcal{X}}$ and ${\tiny y \in \mathcal{Y}}$.

\bigskip

\noindent The notion of computational indistinguishability, through the negligibility function above that is dependent upon the security parameter, is intriguing to connect with PKE. To this end we describe how suitable notions of distance, specifically through the trace distance, between two Quantum states is related to the min-entropy and collision probability.

\bigskip

\noindent \textbf{Definition} \textit{13} (\textit{the trace distance}, [3]). The trace distance, ${\tiny \mathrm{Td} \big[ \cdot , \cdot \big]}$, between two noiseless Quantum states $\rho$ and $\tau$ takes the form,

{\small

\begin{align*}
{\tiny  \mathrm{Td} \big[ \rho , \sigma \big]   =  \big| \big| \rho - \sigma \big| \big|_1 }  = \frac{1}{2} \mathrm{Tr} \bigg\{   \sqrt{ {\tiny \big( \rho - \tau \big)^{\dagger} \big( \rho - \tau \big) } }          \bigg\}   . \\ 
\end{align*}

}

\noindent \textbf{Definition} \textit{14} (\textit{the min-entropy}, [3]). The min-entropy, $H_{\infty} \big[ \cdot \big]$, of a random variable $X$ takes the form,

{\small

\begin{align*}
   H_{\infty} \big[ X \big] = - \mathrm{log} \bigg\{   \underset{x \in \mathcal{X}}{\mathrm{sup}} \textbf{P} \big[ \big\{ X = x \big\}  \big]     \bigg\}  . \\ 
\end{align*}

}

\noindent \noindent \textbf{Definition} \textit{15} (\textit{the averaged conditional min-entropy}, [3]). The averaged conditional min-entropy, $\widetilde{H_{\infty}} \big[ \cdot \big|
 \cdot \big]$, of the random variables $X, Z$ takes the form,

{\small

\begin{align*}
   \widetilde{H_{\infty}} \big[ X \big| Z 
\big] = - \mathrm{log} \bigg\{  \textbf{E}_z \bigg[ \underset{x \in \mathcal{X}}{\mathrm{sup}}     \textbf{P} \big[ \big\{ X = x \big\}  \big| \big\{ Z = z \big\}  \big]     \bigg]   \bigg\}  . \\ 
\end{align*}

}

\noindent \textbf{Definition} \textit{16} (\textit{collision probabilities from hashing functions that are not necessarily two-universal}, [3]). The hashing function $\mathrm{HASH}: \mathcal{X} \longrightarrow \mathcal{Y}$ is said to satisfy a collision probability of $\big| \mathcal{Y} \big|^{-1}$ if,

{\small

\begin{align*}
\textbf{P} \big[  {\tiny\mathrm{HASH} \big( x \big)    = \mathrm{HASH} \big( y \big) : x \neq y  }  \big]   \leq \big| \mathcal{Y} \big|^{-1}  , \\
\end{align*}

}

\noindent where $\textbf{P} \big[ \cdot \big] \equiv \underset{\mathrm{HASH}}{\textbf{P}} \big[ \cdot \big]$ (ie, $\textbf{P} \big[ \cdot \big] \equiv \textbf{P} \big[ \cdot \big] \big|_{\mathrm{HASH}}$) denotes the probability measure supported over hash functions.

\bigskip

\noindent \textbf{Definition} \textit{17} (\textit{PRFs}, [3]). A pseudo random function (PRF) is a mapping of the form,

{\small

\begin{align*}
    {\tiny  \mathrm{PRF} : \big\{ 0 , 1 \big\}^{\lambda} \times \big\{ 0 , 1 \big\}^{\lambda} } \longrightarrow  {\tiny  \big\{ 0 , 1 \big\}^{\lambda}  }     , \\ 
\end{align*}}

\noindent is computationally indistinguishable from a random function $f$.

\bigskip

\noindent \textbf{Definition} \textit{18} (\textit{computational indistinguishability between PRFs and random functions}, [3]). Denote $1^{\lambda}$ as a vector of all $1$s of length $\lambda$. Under a uniformly sampled random function $f$, $f$ is said to be computationally indistinguishable from a PRF if,

{\small

\begin{align*}
   A_{\lambda} \big( 1^{\lambda} \big)^{\mathrm{PRF} [ k , \cdot ] }      \approx   A_{\lambda} \big( 1^{\lambda} \big)^{f [ \cdot ] }        , \\ 
\end{align*}

}

\noindent where $k \leftarrow \big\{ 0 , 1 \big\}^{\lambda}$.

\bigskip

\noindent \textbf{Definition} \textit{19} (\textbf{Definition} \textit{2.5}, [3]: \textit{the noiseless one-time signature scheme}). The one-time signature (OTS) cryptographic scheme is defined with a tuple of algorithms $\big( \mathrm{SGen}, \mathrm{Sign} , \mathrm{Ver} \big)$ (ie, a tuple of protocols corresponding to the generation of Secret Keys, Signature Keys, and Verification Keys, respectively), satisfying:

{\small

\begin{itemize}
    \item[$\bullet$] \textit{The output from the cryptographic protocol for generating the Quantum Secret Key}. Given input $1^{\lambda}$, $\big( \mathrm{vk}, \mathrm{sk} \big) \leftarrow \mathrm{SGen} \big( 1^{\lambda} \big)$, namely that the cryptographic protocol $\mathrm{SGen}$ outputs a tuple of the verification key and secret key, respectively.

    \bigskip

        \item[$\bullet$] \textit{The output from the cryptographic protocol for generating the Signature Key}. Given the tuple of inputs $\big( \mathrm{sk}, m \big)$, ie a candidate secret key generated by $\mathrm{SGen}$ and message $m$ $\sigma \leftarrow \mathrm{Sign} \big( \mathrm{sk} , m \big)$, namely that the cryptographic protocol $\mathrm{Sign}$ outputs a signature key.

          \bigskip

 \item[$\bullet$] \textit{The output from the cryptographic protocol for generating the Verification Key}. Given the signature key $\sigma$ generated by $\mathrm{Sign}$, as well as $\big( m , \sigma \big)$, $\big\{ 0 , 1 \big\} \leftarrow \mathrm{Ver} \big( \mathrm{vk} , m , \sigma \big)$. Namely the polynomial-time verification protocol $\mathrm{Ver}$ outputs $0$ or $1$, which respectively correspond to accepting or rejecting the verification key given a message and signature key.

\end{itemize}

}

\bigskip

\noindent Given the OTS scheme defined above, it is of interest to formulate how correctness of this protocol differs from the notion of correctness provided in \textbf{Definition} \textit{2} for other QKD dependent protocols.

\bigskip

\noindent \textbf{Definition} \textit{20} (\textbf{Definition} \textit{2.5}, [3]: \textit{correctness of the OTS scheme}). The OTS scheme provided in the previous definition above is said to be \textit{correct} if,

{\small

\begin{align*}
 \textbf{P} \big[ {\tiny    1 = \mathrm{Ver} \big( \mathrm{vk} , m ,  \mathrm{Sign} \big(       \mathrm{sk} , m    \big)  \big)  :                \big( \mathrm{vk} , \mathrm{sk} \big) \leftarrow \mathrm{SGen} \big( 1^{\lambda} \big)       }     \big] = 1    , \\ 
\end{align*}

}

\noindent for some $\lambda \in \textbf{R}$ and message $m$.

\bigskip

\noindent Related to the correctness notion provided for the PKE-QKD protocol above is the following quantification of unforgeability.

\bigskip

\noindent \textbf{Definition} \textit{21} (\textbf{Definition} \textit{2.6}, [3]: \textit{the notion of strong existential unforgeability}). The OTS scheme provided in \textbf{Definition} \textit{19} is said to satisfy the strong existential unforgeability property, meaning that, given an instance of the OTS cryptographic scheme, there exists a negligibility function $\mathrm{negl} {\tiny \big( \cdot \big)}$ such that for all QPTs $\big\{ A_{\lambda} \big\}_{\{\lambda \in \textbf{N}\}}$ and messages $m$, one has that,

{\small

\begin{align*}
  \textbf{P} \bigg[  {\tiny 1 = \mathrm{Ver} \big( \mathrm{vk} , m^{*} , \sigma^{*} \big)} \textit{ with } {\tiny \big( m , \sigma \big) \neq \big( m^{*} , \sigma^{*} \big) :                     \big\{ \big( \mathrm{vk} , \mathrm{sk} \big) \leftarrow \mathrm{SGen} \big( 1^{\lambda} \big) \big\} , \big\{ \sigma \leftarrow \mathrm{Sign} \big( \mathrm{sk} , m \big) \big\} , }  \\  {\tiny \big\{        \big( m^{*} , \sigma^{*} \big) \leftarrow A_{\lambda} \big( \mathrm{vk} , m , \sigma \big)       \big\}    }   \bigg] = {\tiny \mathrm{negl} \big( \lambda \big) }  . \\ 
\end{align*}

}

\noindent The above notion of unforgeability for the OTSs is vaguely reminiscent of the conditions enforced for collision probabilities of the two-universal, and ordinary, hash functions provided in \textbf{Definition} \textit{4} and \textbf{Definition} \textit{16}, respectively, in which the output of the evaluation of the hashing function is equal albeit the fact that the inputs from a probability distribution satisfy $ \big\{ x \neq x^{\prime} \big\} $, akin to how the tuple of messages and signatures satisfy $ \big\{ \big( m , \sigma \big) \neq \big( m^{*} , \sigma^{*} \big) \big\}$. To make use of the above unforgeability notion, we demonstrate how Malavolta and Walter in [3] define the PKE protocol below.

\bigskip

\noindent \textbf{Definition} \textit{22} (\textbf{Definition} \textit{3.1}, [3]: \textit{the Quantum PKE cryptographic protocol from the OTS cryptographic protocol}). The Quantum PKE (QPKE) protocol is defined with a quadruple $\big( \mathrm{SKGen}, \mathrm{PKGen}, \mathrm{Enc} , \mathrm{Dec} \big)$ of algorithms such that:

{\small \begin{itemize}
    \item[$\bullet$] \textit{Generation of the Quantum secret key in polynomial time}. Given input $1^{\lambda}$, $\mathrm{sk} \leftarrow \mathrm{SKGen} \big( 1^{\lambda} \big)$, namely the secret key $\mathrm{sk}$ is generated from a vector of $1$s of length $\lambda$. 

    \bigskip

      \item[$\bullet$] \textit{Generation of the Public secret key in Quantum polynomial time}. Given the input from the protocol above, $\big( \rho , \mathrm{pk} \big) \leftarrow \mathrm{PKGen} \big( \mathrm{sk} \big)$, namely a tuple of a Quantum state $\rho$ and public key $\mathrm{pk}$ are generated in Quantum polynomial time. 

      \bigskip

        \item[$\bullet$] \textit{Generation of the ciphertext in Quantum polynomial time}. Given a tuple $\big( \rho , \mathrm{pk} , m \big)$ as input, the Quantum polynomial time algorithm $\mathrm{Enc}$ outputs the ciphertext $\mathrm{ct}$, where $ \mathrm{ct} \leftarrow  \mathrm{Enc} \big( \rho , \mathrm{pk} , m \big) $.

 \bigskip 
 
          \item[$\bullet$] \textit{Generation of the message in Quantum polynomial time}. Given the input of a secret key $\mathrm{sk}$ and ciphertext $\mathrm{ct}$ $m \leftarrow \mathrm{Dec} \big( \mathrm{sk} , \mathrm{ct} \big)$, namely the Quantum polynomial time $\mathrm{Dec}$ output a message $m \in \big\{ 0 , 1 \big\}$.
    
\end{itemize} }

\bigskip

\noindent \textbf{Definition} \textit{23} (\textbf{Definition} \textit{3.1}, [3]: \textit{correctness of the QPKE scheme}). A QPKE scheme is said to be \textit{correct} if,

{\small

\begin{align*}
  \textbf{P} \bigg[ {\tiny  m = \mathrm{Dec} \big( \mathrm{sk} , \mathrm{ct} \big) : \big\{ \mathrm{sk} \leftarrow \mathrm{SKGen} \big( 1^{\lambda} \big) \big\} , \big\{ \big( \rho , \mathrm{pk} \big) \leftarrow \mathrm{PKGen} \big( \mathrm{sk} \big) \big\} , \big\{   \mathrm{ct} \leftarrow \mathrm{Enc} \big( \rho , \mathrm{pk} , m \big)     \big\}       }    \bigg] = 1   , \\ 
\end{align*}

}

\noindent given a Quantum state $\rho$, public key $\mathrm{pk}$, message $m$, secret key $\mathrm{sk}$ and ciphertext $\mathrm{ct}$.

\bigskip

\noindent We conclude with the statement of everlasting security, which stipulates that an adversary Eve can only receive \textit{one} copy of the secret keys that Alice and Bob wish to share.

\bigskip

\noindent \textbf{Definition} \textit{24} (\textbf{Definition} \textit{3.1}, [3]: \textit{QPKE everlasting security}). A QPKE cryptographic protocol is said to satisfy the notion of \textit{everlasting} security if,

{\small

\begin{align*}
  \mathrm{Td} \big[    \mathrm{Exp}^{A_{\lambda}}  \big( 1^{\lambda} , 1 \big)           ,   \mathrm{Exp}^{A_{\lambda}} \big( 1^{\lambda} , 0 \big)        \big]  =  {\tiny \mathrm{negl} \big( \lambda \big) }   , \\ 
\end{align*}

}

\noindent for some $\lambda \in \textbf{R}$ and random experiments $\mathrm{Exp}^{A_{\lambda}}  \big( 1^{\lambda} , 1 \big) $ and $\mathrm{Exp}^{A_{\lambda}} \big( 1^{\lambda} , 0 \big)$, each of which take as input $1^{\lambda}$ in addition to a bit initialize to $1$ and to $0$, respectively.

\bigskip

\noindent The fact that the negligibility function can be expressed with respect to the trace distance is not only extremely informative, but also helpful for establishing comparisons with respects to other norms in Quantum Information theory (as mentioned for multiplayer parallel repetition in [16]). Besides these observations, it is also important to generalize the following characterization as the advantage between two resources with their trace distance.

\bigskip

\noindent \textbf{Definition} \textit{25} (\textit{the trace distance between two Quantum states is equal to the advantage between their respective resources $\textit{Proof of Theorem 3.4}$, [3]}). Fix some $i \in \textbf{N}$. The advantage at $i$, ${\tiny \mathrm{Adv} \big( i \big)}$, is equal to the  following trace variation distance,

{\small

\begin{align*}
\mathrm{Adv} \big( i \big) \equiv         \mathrm{Td} \big[ {\tiny   \mathrm{Hyb}^{\mathcal{A}_{\lambda}}_0     ,   \mathrm{Hyb}^{\mathcal{A}_{\lambda}}_1  }   \big]        , \\ 
\end{align*}

}

\noindent where:

{\small

\begin{itemize}
    \item[$\bullet$] \textit{Initialization}. Initialize the bit $b$ where $b \in \big\{ 0 , 1 \big\}$, 

    \bigskip

     \item[$\bullet$] \textit{Hybrid experiments where $b \equiv 0$}. Define $\mathrm{Hyb}^{\mathcal{A}_{\lambda}}_0$ as the hybrid experiment generated from the QPT machine $\mathcal{A}_{\lambda}$,

     \bigskip

      \item[$\bullet$] \textit{Hybrid experiments where $b \equiv 1$}. Define $\mathrm{Hyb}^{\mathcal{A}_{\lambda}}_1$ as the hybrid experiment generated from the QPT machine $\mathcal{A}_{\lambda}$.

\end{itemize}

}

\noindent \textbf{Definition} \textit{26} (\textit{the condition of privacy expressed through a correspondence between the negiligibility and the trace distance,} \textbf{Definition} \textit{\textit{4.2}, [3]}). A QKD-QPKE protocol is said to satisfy \textit{everlasting} security if the notion of privacy is related to the following characterization between the negilibility and the trace distance,

{\small

\begin{itemize}
    \item[$\bullet$] \textit{Privacy of the QKD-QPKE protocol}. The existence of the following negligible function, given a choice of $\lambda \in \textbf{R}$ and QPT as introduced in previous results above, takes the form,

    {\small
    
    \begin{align*}
  \mathrm{Td} \big[   \big\{            {\tiny  \mathrm{QKDSec}^{\mathcal{A}_{\lambda}}       , k_0 , k_1 } \big\}       ,     \big\{  {\tiny \mathrm{QKDSec}^{\mathcal{A}_{\lambda}}       , \widetilde{k_0} , \widetilde{k_1} }  \big\}      \big]    = {\tiny \mathrm{negl} \big( \lambda \big)  }  . \\
    \end{align*}

    }

\end{itemize}

}

\noindent where:

{\small

\begin{itemize}
    \item[$\bullet$] \textit{Experiment initialization}. Define $k_0 \neq \widetilde{k_0}$ and $k_1 \neq \widetilde{k_1}$ as the noisy collection of parameters initialized in the two above experiments for generating the Quantum secret key.

    \bigskip

     \item[$\bullet$] \textit{Generating the experiment with respect to a Quantum polynomial time machine and an initialization of parameters}. Denote an experiment generated by the QPT $\mathcal{A}_{\lambda}$ as $\mathrm{QKDSec}^{\mathcal{A}_{\lambda}}$. 

\end{itemize}

}

\subsection{The negligibility as the indistinguishability: a constructive resource-theoretic approach }

\noindent We discuss connections between the PKE protocol introduced in the previous subsection with the resource-theoretic approach provided in [9].

\bigskip

\noindent \textbf{Definition} \textit{27} (\textbf{Definition} \textit{1}, [9]: \textit{the indistinguishability metric as a difference between probabilities over disjoint random experiments, as opposed to through the negiligibility function}). The distinguishing advantage (ie, indistinguishability) of a distinguisher $D$ for the systems $U$ and $V$ is defined as,

{\small

\begin{align*}
  \Delta^D \big( U , V \big) = \big| \textbf{P}^{DV} \big[ {\tiny W = 1  } \big]       -       \textbf{P}^{D U} \big[ {\tiny W = 1 }  \big]      \big|   , \\ 
\end{align*}

}

\noindent for the final output bit $W \in \big\{ 0 , 1 \big\}$ of $D$. The distinguishing advantage can be obtained through optimizing over the set of all distinguishers, ie through ${\tiny \Delta^{\mathcal{D}} \big( U , V \big) \equiv \underset{D \in \mathcal{D}}{\mathrm{sup}} \Delta^{D} \big( U , V \big)}$. If $\mathcal{D}_q$ denotes the set of distinguishers which issue at most $q > 0 $ queries, the indistinguishability is given by ${\tiny \Delta^{\mathcal{D}_q} \big( U , V \big) \equiv \underset{D \in \mathcal{D}_q }{\mathrm{sup}} \Delta^{D}_q \big( U , V \big)}$. Finally, denote $\mathcal{E}$ as the set of all distinguishers that are efficient.

\bigskip

\noindent \textbf{Definition} \textit{28} (\textbf{Definition} \textit{2}, [9]: \textit{the notions of security and availability}). A protocol ${\tiny \pi \equiv \big( \pi_A , \pi_B, \pi_E \big)}$ for Alice, Bob and Eve, respectively, is said to construct a resource $R$ from a resource $S$ if it satisfies:

{\small \begin{itemize}
    \item[$\bullet$] \textit{Security}. Fix some $\epsilon > 0$ and a simulator $\sigma_E$ for Eve to use on the ideal environment. The first property of security is given by the condition ${\tiny \Delta^D \big( \pi^A \pi^B R  ,  \sigma^E S  \big) \leq \epsilon}$.

    \bigskip

    \item[$\bullet$] \textit{Availability}. Fix some $\epsilon > 0$. The second property of availability is given by the condition ${\tiny \Delta^D \big( \pi^A \pi^B \bot^E R , \bot^E S  \big) \leq \epsilon         }$.
\end{itemize} }

\bigskip

\noindent \textbf{Theorem} (\textbf{Theorem} \textit{1}, [9]: \textit{composition result for the three party setting}). Let $R,S,T$ be three resources that are used by Alice, Bob, and Eve, respectively, as well as the protocols ${\tiny \pi \equiv \big( \pi_1 , \pi_2 \big)}$ and ${\tiny \psi \equiv \big( \psi_1, \psi_2 \big)}$ where $\pi$ constructs $S$ from $R$ and $\psi$ constructs $T$ from $S$. If $\pi$ performs the desired construction with error ${\tiny\epsilon_{\pi}}$ and $\psi$ performs the desired construction with error $\epsilon_{\psi}$, the class of efficient distinguishers $\mathcal{E}$ is said to satisfy the following composition result:

{\small

\begin{itemize}

\item[$\bullet$] \textit{Closure under composition}. The protocols $\pi$ and $\psi$ are said to be \textit{closed} with respect to the composition operator $\circ$ if ${\tiny \big( \pi_1 \circ \psi_1 , \pi_2 \circ \psi_2 \big)}$ constructs $T$ from $R$ with the additive error $\epsilon_{\pi} + \epsilon_{\psi}$. 

\bigskip

\item[$\bullet$] \textit{Closure under parallelization}. The three aforementioned resources defined above are said to \textit{closed} with respect to the parallelization operator ${\tiny \big| \big|}$ if ${\tiny \big( \pi_1  \big| \big|  \textbf{I}, \pi_2 \big| \big| \textbf{I} \big)}$ performs the desired construction of ${\tiny S \big| \big| U}$ from ${\tiny R \big| \big| U }$ with error ${\tiny \epsilon_{\pi}}$ and if ${\tiny \big(  \textbf{I} \big| \big| \psi_1 ,  \textbf{I} \big| \big| \psi_2  \big)}$ performs the desired constructions of ${\tiny \big| \big| S}$ from ${\tiny U \big| \big| R}$ with error ${\tiny \epsilon_{\psi}}$. 
    
\end{itemize}

}

\bigskip

\noindent \textbf{Definition} \textit{29} (\textit{message authentication codes}, [9]). The message authentication code (MAC) is defined as a tuple of stateless converters (ie, converters that are independent of the state in of the previous times) $\big( \mathrm{tag} , \mathrm{vrf} \big)$, where $\mathrm{tag}$ and $\mathrm{vrf}$ respectively correspond to the tag and verification keys from the  key space. The converter $\mathrm{tag}$ takes as input some key distributed iid from the secret key space through a message $M$, while the converter $\mathrm{vrf}$ takes as input the tuple of pairs distributed iid from the direct product of the message space with the tag space, $M \times T$.

\bigskip

\noindent There are a wide variety of methods to generate authentication tags before a candidate Quantum secret key is agreed upon by Alice and Bob. One such method through evaluation of the hash function is presented in [15]. Below we describe how the notion of strong unforgeability is formulated for chosen message attacks (CMAs).

\bigskip

\noindent \textbf{Definition} \textit{30} (\textit{MACs which satisfy the strong unforgeability property under a chosen message attack from the adversary}, [9]). The MAC is said to satisfy the strong unforgeability property, hence making it a SUF-CMA, if, after having received $q = q_A + q_E$ queries, where $q_A$ and $q_E$ respectively denote the queries from Alice and Eve, if the MAC satisfies:

{\small
\begin{itemize}
    \item[$\bullet$] \textit{Decomposition with respect to $q$}. $S_q$, the with respect to the total number of queries $q = q_A + q_E $, can be decomposed as,

{\small
\begin{align*}
    S_q = S_{q-1} \vee \big(\big\{  {\tiny F_{q_B} = 1  } \big\}  \wedge   \big\{    {\tiny  \forall i \leq q_A : \big( M_i , T_i \big) \neq V_{q_B}  }      \big\}        \big)       , \\ 
\end{align*}

}

\noindent corresponding to the fact that the value of $S_q$ is not only dependent upon the value of some function of Bob's queries $q_B$, but also upon each $M_i$ and $T_i$ taken over Alice's queries $q_A$. 

    \bigskip

    \item[$\bullet$] \textit{Authentication cryptographic protocol for MACs}. Given a MAC the authentication scheme is given by a tuple. $\big( \mathrm{aut}, \mathrm{chk} \big) $, which consists of the steps:

    \begin{itemize}
        \item[$\bullet$] \textit{Generation of a tag}. A tag associated with each message distributed iid over the message space $\mathcal{M} \times \textbf{N}$, key space $\mathcal{K}$ and authentication tag space $\mathcal{T}$ is obtained through the evaluation of the hashing function against unique inputs.

        \bigskip

        \item[$\bullet$] \textit{Counter through the system $\mathrm{Seq}$}. Start an initial counter with $i \equiv 0$. The counter is incremented by $1$ every time there is an additional evaluation of the hashing function at each $M_i$ as described in the previous step of the protocol.

 \bigskip

        \item[$\bullet$] \textit{Counter through the system $\mathrm{Seq}^{\prime}$}. Start an initial counter with $j \equiv 0$. The counter is incremented by $1$ every time there is a message $m_j$ produced at time $t^{\prime}_j$ for which $\big( \big( m^{\prime}_j , j^{\prime} \big) , t^{\prime}_j \big)$ belongs to the set of verification tags for each $m^{\prime}_j$.

    \end{itemize}

     \bigskip

    \item[$\bullet$] \textit{Termination of the protocol}. The protocol is said to terminate if there exists countably many counters for $\mathrm{Seq}$ and $\mathrm{Seq}^{\prime}$ defined above.  
    
\end{itemize}

}

\bigskip

\noindent \textbf{Definition} \textit{31} (\textbf{Definition} \textit{5}, [9]: \textit{malleability of transformations of the plaintext space}). A function $F$ supported over the plaintext space is said to be malleable if:

{\small \begin{itemize}
    \item[$\bullet$] \textit{(1). Forwarding error probability}. The forwarding error probability of $F$ is given by the existence of some ${\tiny \delta \big( q \big) >0}$, for ${\tiny m_{q_A} \neq \widetilde{m}_{q_A} \in \mathcal{M}_{q_A}}$, ${\tiny m^{\prime}_{q_E} \neq \widetilde{m^{\prime}}_{q_E} \in \mathcal{M}_{q_E}}$ with ${\tiny \big| m_{q_A} \big| = \big| \widetilde{m}_{q_A} \big|}$ and ${\tiny \big| m^{\prime}_{q_E} \big| = \big| \widetilde{m^{\prime}}_{q_E} \big|}$ so that,

    {\small
    
    \begin{align*}
         \big|     \textbf{P} \big[ {\tiny  F \big( m^{q_A} , m^{\prime}_{q_E} \big)  = m_{q_E + 1}   }   \big]     -      \textbf{P} \big[  {\tiny  F \big( \widetilde{m}^{q_A} , \widetilde{m^{\prime}}_{q_E} \big)  = \widetilde{m}_{q_E + 1}   }     \big]       \big| \leq \delta \big(    q \big)      . \\ 
    \end{align*}

    }

    \bigskip

       \item[$\bullet$] \textit{(2). Deleting error probability}. The deleting error probability of $F$ is given by the existence of some ${\tiny \delta \big( q \big) >0}$, where ${\tiny m^{q_A} \in \mathcal{M}^{q_A}}$ ${\tiny \forall m^{\prime}_{q_E} \in \mathcal{M}^{q_E}}$,

{\small

\begin{align*}
      \textbf{P} \big[ {\tiny F \big( m_{q_A} , m^{\prime}_{q_E} \big) = m_{q_E + 1}   }   \big] \leq \delta \big( q \big)      . \\ 
\end{align*}

}

       \bigskip

         \item[$\bullet$] \textit{(3). Reconstruction error probability}. The reconstruction error probability of $F$ is given by the existence of some ${\tiny \delta \big( q \big) > 0}$ where ${\tiny m^{q_A} \in \mathcal{M}^{q_A}}$ ${\tiny \forall m^{\prime}_{q_E} \in \mathcal{M}^{q_E}}$,

         {\small

         \begin{align*}
             \big|   \textbf{P} \big[      {\tiny    F \big( m_{q_A} , m^{\prime}_{q_E} \big) = m_{q_E}    }  \big]       -  \textbf{P} \big[   {\tiny  R \big( M \big) = m_{q_E + 1}    }     \big]       \big| \leq \delta \big( q \big)                 .  \\ 
         \end{align*}
         
         }
    
\end{itemize} }

\noindent where the message space ${\tiny M \in \big\{     m_1 , \cdots , m_{q_A} , m^{\prime}_1 , \cdots , m^{\prime}_{q_B} , \widetilde{m}_1 , \cdots , \widetilde{m}_{q_E}      \big\} \backslash \big\{ m_{q_E + 1 } \big\}}$ and an algorithm $R$.

\section{Incorporating noise through malleability of functions on the plaintext}

\noindent We introduce several objects associated with the noisy Quantum public key encryption protocol, beginning from the completely positive trace preserving map. 

\subsection{Noisy projection operators for computational and everlasting securities, beginning from Completely Positive Trace Preserving maps}

\noindent We demonstrate how noise can be introduced through the following objects, beginning with noisy QPTs (NQPTs). To this end fix a noisy security threshold $\lambda^{\prime}$, where $\lambda^{\prime} > \lambda$.

\bigskip

\noindent \textbf{Definition} \textit{32} (\textit{a counterpart to the noiseless QPT introduced in $\textbf{Definition}$ $\textit{11}$}). Define an NQPT as a noisy Quantum machine with polynomial runtime such that:

{\begin{itemize}
    \item[$\bullet$] \textit{Initialization}. Each NQPT requires an initially noisy input state $\ket{\alpha^{\prime}_{\lambda^{\prime}}}$, for $\lambda^{\prime} \in \textbf{N}$.

    \bigskip

  \item[$\bullet$] \textit{Specification of the CPTP map}. For each choiceof $\lambda^{\prime}$ introduced above, the CPTP map for outputting a concatenated state for each NQPT satisfies:

  \begin{itemize}
    \item[$\bullet$] \textit{CPTP positivity}. The image of the CPTP map into ${\tiny \mathcal{B} \big[ \mathcal{H}_A \big]}$, the Hilbert space associated with states that Alice transmits across a noisy Quantum channel given a choice of encoding, is strictly positive iff the input probability itself is strictly positive.

\bigskip

\item[$\bullet$] \textit{CPTP complete positivity}. A CPTP map is completely positive iff the tensor product of the CPTP map with the identity operator is positive.

    \noindent

    \bigskip

     \item[$\bullet$] \textit{CPTP maps are linear}. One has that,

    {\small \begin{align*}
  \Phi \big[ {\tiny \alpha \rho + \beta \sigma  } \big] =  \alpha \Phi \big[ {\tiny \rho} \big]  + \beta  \Phi \big[  {\tiny\sigma }  \big]                ,
     \end{align*} } 

     \noindent for $\alpha, \beta \in \textbf{C}$, and,

       \begin{align*}
      \rho \equiv  \textit{State distributed to Eve with support over $\mathcal{H}_A $}  , \\ \\ \sigma \equiv \textit{State distributed to Eve with support over $\mathcal{H}_B $} . 
    \end{align*}

     \bigskip

      \item[$\bullet$] \textit{CPTP tensor product closure}. Denote $\Phi^{\prime}$ as another CPTP which is not equal to $\Phi$. Then $\Phi^{\prime} \otimes \Phi$ is also CPTP.

\bigskip

      \item[$\bullet$] \textit{CPTP composition closure}. Denote $\Phi^{\prime}$ as another CPTP which is not equal to $\Phi$. Then the composition $\Phi^{\prime} \circ \Phi$ is also CPTP.

      \bigskip

       \item[$\bullet$] \textit{CPTP contractivity with respect to the trace norm}. Eve's error probability is shown to satisfy the above inequality with respect to the binary entropy function as a result of the \textit{CPTP contractivity property}, which states,

       {\small \begin{align*}
        \big| \big| \Phi \big[ {\tiny \rho }  \big] - \Phi \big[ {\tiny \sigma }  \big] \big| \big|_1 \leq \big| \big| {\tiny \rho - \sigma }  \big| \big|_1    .
       \end{align*} } 

\noindent Hence,

{\small \begin{align*}
  P_{\mathrm{FA},\mathrm{E}} \gtrsim  \mathrm{log} \bigg[  {\tiny \frac{1}{N} }  \bigg] \big[ \mathrm{log} M - \chi_E - 1 \big]  .
\end{align*} }

       \noindent as provided in \textbf{Lemma} \textit{3.2.3}, [17], for,

       {\small \begin{align*}
         \rho_0 \equiv  \textit{State distributed to Eve with support over $\mathcal{H}_A $}  , \\ \\ \rho_1 \equiv \textit{State distributed to Eve with support over $\mathcal{H}_B $} . 
       \end{align*} }

    \item[$\bullet$] \textit{CPTP data-processing inequality}. One has that,

    {\small \begin{align*}
       D \big[ \rho \big| \big| \sigma \big] \geq   D \big[ \Phi \big[ \rho \big] \big| \big| \Phi \big[ \sigma \big] \big]  ,
    \end{align*} } 

    \noindent for,

   {\small \begin{align*}
      \rho \equiv  \textit{State distributed to Eve with support over $\mathcal{H}_A $}  , \\ \\ \sigma \equiv \textit{State distributed to Eve with support over $\mathcal{H}_B $} . 
    \end{align*} }

\end{itemize}

      \item[$\bullet$] \textit{Noisy concatenation with CPTP maps}. Lastly, the NQPT is obtained by the concatenation of a Quantum channel with each $\ket{\alpha^{\prime}_{\lambda^{\prime}}}$, output and work registers.

\bigskip

\noindent As mentioned previously in the noiseless setting, for the below noisy setting, as a matter of notation we write,

{\small

\begin{align*}
\big\{ \widetilde{x} \leftarrow \mathcal{X}   \big\}  , \\ \\ \big\{ \widetilde{y} \leftarrow \mathcal{Y}   \big\} ,  \end{align*}

\begin{align*} \big\{  1 \leftarrow A_{\lambda^{\prime}} \big( \widetilde{x} \big)  \big\}      , \\ \\   \big\{  1 \leftarrow A_{\lambda^{\prime}} \big( \widetilde{y} \big)  \big\}       , \\ 
\end{align*}

}

\noindent which correspond to the iid noisy inputs $\widetilde{x}$ and $\widetilde{y}$ from $\mathcal{X}$ and $\mathcal{Y}$, respectively, in addition to the outputs ${\tiny A_{\lambda^{\prime}} \big( \widetilde{x} \big)}$ and ${\tiny A_{\lambda^{\prime}} \big( \widetilde{y} \big)}$, respectively, for $\lambda^{\prime} > \lambda$.

        \bigskip 
    
\end{itemize}}

\noindent \textbf{Definition} \textit{33} (\textit{a counterpart to the noiseless formulation of the indistinguishability provided in $\textbf{Definition}$ $\textit{12}$}). Denote $\mathcal{P} \equiv \mathcal{P} \big[ \cdot \big]$ as the probability measure over the space of NQPTs and fix a security parameter $\lambda^{\prime}$ of the noisy QPKE-QKD protocol. Define,

{\small

\begin{align*}
            \big|      \mathcal{P} \big[    {\tiny  1 \leftarrow A_{ \lambda^{\prime}} \big( \widetilde{x} \big) :   \widetilde{x} \leftarrow \mathcal{X}      } \big]          -      \mathcal{P} \big[    {\tiny 1 \leftarrow A_{ \lambda^{\prime}} \big( \widetilde{y} \big) :   \widetilde{y} \leftarrow \mathcal{Y}    }     \big]          \big| = {\tiny \mathrm{NEGL} \big( \lambda^{\prime} \big)   }                 , \\ 
\end{align*}} 

\noindent given the negligibility function under noise, $\mathrm{NEGL} {\tiny \big( \cdot \big)}$,  corresponding to the noisy indistinguishability, ie the indistinguishability with respect to an NQPT.

\bigskip

\noindent \textbf{Definition} \textit{34} (\textit{a counterpart to the noiseless formulation of a PRF in $\textbf{Definition}$ $\textit{17}$}). Define a noisy PRF (NPRF) as,

{\small

\begin{align*}
       \mathrm{NPRF}   : \big\{ 0 , 1 \big\}^{\lambda^{\prime}} \longrightarrow    \big\{ 0 , 1 \big\}^{\lambda^{\prime}}           , \\ 
\end{align*}

}

\noindent given the same choice of $\lambda^{\prime}$ introduced in $\textbf{Definition}$ $\textit{31}$ above.

\bigskip

\noindent \textbf{Definition} \textit{35} (\textit{the correctness of the signature generation, and verification, protocols under noise as counterparts to the protocols provided in $\textbf{Definition}$ $\textit{20}$}). The noisy OTS (NOTS) scheme provided in the previous definition above is said to be \textit{correct} if,

{\small

\begin{align*}
 \mathcal{P} \big[   {\tiny  1 = \widetilde{\mathrm{Ver}} \big( \widetilde{\mathrm{vk}} , m^{\prime} ,  \widetilde{\mathrm{Sign}} \big(       \widetilde{\mathrm{sk}} , m^{\prime}    \big)  \big)  :                \big( \widetilde{\mathrm{vk}} , \widetilde{\mathrm{sk}} \big) \leftarrow  \widetilde{\mathrm{SGen}} \big( 1^{\lambda^{\prime}} \big)         }   \big] = 1    , \\ 
\end{align*}

}

\noindent for some $\lambda^{\prime} \in \textbf{R}$ and noisy message $m^{\prime}$, where ${\tiny \widetilde{\mathrm{Ver}} \big( \cdot , \cdot , \cdot \big)}$, $\widetilde{\mathrm{vk}}$, ${\tiny \widetilde{\mathrm{Sign}} \big(  \cdot , \cdot \big)}$, $\widetilde{\mathrm{sk}}$ and ${\tiny\widetilde{\mathrm{SGen}} \big( \cdot, \cdot \big)}$ denote the verification protocol, verification key, signature protocol, secret key, and secret key generation protocol under noise, respectively.

\bigskip

\noindent \textbf{Definition} \textit{36} (\textit{the correctness of the QPKE protocols associated with decoding, secret key generation, public key generation, and encoding under noise as counterparts to the protocols provided in $\textbf{Definition}$ $\textit{23}$}). A noisy QPKE (NQPKE) scheme is said to be \textit{correct} if,

{\small

\begin{align*}
  \mathcal{P} \bigg[  {\tiny \widetilde{m^{\prime}}  = \widetilde{\mathrm{Dec}} \big( \widetilde{\mathrm{sk}} , \widetilde{\mathrm{ct}} \big) : \big\{ \widetilde{\mathrm{sk}} \leftarrow \widetilde{\mathrm{SKGen}} \big( 1^{\lambda^{\prime}} \big) \big\} , \big\{ \big( \widetilde{\rho} , \widetilde{\mathrm{pk}} \big) \leftarrow \widetilde{\mathrm{PKGen}} \big( \widetilde{\mathrm{sk}} \big) \big\} , \big\{   \widetilde{\mathrm{ct}} \leftarrow \widetilde{\mathrm{Enc}} \big( \widetilde{\rho} , \widetilde{\mathrm{pk}} , m^{\prime} \big)     \big\}       }    \bigg] = 1   , \\ 
\end{align*}

}

\noindent given a noisy Quantum state $\widetilde{\rho}$, noisy public key $\widetilde{\mathrm{pk}}$ and message $m^{\prime}$, noisy secret key $\widetilde{\mathrm{sk}}$ and ciphertext $\widetilde{\mathrm{ct}}$, in addition to ${\tiny \widetilde{\mathrm{Dec}} \big( \cdot, \cdot \big)}$, ${\tiny \widetilde{\mathrm{SKGen}} \big( \cdot \big)}$, ${\tiny \widetilde{\mathrm{PKGen}} \big( \cdot \big)}$ and ${\tiny \widetilde{\mathrm{Enc}} \big( \cdot , \cdot , \cdot \big)}$ denote the decoding, secret key generation, public key generation, and encoding protocols under noise, respectively.

\bigskip

\noindent \textbf{Definition} \textit{37} (\textit{the trace distance between two noisy random experiments initialized to a 0, or 1, bit can be expressed in terms of the negligibility of the noisy security parameter, as counterparts to the noiseless negligibility function,}  $\mathrm{negl} {\tiny \big( \cdot \big)}$ \textit{, provided in $\textbf{Definition}$ $\textit{24}$}). A noisy QPKE (NQPKE) cryptographic protocol is said to satisfy the notion of \textit{everlasting} security if,

{\small

\begin{align*}
  \mathrm{Td} \big(    {\tiny \widetilde{\mathrm{Exp}}^{A_{\lambda^{\prime}}}  \big( 1^{\lambda^{\prime}} , 1 \big)     }      ,   {\tiny \widetilde{\mathrm{Exp}}^{A_{\lambda^{\prime}}} \big( 1^{\lambda^{\prime}} , 0 \big)     }    \big)  \lesssim  {\tiny \mathrm{NEGL} \big( \lambda^{\prime} \big) }   , \\ 
\end{align*}

}

\noindent for some $\lambda^{\prime} \in \textbf{R}$ and noisy random experiments ${\tiny  \widetilde{\mathrm{Exp}}^{A_{\lambda^{\prime}}}  \big( 1^{\lambda^{\prime}}  , 1 \big) }$ and ${\tiny \widetilde{\mathrm{Exp}}^{A_{\lambda}} \big( 1^{\lambda} , 0 \big)} $, each of which take as input $1^{\lambda^{\prime}}$ in addition to a bit initialized to $1$ and to $0$, respectively.

\bigskip

\noindent \textbf{Definition} \textit{38} (\textit{the trace distance between two noisy Quantum states is equal to the advantage between their respective resources, as a counterpart to the trace distance introduced in $\textit{Proof of Theorem 3.4}$, [3], of $\textbf{Definition}$ $\textit{25}$}). Fix some $i \in \textbf{N}$. The noisy advantage at $i$, ${\tiny \widetilde{\mathrm{Adv}} \big( i \big)}$, is equal to the  following trace variation distance,

{\small

\begin{align*}
\widetilde{\mathrm{Adv}} \big( i \big) \equiv \mathrm{Td} \big[   {\tiny \widetilde{\mathrm{Hyb}^{\mathcal{A}_{\lambda^{\prime}}}_{0,i}}     ,   \widetilde{\mathrm{Hyb}^{\mathcal{A}_{\lambda^{\prime}}}_{1,i} }  }  \big]   =       \mathrm{Td} \big[   {\tiny \widetilde{\mathrm{Hyb}^{\mathcal{A}_{\lambda^{\prime}}}_0}     ,   \widetilde{\mathrm{Hyb}^{\mathcal{A}_{\lambda^{\prime}}}_1 }  }  \big]        , \\ 
\end{align*}

}

\noindent where:

{\small

\begin{itemize}
    \item[$\bullet$] \textit{Initialization}. Initialize the bit $b$ where $b \in \big\{ 0 , 1 \big\}$, 

    \bigskip

     \item[$\bullet$] \textit{Hybrid experiments where $b \equiv 0$}. Define ${\tiny \widetilde{\mathrm{Hyb}^{\mathcal{A}_{\lambda^{\prime}}}_{0,i}} \equiv \widetilde{\mathrm{Hyb}^{\mathcal{A}_{\lambda^{\prime}}}_0}}$ as the noisy hybrid experiment generated from the NQPT machine ${\tiny \widetilde{\mathcal{A}_{\lambda^{\prime}}}}$,

     \bigskip

      \item[$\bullet$] \textit{Hybrid experiments where $b \equiv 1$}. Define ${\tiny \widetilde{\mathrm{Hyb}^{\mathcal{A}_{\lambda^{\prime}}}_{1,i}} \equiv  \widetilde{\mathrm{Hyb}^{\mathcal{A}_{\lambda^{\prime}}}_1}}$ as the noisy hybrid experiment generated from the NQPT machine ${\tiny \widetilde{\mathcal{A}_{\lambda^{\prime}}}}$.

\end{itemize}

}

\bigskip

\noindent \textbf{Definition} \textit{39} (\textit{the condition of noisy privacy expressed through a correspondence between the negiligibility and the trace distance, as a counterpart to the noiseless trace distance provided in \textbf{Definition} \textit{26}}). A noisy QKD-QPKE (NQKD-QPKE) protocol is said to satisfy \textit{everlasting} security if the notion of privacy is related to the following characterization between the negilibility and the trace distance,

{\small

\begin{itemize}
    \item[$\bullet$] \textit{Privacy of the NQKD-QPKE protocol}. The existence of the following negligible function, given a choice of $\lambda^{\prime} \in \textbf{R}$ and NQPT as introduced previously above, takes the form,

    {\small
    
    \begin{align*}
  \mathrm{Td} \big[   \big\{              \widetilde{\mathrm{QKDSec}^{\mathcal{A}_{\lambda^{\prime}}}}       , \widetilde{k_0} , \widetilde{k_1} \big\}       ,     \big\{   \widetilde{\mathrm{QKDSec}^{\mathcal{A}_{\lambda^{\prime}}}}        , \widetilde{k^{\prime}_0} , \widetilde{k^{\prime}_1} \big\}      \big]     \lesssim   {\tiny \mathrm{NEGL} \big( \lambda^{\prime} \big)  }  . \\
    \end{align*}

    }

\end{itemize}

}

\noindent where:

{\small

\begin{itemize}
    \item[$\bullet$] \textit{Experiment initialization}. Define ${\tiny \widetilde{k_0} \neq \widetilde{k^{\prime}_0}}$ and ${\tiny k^{\prime}_1 \neq \widetilde{k^{\prime}_1}}$ , 

    \bigskip

     \item[$\bullet$] \textit{Generating the experiment with respect to a noisy Quantum polynomial time machine and an initialization of parameters}. Denote an experiment generated by the NQPT $\mathcal{A}_{\lambda^{\prime}}$ with ${\tiny \mathrm{QKDSec}^{\mathcal{A}_{\lambda^{\prime}}}}$. 

\end{itemize}

}

\bigskip

\noindent The role of the negligibility function, $\mathrm{NEGL} {\tiny \big( \cdot \big)}$, under what can be viewed as a noising operation to the noiseless Quantum states under the trace distance provided in $\textbf{Definition}$ $\textit{13}$ is of value to generalize through the notion of XOR malleability. As functions of the plaintext introduced in $\textbf{Definition}$ $\textit{31}$ which are dependent upon the number of queries from Alice and Eve,  under the presence of noise one would expect for such transformations to: have higher probabilities of error, as demonstrated through the forwarding error, deleting error, and reconstruction error, probabilities provided in the definition of the plaintext transformation introduced in $\textbf{Definition}$ $\textit{31}$; the manner in which the security threshold gap ${\tiny \big| \lambda - \lambda^{\prime} \big|}$ between the security parameters of the QPKE-QKE and NQPKE-QKD protocols, respectively, is related to the forwarding error, deleting error, and reconstruction error, probabilities under noise; possible implications regarding the structure of encoding and decoding protocols.

\bigskip

\noindent \textbf{Definition} \textit{40} (\textit{alternative formulation of the negligibility function, and hence of the indistinguhishability, from the verification protocol provided in $\textbf{Definition}$ $\textit{21}$}). The NOTS scheme is said to satisfy the strong existential unforgeability property, meaning that, given an instance of the NOTS cryptographic scheme, there exists a negligibility function $\mathrm{NEGL} {\tiny \big( \cdot \big)}$ such that for all NQPTs $\big\{ A_{\lambda^{\prime}} \big\}_{\{{\lambda^{\prime} \in \textbf{N}\}}}$ and messages $\widetilde{m}$, one has that,

{\small

\begin{align*}
  \mathcal{P} \bigg[ {\tiny 1 = \widetilde{\mathrm{Ver}} \big( \widetilde{\mathrm{vk}} , \widetilde{m^{*}} , \widetilde{\sigma^{*}} \big)}  \textit{ with } {\tiny \big( \widetilde{m} , \widetilde{\sigma} \big) \neq \big( \widetilde{m^{*}} , \widetilde{\sigma^{*}} \big) :                     \big\{ \big( \widetilde{\mathrm{vk}} , \widetilde{\mathrm{sk}} \big) \leftarrow \widetilde{\mathrm{SGen}} \big( 1^{\lambda^{\prime}} \big) \big\} , \big\{ \sigma \leftarrow \widetilde{\mathrm{Sign}} \big( \widetilde{\mathrm{sk}} , \widetilde{m} \big) \big\} ,}  \\ {\tiny \big\{        \big( \widetilde{m^{*}} , \widetilde{\sigma^{*}} \big) \leftarrow A_{\lambda^{\prime}} \big( \mathrm{vk} , \widetilde{m} , \widetilde{\sigma} \big)       \big\}  }     \bigg] = {\tiny \mathrm{NEGL} \big( \lambda^{\prime} \big) }   . \\ 
\end{align*}

}

\subsection{Statement of main results}

\noindent While it is straightforward to introduce noise to the objects from the previous subsection with $\textbf{Definition}$ $\textit{32}$ -$\textbf{Definition}$ $\textit{40}$, it is important to discuss potentially adversarial tradeoffs which can increase $\lambda^{\prime}$. To this end introduce expansions in terms of summations for,

{\small

\begin{align*}
 \textit{Span of the polynomial basis} \equiv  \underset{i, j }{\sum} \big\{ {\tiny C \big( i , j \big) \lambda^i \lambda^j }  \big\}    , \end{align*}

 \begin{align*} \textit{Span of the exponential basis} \equiv  \underset{i, j }{\sum} \big\{ {\tiny C^{\prime} \big( i , j \big) \mathrm{exp} \big[ \lambda^i +  \lambda^j }  \big]   \big\}         , \\ 
\end{align*}

}

\noindent over a suitably chosen polynomial basis, where ${\tiny C \big( i , j \big) \neq C^{\prime} \big( i , j \big)}$ for all $i,j$. In each of the items below we compare how the expected runtimes of various protocols, through the 'Run Time' (which is understood as the computational run time of a cryptographic protocol of interest) function (defined as the summation of the expected runtime of all routines within a cryptographic protocol of interest),

{\small

\begin{align*}
\text{Run Time} : \textbf{N} \longrightarrow \textbf{N} ,
\end{align*}

}

\noindent impacts the security thresholds (ie, the ratio between the security threshold with noise and the security threshold without noise). Tradeoffs between the noisy counterparts from the objects introduced in the previous section include:

\begin{itemize}
    \item[$\bullet$] \textit{Computational runtime of the NQPT compared to that of QPT}. If the computational runtime associated with outputting a result from a NQPT, from a choice of $\lambda^{\prime}$, is significantly larger than that from a QPT, from a choice of $\lambda$, then $\lambda^{\prime} >> \lambda$. Quantitatively speaking, the amplification gap between $\lambda^{\prime}$ and $\lambda$ between the NQPKE-QKD, and QPKE-QKD, protocols, respectively, takes the form,

{\small

\begin{align*}
  \frac{\mathrm{Run \text{ } Time} \big[   \mathrm{NQPKE-QKD}   \big] }{\mathrm{Run \text{ } Time} \big[   \mathrm{QPKE-QKD}      \big] } > > 0         , \\ 
\end{align*}

}

\noindent then,

{\small

\begin{align*}
 \frac{\lambda \big|_{\mathrm{NQPKE-QKD}}}{\lambda \big|_{\mathrm{QPKE-QKD}} }   \equiv  \frac{\lambda^{\prime}}{\lambda} \propto {\tiny \mathrm{exp} \big[ - \lambda^{\prime} \big] \gtrsim   \mathrm{exp} \big[ - \big(  \lambda^{\prime}  - \lambda \big)  \big]     }     .  \\ 
\end{align*}

}

    \item[$\bullet$] \textit{Computational runtime of the noiseless versus noisy encoding protocols before information is transmitted}. Given the same choice of $\lambda^{\prime} > \lambda$ above, if,

{\small

\begin{align*}
   \frac{\mathrm{Run \text{ } Time} \big[   \textit{Encoding of } \mathrm{NQPKE-QKD}   \big] }{\mathrm{Run \text{ } Time} \big[  \textit{Encoding of }   \mathrm{QPKE-QKD}      \big] } > > 0        , \\ 
\end{align*}

}

    then,

{\small

\begin{align*}
 \frac{\lambda \big|_{\mathrm{NQPKE-QKD \text{ }Encoding}}}{\lambda \big|_{\mathrm{QPKE-QKD \text{ }Encoding}} }   \equiv   \frac{\lambda^{\prime}_{\mathrm{Encoding}}}{\lambda_{\mathrm{Encoding}}} \propto   {\tiny        \mathrm{exp} \big[ - \lambda^{\prime} \big] \gtrsim   \mathrm{exp} \big[ - \big(  \lambda^{\prime}  - \lambda \big) \big]      }           .  \\ 
\end{align*}

}

    \item[$\bullet$] \textit{Sharpening the security threshold gap of $\frac{\lambda^{\prime}}{\lambda}$ through more stringent assumptions on the runtime gap between the noisy and noiseless encoding protocols}. As opposed to the supposition,

{\small

\begin{align*}
   \frac{\mathrm{Run \text{ } Time} \big[   \textit{Encoding of } \mathrm{NQPKE-QKD}   \big] }{\mathrm{Run \text{ } Time} \big[  \textit{Encoding of }   \mathrm{QPKE-QKD}      \big] } > > 0        , \\ 
\end{align*}

}

    \noindent if one instead supposes that,

{\small

\begin{align*}
   \frac{\mathrm{Run \text{ } Time} \big[   \textit{Encoding of } \mathrm{NQPKE-QKD}   \big] }{\mathrm{Run \text{ } Time} \big[  \textit{Encoding of }   \mathrm{QPKE-QKD}      \big] }       \gtrsim    {\tiny \mathrm{Poly} \big[ \lambda^{\prime} \big]    \gtrsim     \mathrm{Poly} \big[ \lambda^{\prime} - \lambda \big]  }            , \\ 
\end{align*}

}

   \noindent for polynomial functions Poly in $\lambda^{\prime}$, and $\lambda^{\prime} - \lambda$, respectively, then,

{\small

\begin{align*}
  \frac{\lambda^{\prime}_{\mathrm{Encoding}}}{\lambda_{\mathrm{Encoding}}}   \propto          {\tiny \big(  \lambda^{\prime} - \lambda \big)  \mathrm{Poly} \big[  - \lambda^{\prime}  \big]           \lesssim   \lambda^{\prime} \mathrm{Poly} \big[  - \lambda^{\prime}  \big]    \lesssim    \lambda   \mathrm{exp} \big[  -  \lambda   \big]  \lesssim    \lambda^{\prime}   \mathrm{exp} \big[  -  \lambda   \big]        \lesssim        \big( \lambda^{\prime} - \lambda \big) }  \\ \times  {\tiny \mathrm{exp} \big[  - \big( \lambda^{\prime}   - \lambda \big)  \big]               \lesssim  \lambda^{\prime} \mathrm{exp} \big[  - \big( \lambda^{\prime} - \lambda \big)  \big]            \lesssim  \lambda^{\prime} \mathrm{exp} \big[  - \lambda^{\prime}  \big]   }                   , \\ 
\end{align*}

}

\noindent where,

{\small

\begin{align*}
  \frac{\lambda \big|_{\mathrm{NQPKE-QKD \text{ }Encoding}}}{\lambda \big|_{\mathrm{QPKE-QKD \text{ }Encoding}} }   \equiv    \frac{\lambda^{\prime}_{\mathrm{Encoding}}}{\lambda_{\mathrm{Encoding}}}       . \\ 
\end{align*}

}

    \item[$\bullet$] \textit{Computational runtime of the decoding portion of the NQPKE-QKD and QPKE-QKD protocols from the polynomial threshold gap between $\lambda^{\prime}$ and $\lambda^{\prime}$}. Suppose,

    {\small

\begin{align*}
  \frac{\lambda \big|_{\mathrm{NQPKE-QKD \text{ }Decoding}}}{\lambda \big|_{\mathrm{QPKE-QKD \text{ }Decoding}} }   \equiv    \frac{\lambda^{\prime}_{\mathrm{Decoding}}}{\lambda_{\mathrm{Decoding}}} \propto      {\tiny    \big(  \lambda^{\prime} - \lambda \big)  \mathrm{Poly} \big[  - \lambda^{\prime}  \big]   }                 ,  \\ 
\end{align*}

}

 \noindent    then,

{\small

\begin{align*}
      \frac{\mathrm{Run \text{ } Time} \big[   \textit{Decoding of } \mathrm{NQPKE-QKD}   \big] }{\mathrm{Run \text{ } Time} \big[  \textit{Decoding of }   \mathrm{QPKE-QKD}      \big] }        \gtrsim                           {\tiny \big(  \lambda^{\prime} - \lambda \big)  \mathrm{Poly} \big[  - \lambda^{\prime}  \big]      }              ,  \\ 
\end{align*}

}

\noindent and hence,

{\small

\begin{align*}
  \frac{\lambda^{\prime}_{\mathrm{Decoding}}}{\lambda_{\mathrm{Decoding}}}  \propto              {\tiny \lambda^{\prime}   \mathrm{exp} \big[  - \lambda^{\prime}  \big]                              \gtrsim                         \big(  \lambda^{\prime} - \lambda \big)  \mathrm{exp} \big[  - \lambda^{\prime}  \big]                         \gtrsim    \lambda^{\prime}  \mathrm{Poly} \big[  - \lambda^{\prime}  \big]                    \gtrsim                            \big(  \lambda^{\prime} - \lambda \big)  \mathrm{Poly} \big[  - \lambda^{\prime}  \big]      }           .  \\ 
\end{align*}

}

\item[$\bullet$] \textit{Polynomial security threshold gap from the computational runtime of the Secret Key generation protocol in the NQPKE-QKD and QPKE-QKD settings, respectively}. Suppose,

{\small

\begin{align*}
   \frac{\mathrm{Run \text{ } Time} \big[   \textit{Secret Key Generation of } \mathrm{NQPKE-QKD}   \big] }{\mathrm{Run \text{ } Time} \big[  \textit{Secret Key Generation of }   \mathrm{QPKE-QKD}      \big] } > > 0        , \\ 
\end{align*}

}

\noindent then,

{\small

\begin{align*}
      \frac{\lambda \big|_{\text{NQPKE-QKD Secret Key Generation}}}{\lambda \big|_{\text{QPKE-QKD Secret Key Generation}} }   \equiv    \frac{\lambda^{\prime}_{\text{SKG}}}{\lambda_{\text{SKG}}} \propto   \mathrm{Td} \bigg[   \ket{c , \sigma_c}     \big/  \ket{\widetilde{c} , \widetilde{\sigma_c}}  ,     \bra{c , \sigma_c}     \big/  \bra{\widetilde{c} , \widetilde{\sigma_c}}     \bigg]            . \\ 
\end{align*}

}

\item[$\bullet$] \textit{Subsequentiality}. Suppose,

{\small

\begin{align*}
  \big\{ \lambda_i \big\}_{1 \leq i \leq C}          \equiv   {\tiny \big\{ \lambda_{i,\mathrm{Encoding}} \big\}_{1 \leq i \leq C}   +  \big\{ \lambda_{i,\mathrm{Decoding}} \big\}_{1 \leq i \leq C}   +  \big\{ \lambda_{i,\text{Secret Key Generation}}  \big\}_{1 \leq i \leq C}    }  \\      \\  +  {\tiny \big\{ \lambda_{i,\text{Public Key Generation}}  \big\}_{1 \leq i \leq C}  }   , \\ 
\end{align*}

}

\noindent for $C>0$ taken sufficiently large, satisfies,

{\small

\begin{align*}
      {\tiny   {\underset{i \equiv ( i_1 , \cdots , i_n) :  \{ i_j \in \mathcal{U}_j \textit{ for every $1 \leq j \leq n$} \} }{\prod}} \lambda_i } \bigg/           {\tiny            {\underset{ j \equiv ( j_1 , \cdots , j_n) : \{ j_k \in \mathcal{U}^{\prime}_k \textit{ for every $1 \leq k \leq n$} \}  }{\prod}} \lambda^{-1}_j      \lesssim   1    }      , \\ 
\end{align*}

}

\noindent for the collection of $i$ for which there are permutations along one $i$ for $1 \leq i \leq n$ (corresponding to $\mathcal{U}_1$), up to permutations along all $i$ for $1 \leq i \leq n$ (corresponding to $\mathcal{U}_N$). Then,

{\small

\begin{align*}
\bigg\{ {\tiny \frac{\lambda_i}{\lambda_j} }  \bigg\}_{\{ i \neq j, 1 \leq i \leq j \leq N \} }  \propto      {\tiny   \big(  \lambda^{\prime}  - \lambda \big) \mathrm{Poly} \big[ \lambda^{\prime} - \lambda \big]        \lesssim      \lambda^{\prime}  \mathrm{Poly} \big[ \lambda^{\prime} - \lambda \big]          \lesssim        \lambda^{\prime}  \mathrm{Poly} \big[ \lambda^{\prime} \big]    \lesssim      \big(   \lambda^{\prime} - \lambda \big) }  \\ \times {\tiny \mathrm{exp} \big[ \lambda^{\prime}   - \lambda  \big]  \lesssim       \lambda^{\prime}   \mathrm{exp} \big[ \lambda^{\prime}  \big]   }         ,       \\ 
\end{align*}

}

\noindent where,

{\small

\begin{align*}
  \bigg\{ {\tiny \frac{\lambda_i}{\lambda_j}  } \bigg\}_{\{ i \neq j, 1 \leq i \leq j \leq N \} }  \equiv  \frac{ {\tiny   \big\{ \lambda_{i,\mathrm{Encoding}} \big\}_{1 \leq i \leq C}   +  \big\{ \lambda_{i,\mathrm{Decoding}} \big\}_{1 \leq i \leq C}   +  \big\{ \lambda_{i,\text{Secret Key Generation}}  \big\}_{1 \leq i \leq C}         \cdots }}{ {\tiny \big\{ \lambda_{j,\mathrm{Encoding}} \big\}_{1 \leq j \leq C, j \neq i }   +  \big\{ \lambda_{j,\mathrm{Decoding}} \big\}_{1 \leq i \leq C, j \neq i }   +     \cdots }  } \\  \end{align*}
  
  \begin{align*} \frac{ {\tiny +  \big\{   \lambda_{i,\text{Public Key Generation}}      \big\}_{1 \leq i \leq C}  } }{ {\tiny +   \big\{ \lambda_{j,\text{Secret Key Generation}}  \big\}_{1 \leq i \leq C, j \neq i }      +   \big\{   \lambda_{j,\text{Public Key Generation}}   \big\}_{1 \leq j \leq C, j \neq i } }}. \\ 
\end{align*}

}

\end{itemize}

\noindent \textbf{Lemma} (\textit{plaintext transformations under noise in the NQPKE-QKD protocol are related to the negligibility function of the security threshold $\lambda^{\prime}$}). Denote $\widetilde{\rho}$ and $\widetilde{\tau}$ as noisy Quantum states counterpart to $\rho$ and $\tau$, respectively. For noisy transformations $\widetilde{F}$ of the plaintext satisfying:

{\small \begin{itemize}
    \item[$\bullet$] \textit{(1). Noisy forwarding error probability}. The forwarding error probability of $\widetilde{F}$ is given by the existence of some ${\tiny \widetilde{\delta_1} \big( q \big) >0}$, for ${\tiny \widetilde{m_{q_A}} \neq \widetilde{{m}_{q_A}} \in \widetilde{\mathcal{M}_{q_A}}}$, ${\tiny \widetilde{m^{\prime}_{q_E} } \neq \widetilde{{m^{\prime}}_{q_E}} \in \widetilde{\mathcal{M}_{q_E}}}$ with ${\tiny \big| \widetilde{m_{q_A}} \big| = \big| \widetilde{{m}_{q_A}}  \big|}$ and ${\tiny \big| \widetilde{m^{\prime}_{q_E}} \big| = \big| \widetilde{{m^{\prime}}_{q_E}}  \big|}$ so that,

    {\small
    
    \begin{align*}
         \big|     \mathcal{P} \big[  {\tiny \widetilde{F} \big( \widetilde{m^{q_A}}  , \widetilde{m^{\prime}_{q_E}} \big)  = \widetilde{m_{q_E + 1}}      \big] }     -      \mathcal{P} \big[   {\tiny \widetilde{F} \big( \widetilde{{m}^{q_A}} , \widetilde{{m^{\prime}}_{q_E}}  \big)  = \widetilde{{m}_{q_E + 1}  }   }    \big]       \big| \leq \widetilde{\delta_1}  \big(    q \big)      . \\ 
    \end{align*}

    }

    \bigskip

       \item[$\bullet$] \textit{(2). Noisy deleting error probability}. The deleting error probability of $\widetilde{F}$ is given by the existence of some ${\tiny \widetilde{\delta_2} \big( q \big) >0}$, where ${\tiny \widetilde{m^{q_A}}  \in \widetilde{\mathcal{M}^{q_A}}}$ ${\tiny \forall \widetilde{m^{\prime}_{q_E}} \in \widetilde{\mathcal{M}^{q_E}}}$,

{\small

\begin{align*}
      \mathcal{P} \big[ {\tiny \widetilde{F} \big( \widetilde{m_{q_A}} , \widetilde{m^{\prime}_{q_E}} \big) = \widetilde{m_{q_E + 1}}   }   \big] \leq \widetilde{\delta_2}  \big( q \big)      . \\ 
\end{align*}

}

       \bigskip

         \item[$\bullet$] \textit{(3). Noisy reconstruction error probability}. The reconstruction error probability of $F$ is given by the existence of some ${\tiny \widetilde{\delta_3} \big( q \big) > 0}$ where ${\tiny \widetilde{m^{q_A}} \in \widetilde{\mathcal{M}^{q_A}}}$ ${\tiny \forall \widetilde{m^{\prime}_{q_E}} \in \widetilde{\mathcal{M}^{q_E}}}$,

         {\small

         \begin{align*}
             \big|   \mathcal{P} \big[     {\tiny     \widetilde{F} \big( \widetilde{m_{q_A}} , \widetilde{m^{\prime}_{q_E}} \big) = \widetilde{m_{q_E}} }      \big]       -  \mathcal{P} \big[  {\tiny    R \big( \widetilde{M} \big) = \widetilde{m_{q_E + 1}}   }      \big]       \big| \leq \widetilde{\delta_3}  \big( q \big)                 .  \\ 
         \end{align*}
         
         }
    
\end{itemize} }

\noindent where the message space ${\tiny \widetilde{M} \in \big\{     \widetilde{m_1} , \cdots , \widetilde{m_{q_A}} , \widetilde{m^{\prime}_1} , \cdots , \widetilde{m^{\prime}_{q_B}} , \widetilde{{m}_1 } , \cdots , \widetilde{{m}_{q_E} }      \big\} \backslash \big\{ \widetilde{m_{q_E + 1 }}  \big\}}$ in NQPKE-QKD protocol and an algorithm $R$ then:

{\small

\begin{itemize}
    \item[$\bullet$] \textit{(A). The ratio between the forwarding error probability of XOR malleable transformations independently, and dependently, of noise given the total number of queries $q$ issued by Alice, and by Eve}. The forwarding error probability ${\tiny \delta_{\mathrm{SUF-CMA}} \big( q \big) \equiv \delta \big( q \big)}$ previously formulated for XOR malleable functions on ciphertexts produced in the SUF-CMA setting, with respect to its noisy counterpart ${\tiny \widetilde{\delta_{\mathrm{NQPKE-QKD}}} \big( q \big) \equiv \widetilde{\delta} \big( q \big)}$ in the NQPKE-QKD setting, from,

    {\small
    
    \begin{align*}
    \frac{\pi_{\mathrm{SUF-CMA}}}{\pi_{\mathrm{NQPKE-QKD}}}   \sim   \frac{\delta\big( \mathcal{Q}  \big)}{ \widetilde{\delta} \big( \mathcal{Q} \big)}      =    \delta\big( \underset{Q \in \mathcal{Q}}{\bigcup} Q   \big) \bigg/  \widetilde{\delta} \big(\underset{Q \in \mathcal{Q}}{\bigcup} Q    \big)               , \\ 
    \end{align*}

    }

    \noindent for the probability distributions $\pi_{\mathrm{SUF-CMA}}$ and $\pi_{\mathrm{NQPKE-QKD}}$ in the SUF-CMA and NQPKE-QKD settings, respectively, and the ratio,

    {\small
    
    \begin{align*}
   \underset{\textit{and Eve}}{\underset{\textit{by Alice}}{\underset{\textit{queries} \text{ } q}{\bigcup}}} \bigg\{       \frac{\delta\big( q \big)}{ \widetilde{\delta} \big( q \big)}  \bigg\}    
   \equiv    \frac{\delta\big( q \big)}{ \widetilde{\delta} \big( q \big)}      , 
    \end{align*}

    }

    \noindent satisfies,

{\small \begin{align*}
  \\  \frac{\delta\big( q \big)}{ \widetilde{\delta} \big( q \big)} \propto   \underset{0 <  j \leq q}{\sum}       \bigg\{       \bigg\{  {\tiny \ket{\frac{c_j   }{\widetilde{c_j}}}   }    \bigotimes  {\tiny \ket{\frac{\sigma_j    }{\widetilde{\sigma_j}}} }  \bigg\}  \bigg\{   {\tiny \bra{\frac{c_j   }{\widetilde{c_j}}} }    \bigotimes       {\tiny  \bra{\frac{\sigma_j    }{\widetilde{\sigma_j}}}  }   \bigg\}                             \bigg\}    \lesssim              \bigg\{   \underset{0 < j^{\prime} \leq q}{\underset{0 \leq j \leq q}{\sum}}            \bigg\{ {\tiny \frac{\ket{c_j}}{\ket{\widetilde{c_{j^{\prime}}}}}       }   \bigg\}        \bigotimes                  \underset{0 < j^{\prime} \leq q}{\underset{0 \leq j \leq q}{\sum}}            \bigg\{      {\tiny \frac{\ket{\sigma_j}}{\ket{\widetilde{\sigma_{j^{\prime}}}}}   }           \bigg\}    \bigg\}                \\ \\ \times \bigg\{    \underset{0 < j^{\prime} \leq q}{\underset{0 \leq j \leq q}{\sum}}   \bigg\{  {\tiny \frac{\bra{c_j}}{\bra{\widetilde{c_{j^{\prime}}}}}     }     \bigg\}     \bigotimes                  \underset{0 < j^{\prime} \leq q}{\underset{0 \leq j \leq q}{\sum}}            \bigg\{      {\tiny \frac{\bra{\sigma_j}}{\bra{\widetilde{\sigma_{j^{\prime}}}}}    }          \bigg\}             \bigg\}   , \end{align*} }

\noindent where,

   {\small

    \begin{align*}
       \Pi \rho^* \Pi / \widetilde{\Pi} \widetilde{\rho^*} \widetilde{\Pi}      ,  \end{align*}

       \begin{align*}  \tag{\textit{Ratio of noiseless and noisy projection operators}} \end{align*}

       \begin{align*}   \mathrm{Tr} \big[        \Pi \rho^*   \big] /     \mathrm{Tr} \big[  \widetilde{\Pi } \widetilde{\rho^{*}}    \big]             ,  \end{align*}

          \begin{align*}  \tag{\textit{Ratio of noiseless and noisy trace operations}} \end{align*} 
       
       \begin{align*} \frac{\ket{c, \sigma_c}}{\ket{\widetilde{c}, \widetilde{\sigma_c}}}                         , \end{align*}

          \begin{align*}  \tag{\textit{Ratio of noiseless and noisy ket states for the ciphertext}} \end{align*}

       \begin{align*} \frac{\bra{c, \sigma_c}}{\bra{\widetilde{c}, \widetilde{\sigma_c}}}                  , 
    \end{align*}

          \begin{align*}  \tag{\textit{Ratio of noiseless and noisy bra states for the ciphertext}} \end{align*}

    }

    \bigskip

\noindent are individually upper bounded, up to a constant, with the generalization of the notion of unforgeability under noise stated below.

\bigskip

    \item[$\bullet$] \textit{(B). The noisy unforgeability of the OTS scheme for a NQPKE-QKD protocol can be related to the noiseless unforgeability of the OTS scheme of the QPKE-QKD protocol through a suitably chosen function dependent upon the ratio of forwarding error probabilities}. One has that the desired up to constants bound for,

    {\small
    
    \begin{align*} 
 \mathrm{Td} \bigg[    \frac{ \big\{ \Pi \rho^* \Pi / \widetilde{\Pi} \widetilde{\rho^*} \widetilde{\Pi}  \big\}            }{  \big\{  \mathrm{Tr} \big[        \Pi \rho^*   \big] /     \mathrm{Tr} \big[  \widetilde{\Pi } \widetilde{\rho^{*}}    \big]    \big\}                           }  ,          \frac{ \big\{ \ket{c , \sigma_c}  \bra{c , \sigma_c }   \big\}     }{ \big\{ \ket{\widetilde{c} , \widetilde{\sigma_c}}   \bra{\widetilde{c}  , \widetilde{\sigma_c }} \big\}   }             \bigg]        , \\ 
    \end{align*}

    }

    \noindent can be approximated from the following series of approximations given through items $\textbf{(1)}$-$\textbf{(4)}$ below on the trace distance:

    {\small

    \begin{align*}
  \tag{\textbf{1}. Upper bounding, up to constants, the ratio between the noiseless and noisy} \\ \tag{projection operators.}  \end{align*}
  
  \begin{align*}   \Pi \rho^* \Pi / \widetilde{\Pi} \widetilde{\rho^*} \widetilde{\Pi}      \\ \end{align*}

  \begin{align*} 
   \overset{(\mathrm{B.1})}{\lesssim}  {\tiny   \underset{0 < j^{\prime} \leq q}{\underset{0 \leq j \leq q}{\sum}}} \bigg[ {\tiny           \underset{\sigma_0 \in \Sigma_0}{\sum}  \bigg\{        \ket{0} \bigotimes   \ket{\frac{\sigma_0}{\sigma_0 + \textit{noise}}}                             \bigg\}     \bigg\{   \bra{\frac{\sigma_0}{\sigma_0 + \textit{noise}}}   \bigotimes    \bra{0}                 \bigg\}   \rho^* / \widetilde{\rho^*}   +  \underset{\sigma_1 \in \Sigma_1}{\sum}     \bigg\{        \ket{1}   \bigotimes   \ket{\frac{\sigma_1}{\sigma_1 + \textit{noise}}}                             \bigg\}  } \\ {\tiny \times   \bigg\{   \bra{\frac{\sigma_1}{\sigma_1 + \textit{noise}}}   \bigotimes    \bra{1}                 \bigg\}    \rho^* / \widetilde{\rho^*} }  \bigg]                      ,   \end{align*}
       
        \begin{align*}
  \tag{\textbf{2}. Upper bounding, up to constants, the ratio of noiseless and noisy trace} \\ \tag{ operations.}  \end{align*}

       \begin{align*}   \mathrm{Tr} \big[        \Pi \rho^*   \big] /     \mathrm{Tr} \big[  \widetilde{\Pi } \widetilde{\rho^{*}}    \big]                     \\  \end{align*}

       \begin{align*} \overset{\mathrm{(B.2)}}{\lesssim }          {\tiny \underset{\sigma_0 \in \Sigma_0}{\underset{0 < j^{\prime} \leq q}{\underset{0 \leq j \leq q}{\sum}}}      \bigg\{ \bigg[    \mathrm{Tr}  \big\{  \ket{0} \big\}                \bigotimes  \mathrm{Tr}  \bigg\{    \ket{\frac{\sigma_0}{\sigma_0 + \textit{noise}}} \bigg\}  \bigg]   \bigg[ \mathrm{Tr}  \bigg\{    \bra{\frac{\sigma_0}{\sigma_0 + \textit{noise}}} \bigg\}              \bigotimes  \mathrm{Tr}  \big\{  \bra{0} \big\}              \bigg]           \bigg\}   + \underset{\sigma_1 \in \Sigma_1}{\underset{0 < j^{\prime} \leq q}{\underset{0 \leq j \leq q}{\sum}}}      \bigg\{   \bigg[   \mathrm{Tr}  \big\{  \ket{1} \big\}             }  \\ {\tiny   \bigotimes        \mathrm{Tr}  \bigg\{    \ket{\frac{\sigma_2}{\sigma_1 + \textit{noise}}} \bigg\}                 \bigg]  \bigg[  \mathrm{Tr}  \bigg\{    \bra{\frac{\sigma_1}{\sigma_1 + \textit{noise}}} \bigg\}              \bigotimes  \mathrm{Tr}  \big\{  \bra{1} \big\}          \bigg]        \bigg\}  }  .      \\              \end{align*}

    \begin{align*}
  \tag{\textbf{3}. Upper bounding, up to constants, the ratio of ket ciphertext states.} \\ \end{align*}

       \begin{align*} \frac{\ket{c, \sigma_c}}{\ket{\widetilde{c}, \widetilde{\sigma_c}}}         \overset{\mathrm{(B.3)}}{\lesssim}  \underset{0 < j^{\prime} \leq q}{\underset{0 \leq j \leq q}{\sum}}           \bigg\{   \frac{\ket{c}}{\ket{\widetilde{c}}}  \bigg\}  \bigotimes     \underset{0 < j^{\prime} \leq q}{\underset{0 \leq j \leq q}{\sum}}           \bigg\{   \frac{\ket{\sigma_c }}{\ket{\widetilde{\sigma_c}}}    \bigg\}      . \\ \end{align*}

          \begin{align*}
  \tag{\textbf{4}. Upper bounding, up to constants, the ratio of bra ciphertext states.} \\ \end{align*}

       \begin{align*} \frac{\bra{c, \sigma_c}}{\bra{\widetilde{c}, \widetilde{\sigma_c}}}           \overset{\mathrm{(B.4)}}{\lesssim}  \underset{0 < j^{\prime} \leq q}{\underset{0 \leq j \leq q}{\sum}}           \bigg\{   \frac{\bra{c}}{\bra{\widetilde{c}}}  \bigg\}  \bigotimes     \underset{0 < j^{\prime} \leq q}{\underset{0 \leq j \leq q}{\sum}}           \bigg\{   \frac{\bra{\sigma_c }}{\bra{\widetilde{\sigma_c}}}    \bigg\}            , \\ \\ 
    \end{align*}

    }

    \noindent where,

{\small

\begin{align*}
 \rho^{*} \mapsto \widetilde{\rho^{*} } = \rho^{*} \otimes \textit{noise}   , \\ \\ \sigma_c  \mapsto \widetilde{\sigma_c} = \sigma_c  \otimes \textit{noise}  , \\ 
\end{align*}

}

\noindent correspond to the noisy states $\widetilde{\rho}$ and $\widetilde{\sigma_c}$, respectively. Altogether, by the associativity property of the standard multiplication operation $\cdot$ over $\textbf{R}$, the resultant superposition associated with the product,

{\small

\begin{align*}
\bigg[  \frac{\ket{c, \sigma_c}}{\ket{\widetilde{c}, \widetilde{\sigma_c}}} \bigg] \bigg[   \frac{\bra{c, \sigma_c}}{\bra{\widetilde{c} , \widetilde{\sigma_c}}}                 \bigg] = \bigg[  \frac{ \big\{ \ket{c, \sigma_c} \bra{c, \sigma_c}  \big\} }{\ket{\widetilde{c}, \widetilde{\sigma_c}}} \bigg] \bigg[   \frac{1}{\bra{\widetilde{c}, \widetilde{\sigma_c}}}                 \bigg] =    \bigg[   \frac{ \big\{ \ket{c, \sigma_c} \bra{c, \sigma_c}  \big\} }{ \big\{ \ket{\widetilde{c}, \widetilde{\sigma_c}} \bra{\widetilde{c}, \widetilde{\sigma_c}} \big\}  }    \bigg] =    \frac{ \big[ \big\{ \ket{c, \sigma_c} \bra{c, \sigma_c}  \big\} \big]  }{ \big[ \big\{ \ket{\widetilde{c}, \widetilde{\sigma_c}} \bra{\widetilde{c}, \widetilde{\sigma_c}} \big\} \big]  }     , \\ 
\end{align*}

}

\noindent implies,

{\small

\begin{align*}
 \mathrm{Td} \bigg[    \frac{ \big\{ \Pi \rho^* \Pi / \widetilde{\Pi} \widetilde{\rho^*} \widetilde{\Pi}  \big\}            }{  \big\{  \mathrm{Tr} \big[        \Pi \rho^*   \big] /     \mathrm{Tr} \big[  \widetilde{\Pi } \widetilde{\rho^{*}}    \big]    \big\}                           }  ,          \frac{ \big\{ \ket{c , \sigma_c}  \bra{c , \sigma_c }   \big\}     }{ \big\{ \ket{\widetilde{c} , \widetilde{\sigma_c}}   \bra{\widetilde{c} , \widetilde{\sigma_c }} \big\}   }             \bigg]       \lesssim   \big[  \textit{$(\mathrm{B.1})$} +  \textit{ $(\mathrm{B.2})$,} +   \textit{$(\mathrm{B.3})$, $(\mathrm{B.4})$}   \big]   , \\ 
 \end{align*}
 }

 \noindent from an application of the GML:

 \bigskip

 \noindent \textbf{Lemma} (\textit{the Gentle Measurement Lemma of Quantum information theory: an upper bound on the trace distance can be obtained from a suitably defined transformation on the density matrix corresponding to $\rho$ through an up to constants lower bound on the trace}, [18]). For the density operator $\rho$, a parameter $\epsilon$ for which,
 
 {\small

 \begin{align*}
    0 < \epsilon \leq 1 ,
 \end{align*}

 }

 \noindent and a measurement operator $\Lambda$ for which,

  {\small

 \begin{align*}
    0 < \Lambda \leq \textbf{I} ,
 \end{align*}

 }

 \noindent the following estimate on the trace,

{\small

\begin{align*}
 \mathrm{Tr} \big\{ \Lambda \rho \big\} \geq  1 - \epsilon   , \\ 
\end{align*}

}

\noindent corresponding to detecting $\rho$ implies the upper bound,

{\small

\begin{align*}
  \mathrm{Td} \big[ \rho , \rho^{\prime} \big] \leq 2 \sqrt{\epsilon}  , 
\end{align*}

}

\noindent on the trace distance, where,

{\small

\begin{align*}
 \rho^{\prime} \equiv  \big\{   \sqrt{\Lambda} \rho \sqrt{\Lambda} \big\}  / \mathrm{Tr} \big\{ \Lambda \rho \big\}     .
\end{align*}

}

\noindent To argue that an estimate of the form,

{\small

\begin{align*}
  \mathrm{Td} \bigg[ \frac{ \big\{ \Pi \rho^* \Pi / \widetilde{\Pi} \widetilde{\rho^*} \widetilde{\Pi}  \big\}            }{  \big\{  \mathrm{Tr} \big[        \Pi \rho^*   \big] /     \mathrm{Tr} \big[  \widetilde{\Pi } \widetilde{\rho^{*}}    \big]    \big\}                           }  ,          \frac{ \big\{ \ket{c , \sigma_c}  \bra{c , \sigma_c }   \big\}     }{ \big\{ \ket{\widetilde{c} , \widetilde{\sigma_c}}   \bra{\widetilde{c} , \widetilde{\sigma_c }} \big\}   }      \bigg]       \leq  2 \sqrt{\epsilon } , \\ 
\end{align*}

}

\noindent holds for the trace distance it suffices to argue,

{\small

\begin{align*}
  \mathrm{Td} \bigg[ \frac{ \big\{ \Pi \rho^* \Pi / \widetilde{\Pi} \widetilde{\rho^*} \widetilde{\Pi}  \big\}            }{  \big\{  \mathrm{Tr} \big[        \Pi \rho^*   \big] /     \mathrm{Tr} \big[  \widetilde{\Pi } \widetilde{\rho^{*}}    \big]    \big\}                           }  ,          \frac{ \big\{ \ket{c , \sigma_c}  \bra{c , \sigma_c }   \big\}     }{ \big\{ \ket{\widetilde{c} , \widetilde{\sigma_c}}   \bra{\widetilde{c} , \widetilde{\sigma_c }} \big\}   }      \bigg]  \leq \mathrm{Td} \bigg[      \frac{ \big\{ \Pi \rho^* \Pi / \widetilde{\Pi} \widetilde{\rho^*} \widetilde{\Pi}  \big\}            }{  \big\{  \mathrm{Tr} \big[        \Pi \rho^*   \big] /     \mathrm{Tr} \big[  \widetilde{\Pi } \widetilde{\rho^{*}}    \big]    \big\}                           }       ,                       \rho^* \big/ \widetilde{\rho^*}          \bigg]   , 
\end{align*}

}

\noindent implies the desired upper bound of $2 \sqrt{\epsilon}$ from the fact that,

{\small

\begin{align*}
  \mathrm{Td} \bigg[      \frac{ \big\{ \Pi \rho^* \Pi / \widetilde{\Pi} \widetilde{\rho^*} \widetilde{\Pi}  \big\}            }{  \big\{  \mathrm{Tr} \big[        \Pi \rho^*   \big] /     \mathrm{Tr} \big[  \widetilde{\Pi } \widetilde{\rho^{*}}    \big]    \big\}                           }       ,                     \rho^* \big/ \widetilde{\rho^*}               \bigg]  \leq 2 \sqrt{\epsilon }          , 
\end{align*}

}

\noindent for which we make use of the above lemma through manipulation of the following objects given the choice of constants ${\tiny \mathcal{C} \neq \mathcal{C}^{\prime} \neq \mathcal{C}^{\prime\prime} \neq C \neq C^{\prime} \neq C^{\prime\prime}}$ taken sufficiently small:

{\small

\begin{itemize}

    \item[$\bullet$] \textbf{(1)}. $\text{Ratio of states in the QPKE-QKD and NQPKE-QKD protocols}  \equiv \boxed{ \rho^* / \tau^*  } = \rho^* / \widetilde{\rho^*}                       $

    \bigskip

  \item[$\bullet$] \textbf{(2)}. $\text{Ratio of Projection operators} \equiv \boxed{\Pi / \widetilde{\Pi}} $

    \bigskip 

     \item[$\bullet$] \textbf{(3)}. $\text{Measurement} \equiv \mathcal{M} \big( \cdot , \cdot^{\prime} \big)  \equiv  \boxed{ \sqrt{ \frac{{\tiny \big( \widetilde{\cdot - \cdot^{\prime}} \big)^{\dagger} \big( \widetilde{\cdot - \cdot^{\prime}} \big) }    }{{\tiny  \big( \cdot - \cdot^{\prime} \big)^{\dagger} \big( \cdot - \cdot^{\prime} \big)    } }             } }            $

     \bigskip

\item[$\bullet$] \textbf{(4)}. $\mathcal{M} \cdot \text{State} \equiv \mathcal{M} \big( \cdot , \cdot^{\prime} \big) \cdot \big( \rho^* / \widetilde{\rho^*} \big) =  \mathcal{M} \big(  \rho^* / \widetilde{\rho^*}    \big) =  \mathcal{M} \big(  \rho^* / \tau^*     \big)   =   \boxed{\sqrt{ \frac{{\tiny \big( \widetilde{\rho - \tau } \big)^{\dagger} \big( \widetilde{\rho - \tau^{\prime}} \big) }    }{{\tiny  \big( \rho - \tau^{\prime} \big)^{\dagger} \big( \rho - \tau \big)    } }             }  }      $

     \bigskip 

      \item[$\bullet$] \textbf{(5)}. $\text{Trace} \equiv  \mathrm{Tr} \big\{  \mathcal{M} \big(  \rho^* / \widetilde{\rho^*}    \big)    \big\}    = \boxed{ \mathrm{Tr} \bigg\{    \sqrt{ \frac{{\tiny \big( \widetilde{\rho - \tau } \big)^{\dagger} \big( \widetilde{\rho - \tau^{\prime}} \big) }    }{{\tiny  \big( \rho - \tau^{\prime} \big)^{\dagger} \big( \rho - \tau \big)    } }             }        \bigg\} }$

      \bigskip 

       \item[$\bullet$] \textbf{(6)}. $\text{trace distance} \equiv \boxed{\mathrm{Td} \bigg[         \frac{ \big\{ \Pi \rho^* \Pi / \widetilde{\Pi} \widetilde{\rho^*} \widetilde{\Pi}  \big\}            }{  \big\{  \mathrm{Tr} [        \Pi \rho^*   ] /     \mathrm{Tr} [  \widetilde{\Pi } \widetilde{\rho^{*}}    ]    \big\}                           }  ,         \rho^* \big/ \widetilde{\rho^*}                       \bigg]}$ 

       \bigskip

        \item[$\bullet$] \textbf{(7)}. $\text{Td upper bound} \equiv \epsilon \propto    \boxed{  \big\{ {\tiny  {2 \mathcal{C} \mathcal{C}^{\prime}  \big( \mathcal{C}^{\prime\prime} \big)^3  \big( C C^{\prime} C^{\prime\prime} \big)^2 } } \big\}^{-2}  }      $

\end{itemize}

}

\bigskip 

\noindent We describe how the series of up to constants estimates that are used simultaneously for obtaining the above desired estimate on the trace distance $\mathrm{Td}$.

    \bigskip

    \item[$\bullet$] \textit{(C). The ratio of noisy Quantum states to noiseless Quantum states satisfies an inequality with the negligibilty function}. Under the assumption,

  {\small  \begin{align*}  \frac{1}{2} \mathrm{Tr} \bigg\{   \sqrt{ \frac{ {\tiny \big( \widetilde{\rho - \tau} \big)^{\dagger}}}{{\tiny \big( \rho - \tau \big)^{\dagger } }} \frac{{\tiny \big( \widetilde{\rho - \tau } \big)}}{{\tiny \big( \rho - \tau \big) } }}          \bigg\}  \gtrsim 1 -   {\tiny \mathrm{NEGL} \big(      \lambda^{\prime}  - \lambda     \big) }  
 ,  \end{align*} }

    \noindent one has, by the GML,

{\small

\begin{align*}
 \mathrm{Td} \bigg[    \frac{ \big\{ \Pi \rho^* \Pi / \widetilde{\Pi} \widetilde{\rho^*} \widetilde{\Pi}  \big\}            }{  \big\{  \mathrm{Tr} \big[        \Pi \rho^*   \big] /     \mathrm{Tr} \big[  \widetilde{\Pi } \widetilde{\rho^{*}}    \big]    \big\}                           }  ,          \frac{ \big\{ \ket{c , \sigma_c}  \bra{c , \sigma_c }   \big\}     }{ \big\{ \ket{\widetilde{c} , \widetilde{\sigma_c}}   \bra{\widetilde{c} , \widetilde{\sigma_c }} \big\}   }             \bigg]      \lesssim  \sqrt{ \bigg\{  {\tiny \frac{\mathrm{negl} \big(    \lambda    \big)}{\mathrm{NEGL} \big(      \lambda^{\prime}      \big)}} \bigg\}^{-1}}     \overset{(\mathrm{GM})}{\leq}    2 \sqrt{ {\tiny  \mathrm{NEGL} \big(      \lambda^{\prime}      \big) }}  \\    \lesssim                   {\tiny \mathrm{exp} \big[   \lambda - \lambda^{\prime}        \big] }   ,  \\ 
\end{align*}

}

\noindent applied in the second inequality sign above.

\end{itemize}

\noindent We state the main result, which will be shown to hold by making use of the series of estimates stated in the result \textbf{Lemma} above.

\bigskip 

\noindent \textbf{Theorem} (\textit{the notion of everlasting security in the NQPKE-QKD setting is captured through an inequality between the trace distance and the negligibility function under noise}). According to the series of protocols,

{\small \[ \left\{\!\begin{array}{ll@{}l} 
\textbf{(1).}  \text{ } \textit{Secret key generation independently of noise in the QPKE-QKD protocol}. \text{ Denote $\mathrm{sk}$ as the can-} \\ \text{didate secret key for the QPKE-QKD protocol which is generated according to the noiseless secret key} \\ \text{generation scheme}  . \\ \\   \textbf{(2).}  \text{ } \textit{Extracting the desired noiseless secret key from the noiseless public key generation scheme}. \text{ Denote} \\ \text{ $\mathrm{pk}$ as the candidate public key for the QPKE-QKD protocol, from which $\mathrm{sk}$ above is constructed} \\  \text{through the following steps.} \\ \\       \textbf{(3).}  \text{ } \textit{Noising the secret key generation scheme}. \text{ Denote the noised secret key $\widetilde{\mathrm{sk}}$ as the noisy secret key} \\ \text{in the NQPKE-QKD protocol obtained from noising the public key $\mathrm{pk}$ with $\widetilde{\mathrm{pk}}$ obtained from the protocol} \\ \text{described in the second step. }        \\ \\ \textbf{(4).}  \text{ } \textit{Obtaining the noisy ciphertext from $\widetilde{\mathrm{sk}}$}. \text{ Denote the ciphertext under noise as $\ket{\widetilde{\sigma_c}}$ obtained through} \\ \text{the direct summand operation $\widetilde{m} \oplus \widetilde{\mathrm{sk}}$.}     
\end{array}\right. 
\]  }

\bigskip

\noindent suppose that the NQPKE-QKD protocol is correct, ie that the protocol satisfies the assumption,

{\small

\begin{align*}
 \mathcal{P} \bigg[ {\tiny \bot = \widetilde{\mathrm{Dec}} \big( \widetilde{\mathrm{sk}} ,  \widetilde{\mathrm{ct}} \big)  :       \big\{ \widetilde{\mathrm{sk}} \leftarrow \widetilde{\mathrm{SKGen}} \big( 1^{\lambda^{\prime}} \big) \big\} , \big\{ \big( \widetilde{\rho} , \widetilde{\mathrm{pk}} \big) \leftarrow \widetilde{\mathrm{PKGen}} \big( \widetilde{\mathrm{sk}} \big) \big\} , \big\{   \widetilde{\mathrm{ct}} \leftarrow \widetilde{\mathrm{Enc}} \big( \widetilde{\rho} , \widetilde{\mathrm{pk}} , m^{\prime} \big)     \big\}      ,    } \\ \\  {\tiny \big\{  \mathrm{sk}   \leftarrow   \mathrm{SKGen}  \big( 1^{\lambda^{\prime}} \big) \big\}   }   \bigg] = 0   , \\ \\  \mathcal{P} \bigg[  {\tiny \widetilde{k}   = \widetilde{\mathrm{Dec}}  \big( \widetilde{\mathrm{sk}} ,  \widetilde{\mathrm{ct}} \big)  :        \big\{ \widetilde{\mathrm{sk}} \leftarrow \widetilde{\mathrm{SKGen}} \big( 1^{\lambda^{\prime}} \big) \big\} , \big\{ \big( \widetilde{\rho} , \widetilde{\mathrm{pk}} \big) \leftarrow \widetilde{\mathrm{PKGen}} \big( \widetilde{\mathrm{sk}} \big) \big\} , \big\{   \widetilde{\mathrm{ct}} \leftarrow \widetilde{\mathrm{Enc}} \big( \widetilde{\rho} , \widetilde{\mathrm{pk}} , m^{\prime} \big)     \big\}      , } \\ \\   {\tiny  \big\{  \mathrm{sk}   \leftarrow   \mathrm{SKGen}  \big( 1^{\lambda^{\prime}} \big) \big\}    }   \bigg]  = 1    , \\ 
\end{align*}

}

\noindent for some $\widetilde{k}$, from the fact that, by construction,

{\small

\begin{align*}
  {\tiny \widetilde{\mathrm{SKGen}}  }  \longleftrightarrow  {\tiny \big\{ \mathrm{SKGen}}  \otimes \textit{noise}   {\tiny \big\} }    , 
\end{align*}

}

\noindent and the symbol '$\bot$' denotes an instance of failure of the decoding algorithm, from which the NQPKE-QKD protocol aborts. Then one has that the NQPKE-QKD protocol satisfies,

{\small

\begin{align*}
  \mathrm{Td} \big(    {\tiny \widetilde{\mathrm{Exp}}^{A_{\lambda^{\prime}}}  \big( 1^{\lambda^{\prime}} , 1 \big)     }      ,   {\tiny \widetilde{\mathrm{Exp}}^{A_{\lambda^{\prime}}} \big( 1^{\lambda^{\prime}} , 0 \big)     }    \big)  \lesssim   {\tiny \mathrm{NEGL} \big( \lambda^{\prime} \big) }   , \\ 
\end{align*}

}

\noindent in the noisy setting for the security threshold $\lambda^{\prime}$, as opposed to the statement of everlasting security, 

{\small

\begin{align*}
    \mathrm{Td} \big(    {\tiny {\mathrm{Exp}}^{A_{\lambda}}  \big( 1^{\lambda} , 1 \big)     }      ,   {\tiny {\mathrm{Exp}}^{A_{\lambda}} \big( 1^{\lambda} , 0 \big)     }    \big) =  {\tiny \mathrm{negl} \big( \lambda \big) }   , \\     
\end{align*}

}

\noindent in the noiseless setting for the security threshold $\lambda$.

\bigskip

\noindent Given the two previous results above with the $\textbf{Lemma}$ and $\textbf{Theorem}$ we conclude the statement of the main results below with the following corollary.

\bigskip

\noindent \textbf{Corollary} (\textit{the negligibility function in the up to constants lower bound in the} $\mathrm{\textbf{Lemma}}$ \textit{can be used as an up to constants upper bound for three noisy advantage functions}). For the collection of noisy hybrid experiments,

{\small

\begin{align*}
   \big\{        {\tiny \widetilde{\mathrm{Hyb}^{\mathcal{A}_{\lambda^{\prime}}}_{0,0}}     ,   \widetilde{\mathrm{Hyb}^{\mathcal{A}_{\lambda^{\prime}}}_{1,0} }  }       \big\}  , \\ \\   \big\{        {\tiny \widetilde{\mathrm{Hyb}^{\mathcal{A}_{\lambda^{\prime}}}_{0,1}}     ,   \widetilde{\mathrm{Hyb}^{\mathcal{A}_{\lambda^{\prime}}}_{1,1} }  }        \big\}  , \\ \\   \big\{        {\tiny \widetilde{\mathrm{Hyb}^{\mathcal{A}_{\lambda^{\prime}}}_{0,2}}     ,   \widetilde{\mathrm{Hyb}^{\mathcal{A}_{\lambda^{\prime}}}_{1,2} }  }       \big\}  , \\ 
\end{align*}

}

\noindent for $i=0$, $1$ and $2$, respectively, one has that,

{\small

\begin{align*}
{\tiny \widetilde{\mathrm{Adv}} \big( 0 \big) \lesssim  \widetilde{\mathrm{Adv}} \big( 1 \big) \lesssim  \widetilde{\mathrm{Adv}} \big( 2 \big) \lesssim \mathrm{NEGL} \big( \lambda^{\prime} \big)     }   , \\ 
\end{align*}

}

\noindent given the noisy advantage function,

{\small

\begin{align*}
{   \widetilde{\mathrm{Adv}} \big( i \big) \equiv \mathrm{Td} \big[   {\tiny \widetilde{\mathrm{Hyb}^{\mathcal{A}_{\lambda^{\prime}}}_{0,i}}     ,   \widetilde{\mathrm{Hyb}^{\mathcal{A}_{\lambda^{\prime}}}_{1,i} }  }  \big]   =       \mathrm{Td} \big[   {\tiny \widetilde{\mathrm{Hyb}^{\mathcal{A}_{\lambda^{\prime}}}_0}     ,   \widetilde{\mathrm{Hyb}^{\mathcal{A}_{\lambda^{\prime}}}_1 }  }  \big]   }      , \\ 
\end{align*}

}

\noindent introduced in \textbf{Definition} \textit{38}.

\section{Arguments of main results}

\noindent We demonstrate that each of the three results stated in the previous section hold with the following arguments.

\subsection{Proof of Lemma}

\noindent \textit{Proof of Lemma}. To argue that (A),

{\small \begin{align*}
  \\  \frac{\delta\big( q \big)}{ \widetilde{\delta} \big( q \big)}   \lesssim              \bigg\{   \underset{0 < j^{\prime} \leq q}{\underset{0 \leq j \leq q}{\sum}}            \bigg\{  \frac{\ket{c_j}}{\ket{\widetilde{c_{j^{\prime}}}}}         \bigg\}        \bigotimes                  \underset{0 < j^{\prime} \leq q}{\underset{0 \leq j \leq q}{\sum}}            \bigg\{      \frac{\ket{\sigma_j}}{\ket{\widetilde{\sigma_{j^{\prime}}}}}             \bigg\}    \bigg\}  \bigg\{    \underset{0 < j^{\prime} \leq q}{\underset{0 \leq j \leq q}{\sum}}            \bigg\{  \frac{\bra{c_j}}{\bra{\widetilde{c_{j^{\prime}}}}}         \bigg\}        \bigotimes                  \underset{0 < j^{\prime} \leq q}{\underset{0 \leq j \leq q}{\sum}}            \bigg\{      \frac{\bra{\sigma_j}}{\bra{\widetilde{\sigma_{j^{\prime}}}}}             \bigg\}             \bigg\}   , \\ \end{align*} }

  \noindent for bounding the ratio, 

   \begin{align*}
   \underset{\textit{and Eve}}{\underset{\textit{by Alice}}{\underset{\textit{queries} \text{ } q}{\bigcup}}} \bigg\{       \frac{\delta\big( q \big)}{ \widetilde{\delta} \big( q \big)}  \bigg\}    
   \equiv    \frac{\delta\big( q \big)}{ \widetilde{\delta} \big( q \big)}      ,       \\
    \end{align*}

    }

    \noindent holds it suffices to argue that rearrangements of the form,

{\small \begin{align*}
  \\  \frac{\delta\big( q \big)}{ \widetilde{\delta} \big( q \big)} \propto   \underset{0 <  j \leq q}{\sum}       \bigg\{       \bigg\{   \ket{\frac{c_j   }{\widetilde{c_j}}}       \bigotimes \ket{\frac{\sigma_j    }{\widetilde{\sigma_j}}}  \bigg\}  \bigg\{   \bra{\frac{c_j   }{\widetilde{c_j}}}    \bigotimes         \bra{\frac{\sigma_j    }{\widetilde{\sigma_j}}}    \bigg\}                             \bigg\}   \leq   \underset{0 <  j^{\prime} \leq q}{\underset{0 \leq   j \leq q}{\sum}}       \bigg\{       \bigg\{   \ket{\frac{c_j   }{\widetilde{c_{j^{\prime}}}}}       \bigotimes \ket{\frac{\sigma_j    }{\widetilde{\sigma_{j^{\prime}}}}}  \bigg\}  \bigg\{   \bra{\frac{c_j   }{\widetilde{c_{j^{\prime}}}}}    \\  \bigotimes         \bra{\frac{\sigma_j    }{\widetilde{\sigma_{j^{\prime}}}}}    \bigg\}                             \bigg\} \end{align*}

  \begin{align*} \lesssim                \underset{0 < j^{\prime} \leq q}{\underset{0 \leq j \leq q}{\sum}}            \bigg\{      \frac{\ket{c_j } \otimes    \ket{\sigma_j } \bra{c_j } \otimes   \bra{\sigma_j }}{\ket{\widetilde{c_{j^{\prime}}}} \otimes   \ket{\widetilde{\sigma_{j^{\prime}}}} \bra{\widetilde{c_{j^{\prime}}}} \otimes    \bra{\widetilde{\sigma_{j^{\prime}} }}}                                \bigg\}  \\ \\   \lesssim              \bigg\{   \underset{0 < j^{\prime} \leq q}{\underset{0 \leq j \leq q}{\sum}}            \bigg\{  \frac{\ket{c_j}}{\ket{\widetilde{c_{j^{\prime}}}}}         \bigg\}        \bigotimes                  \underset{0 < j^{\prime} \leq q}{\underset{0 \leq j \leq q}{\sum}}            \bigg\{      \frac{\ket{\sigma_j}}{\ket{\widetilde{\sigma_{j^{\prime}}}}}             \bigg\}    \bigg\} \\  \\ \times \bigg\{    \underset{0 < j^{\prime} \leq q}{\underset{0 \leq j \leq q}{\sum}}            \bigg\{  \frac{\bra{c_j}}{\bra{\widetilde{c_{j^{\prime}}}}}         \bigg\}        \bigotimes                  \underset{0 < j^{\prime} \leq q}{\underset{0 \leq j \leq q}{\sum}}            \bigg\{      \frac{\bra{\sigma_j}}{\bra{\widetilde{\sigma_{j^{\prime}}}}}             \bigg\}             \bigg\}   , \\ \end{align*} }

    \noindent imply the existence of a transformation on the ciphertext, $\widetilde{F}$, which satisfies:

{\small \begin{itemize}
    \item[$\bullet$] \textit{(1). Noisy forwarding error probability}. One has that,

    {\small
    
    \begin{align*}
         \big|     \mathcal{P} \big[  {\tiny \widetilde{F} \big( \widetilde{m^{q_A}}  , \widetilde{m^{\prime}_{q_E}} \big)  = \widetilde{m_{q_E + 1}}    }  \big]     -      \mathcal{P} \big[ {\tiny  \widetilde{F} \big( \widetilde{{m}^{q_A}} , \widetilde{{m^{\prime}}_{q_E}}  \big)  = \widetilde{{m}_{q_E + 1}  }     }  \big]       \big| \leq \widetilde{\delta_1}  \big(    q \big)      . \\ 
    \end{align*}

    }

    \bigskip

       \item[$\bullet$] \textit{(2). Noisy deleting error probability}. One has that,

{\small

\begin{align*}
      \mathcal{P} \big[ {\tiny \widetilde{F} \big( \widetilde{m_{q_A}} , \widetilde{m^{\prime}_{q_E}} \big) = \widetilde{m_{q_E + 1}}    }  \big] \leq \widetilde{\delta_2}  \big( q \big)      . \\ 
\end{align*}

}

       \bigskip

         \item[$\bullet$] \textit{(3). Noisy reconstruction error probability}. One has that,

         {\small

         \begin{align*}
             \big|   \mathcal{P} \big[         {\tiny \widetilde{F} \big( \widetilde{m_{q_A}} , \widetilde{m^{\prime}_{q_E}} \big) = \widetilde{m_{q_E}}   }   \big]       -  \mathcal{P} \big[ {\tiny     R \big( \widetilde{M} \big) = \widetilde{m_{q_E + 1}}      }   \big]       \big| \leq \widetilde{\delta_3}  \big( q \big)                 .  \\ 
         \end{align*}
         
         }
    
\end{itemize} }

\noindent Specifically,

{\small   \begin{align*}  {\tiny \Pi \rho^* \Pi / \widetilde{\Pi} \widetilde{\rho^*} \widetilde{\Pi}       }  \overset{\mathrm{(B.1.1)}}{\lesssim}  {\tiny \big\{ \Pi / \widetilde{\Pi}  \big\}  \rho^* / \widetilde{\rho^*}         \big\{  \Pi / \widetilde{\Pi}    \big\}     }         \\ \end{align*}

\begin{align*} \overset{\mathrm{(B.1.2)}}{\lesssim}    {\tiny \underset{\sigma_1 \in \Sigma_1}{\underset{\sigma_0 \in \Sigma_0}{\sum}}     \bigg\{        \big\{ \Pi \big( \sigma_0 , \sigma_1  \big)  / \widetilde{\Pi} \big( \sigma_0 , \sigma_1  \big)  \big\}  \rho^* / \widetilde{\rho^*}         \big\{  \Pi \big( \sigma_0 ,  \sigma_1 \big)  / \widetilde{\Pi} \big( \sigma_0 , \sigma_1 \big)     \big\}  \bigg\}    } \\ \\      \overset{\mathrm{(B.1.3)}}{\lesssim}  \bigg[ {\tiny \underset{\sigma_1 \in \Sigma_1}{\underset{\sigma_0 \in \Sigma_0 }{\sum }}  \bigg\{         \Pi \big( \sigma_0 , \sigma_1  \big)  / \widetilde{\Pi} \big( \sigma_0 , \sigma_1  \big)                      \bigg\}  }     \bigg]  {\tiny   \rho^* / \widetilde{\rho^*}  }                  \bigg[   {\tiny \underset{\sigma_1 \in \Sigma_1}{\underset{\sigma_0 \in \Sigma_0 }{\sum }}    \bigg\{                       \Pi \big(\sigma_0 ,  \sigma_1 \big)  / \widetilde{\Pi} \big( \sigma_0 ,  \sigma_1 \big)        \bigg\} }  \bigg]           \\  \end{align*}

\begin{align*} =  {\tiny   \underset{\sigma_1 \in \Sigma_1}{\underset{\sigma_0 \in \Sigma_0 }{\sum }}  \big\{         \Pi \big( \sigma_0 , \sigma_1  \big)    \rho^*  / \widetilde{\Pi} \big( \sigma_0 , \sigma_1 \big)                      \big\}  \cdot      \big\{                        \Pi \big( \sigma_0 ,  \sigma_1 \big)  / \widetilde{\Pi} \big( \sigma_0 , \sigma_1 \big)  \widetilde{\rho^*}          \big\}  }    \\     \end{align*}

\begin{align*}
  \overset{\mathrm{(B.1.4)}}{\lesssim}  \bigg[    \underset{\sigma_1 \in \Sigma_1}{\underset{\sigma_0 \in \Sigma_0 }{\sum }} \bigg\{   {\tiny \frac{ \ket{0} \otimes \ket{\sigma_0} \bra{\sigma_0} \otimes  \bra{0}}{ \ket{0} \otimes \ket{\sigma_0 + \textit{noise} } \bra{\sigma_0 + \textit{noise} } \otimes  \bra{0}    +   \ket{1} \otimes   \ket{\sigma_1 + \textit{noise} } \bra{\sigma_1 + \textit{noise} }  \otimes \bra{1}  }   }    \bigg\} \rho^*   \\ +           \underset{\sigma_1 \in \Sigma_1}{\underset{\sigma_0 \in \Sigma_0 }{\sum }} \bigg\{   {\tiny \frac{ \ket{1} \otimes \ket{\sigma_1} \bra{\sigma_1} \otimes  \bra{1}}{ \ket{0} \otimes \ket{\sigma_0 + \textit{noise} } \bra{\sigma_0 + \textit{noise} } \otimes  \bra{0}    +   \ket{1} \otimes   \ket{\sigma_1 + \textit{noise} } \bra{\sigma_1 + \textit{noise} }  \otimes \bra{1}  }   }    \bigg\} \rho^*  \bigg]   \\  \cdot \bigg[       \bigg\{   \underset{\sigma_1 \in \Sigma_1}{\underset{\sigma_0 \in \Sigma_0 }{\sum }} \bigg\{   {\tiny \frac{ \ket{0} \otimes \ket{\sigma_0} \bra{\sigma_0} \otimes  \bra{0}}{ \ket{0} \otimes \ket{\sigma_0 + \textit{noise} } \bra{\sigma_0 + \textit{noise} } \otimes  \bra{0}    +   \ket{1} \otimes   \ket{\sigma_1 + \textit{noise} } \bra{\sigma_1 + \textit{noise} }  \otimes \bra{1}  }   }    \bigg\} / \widetilde{\rho^*}   \\ +           \underset{\sigma_1 \in \Sigma_1}{\underset{\sigma_0 \in \Sigma_0 }{\sum }} \bigg\{   {\tiny \frac{ \ket{1} \otimes \ket{\sigma_1} \bra{\sigma_1} \otimes  \bra{1}}{ \ket{0} \otimes \ket{\sigma_0 + \textit{noise} } \bra{\sigma_0 + \textit{noise} } \otimes  \bra{0}    +   \ket{1} \otimes   \ket{\sigma_1 + \textit{noise} } \bra{\sigma_1 + \textit{noise} }  \otimes \bra{1}  }   }    \bigg\}  / \widetilde{\rho^*}  \bigg\}       \bigg]                  \\ 
\end{align*}

  \begin{align*} \overset{\mathrm{(B.1.5)}}{\lesssim}      \underset{\sigma_0 \in \Sigma_0}{\sum} {\tiny \bigg\{        \ket{0} \bigotimes   \ket{\frac{\sigma_0}{\sigma_0 + \textit{noise}}}                             \bigg\}     \bigg\{   \bra{\frac{\sigma_0}{\sigma_0 + \textit{noise}}}   \bigotimes    \bra{0}                 \bigg\}    \rho^* / \widetilde{\rho^*}  +  \underset{\sigma_1 \in \Sigma_1}{\sum}     \bigg\{        \ket{1} \bigotimes   \ket{\frac{\sigma_1}{\sigma_1 + \textit{noise}}}                             \bigg\}   }   \\ {\tiny \times \bigg\{   \bra{\frac{\sigma_1}{\sigma_1 + \textit{noise}}}   \bigotimes    \bra{1}                 \bigg\}      \rho^* / \widetilde{\rho^*}   }   \\ \end{align*}

  \begin{align*} 
   \overset{\mathrm{(B.1.6)}}{\lesssim}    \underset{0 < j^{\prime} \leq q}{\underset{0 \leq j \leq q}{\sum}} \bigg[     {\tiny       \underset{\sigma_0 \in \Sigma_0}{\sum}  \bigg\{        \ket{0} \bigotimes   \ket{\frac{\sigma_0}{\sigma_0 + \textit{noise}}}                             \bigg\}     \bigg\{   \bra{\frac{\sigma_0}{\sigma_0 + \textit{noise}}}   \bigotimes    \bra{0}                 \bigg\}   \rho^* / \widetilde{\rho^*}   }   +  \underset{\sigma_1 \in \Sigma_1}{\sum}          {\tiny  \bigg\{   \ket{1} }  \\  {\tiny \bigotimes   \ket{\frac{\sigma_1}{\sigma_1 + \textit{noise}}}                             \bigg\}    \bigg\{   \bra{\frac{\sigma_1}{\sigma_1 + \textit{noise}}}   \bigotimes    \bra{1}                 \bigg\}  \rho^* / \widetilde{\rho^*} }  \bigg]  \\ \end{align*}

   \begin{align*} \Longrightarrow                         \Pi \rho^* \Pi / \widetilde{\Pi} \widetilde{\rho^*} \widetilde{\Pi}        \overset{\mathrm{(B.1)}}{\lesssim}     \underset{0 < j^{\prime} \leq q}{\underset{0 \leq j \leq q}{\sum}} \bigg[       {\tiny    \underset{\sigma_0 \in \Sigma_0}{\sum}  \bigg\{        \ket{0} \bigotimes   \ket{\frac{\sigma_0}{\sigma_0 + \textit{noise}}}                             \bigg\}     \bigg\{   \bra{\frac{\sigma_0}{\sigma_0 + \textit{noise}}}   \bigotimes    \bra{0}                 \bigg\} }   \\ {\tiny \times \rho^* / \widetilde{\rho^*} }  \\   \\ +  {\tiny \underset{\sigma_1 \in \Sigma_1}{\sum}     \bigg\{        \ket{1}  \bigotimes   \ket{\frac{\sigma_1}{\sigma_1 + \textit{noise}}}                             \bigg\}   \bigg\{   \bra{\frac{\sigma_1}{\sigma_1 + \textit{noise}}}   \bigotimes    \bra{1}                 \bigg\} }   \\ {\tiny \times \rho^* / \widetilde{\rho^*}  }  \bigg]         , \\  \end{align*} }

\noindent corresponding to \textbf{(1)},

   {\small  \begin{align*}   \mathrm{Tr} \big[        \Pi \rho^*   \big] /     \mathrm{Tr} \big[  \widetilde{\Pi } \widetilde{\rho^{*}}    \big]              \overset{\mathrm{(B.2.1)}}{\lesssim}   \underset{0 < j^{\prime} \leq q}{\underset{0 \leq j \leq q}{\sum}}      \mathrm{Tr}  \bigg\{  \bigg[              {\tiny       \underset{\sigma_0 \in \Sigma_0}{\sum}  \bigg\{        \ket{0} \bigotimes   \ket{\frac{\sigma_0}{\sigma_0 + \textit{noise}}}                             \bigg\}     \bigg\{   \bra{\frac{\sigma_0}{\sigma_0 + \textit{noise}}}   \bigotimes    \bra{0}                 \bigg\} }  \\     +  \underset{\sigma_1 \in \Sigma_1}{\sum}     {\tiny \bigg\{        \ket{1} \bigotimes   \ket{\frac{\sigma_1}{\sigma_1 + \textit{noise}}}                             \bigg\}   \bigg\{   \bra{\frac{\sigma_1}{\sigma_1 + \textit{noise}}}   \bigotimes    \bra{1}                 \bigg\}             }     \bigg] \\ \times                       \bigg\{  {\tiny \ket{\frac{\big( \rho^* \big)_j}{\big( \widetilde{\rho_*} \big)_{j^{\prime}}}}   \bra{\frac{\big( \rho^* \big)_j}{\big( \widetilde{\rho_*} \big)_{j^{\prime}}}} }   \bigg\}     \bigg\}  \\  \end{align*}

      \begin{align*} \overset{\mathrm{(B.2.2)}}{\lesssim}   \underset{0 < j^{\prime} \leq q}{\underset{0 \leq j \leq q}{\sum}}         \mathrm{Tr}  \bigg\{  \bigg[                 {\tiny \underset{\sigma_0 \in \Sigma_0}{\sum}  \bigg\{        \ket{0} \bigotimes   \ket{\frac{\sigma_0}{\sigma_0 + \textit{noise}}}                             \bigg\}     \bigg\{   \bra{\frac{\sigma_0}{\sigma_0 + \textit{noise}}}   \bigotimes    \bra{0}                 \bigg\}  }      +  \underset{\sigma_1 \in \Sigma_1}{\sum}     \bigg\{     {\tiny   \ket{1} } \\ {\tiny \bigotimes   \ket{\frac{\sigma_1}{\sigma_1 + \textit{noise}}}    }                         \bigg\}   \bigg\{  {\tiny  \bra{\frac{\sigma_1}{\sigma_1 + \textit{noise}}}   \bigotimes    \bra{1}      }            \bigg\}                 \bigg]           \bigg\{   {\tiny \frac{\ket{\big( \rho^* \big)_j}}{\ket{\big( \widetilde{\rho_*} \big)_{j^{\prime}}}}  \frac{\bra{\big( \rho^* \big)_j}}{\bra{\big( \widetilde{\rho_*} \big)_{j^{\prime}}}}              }   \bigg\}     \bigg\}     \\   \end{align*}

      \begin{align*} \\   \overset{\mathrm{(B.2.3)}}{\lesssim}          \underset{0 < j^{\prime} \leq q}{\underset{0 \leq j \leq q}{\sum}}         \mathrm{Tr}  \bigg\{                  \underset{\sigma_0 \in \Sigma_0}{\sum}  \bigg\{    {\tiny    \ket{0} \bigotimes   \ket{\frac{\sigma_0}{\sigma_0 + \textit{noise}}}      }                       \bigg\}     \bigg\{ {\tiny  \bra{\frac{\sigma_0}{\sigma_0 + \textit{noise}}}   \bigotimes    \bra{0}          }     \bigg\}        \bigg\{   {\tiny \frac{\ket{\big( \rho^* \big)_j}}{\ket{\big( \widetilde{\rho_*} \big)_{j^{\prime}}}}  \frac{\bra{\big( \rho^* \big)_j}}{\bra{\big( \widetilde{\rho_*} \big)_{j^{\prime}}}}      }           \bigg\}       \bigg\}            \\   +   \underset{0 < j^{\prime} \leq q}{\underset{0 \leq j \leq q}{\sum}}         \mathrm{Tr}  \bigg\{  \underset{\sigma_1 \in \Sigma_1}{\sum}     \bigg\{    {\tiny    \ket{1}  \bigotimes   \ket{\frac{\sigma_1}{\sigma_1 + \textit{noise}}}                        }     \bigg\}   \bigg\{   {\tiny \bra{\frac{\sigma_1}{\sigma_1 + \textit{noise}}}   \bigotimes    \bra{1}             }     \bigg\}                                 \bigg\{  {\tiny \frac{\ket{\big( \rho^* \big)_j}}{\ket{\big( \widetilde{\rho_*} \big)_{j^{\prime}}}} }  \\ \times   {\tiny \frac{\bra{\big( \rho^* \big)_j}}{\bra{\big( \widetilde{\rho_*} \big)_{j^{\prime}}}}              }   \bigg\}     \bigg\}                              \\  \\        \\    \overset{\mathrm{(B.2.4)}}{\lesssim}          \underset{\sigma_0 \in \Sigma_0}{\underset{0 < j^{\prime} \leq q}{\underset{0 \leq j \leq q}{\sum}}}         \mathrm{Tr}  \bigg\{                   \bigg\{      {\tiny  \ket{0} \bigotimes   \ket{\frac{\sigma_0}{\sigma_0 + \textit{noise}}}                       }      \bigg\}     \bigg\{ {\tiny  \bra{\frac{\sigma_0}{\sigma_0 + \textit{noise}}}   \bigotimes    \bra{0}          }     \bigg\}        \bigg\{  {\tiny \frac{\ket{\big( \rho^* \big)_j}}{\ket{\big( \widetilde{\rho_*} \big)_{j^{\prime}}}}  \frac{\bra{\big( \rho^* \big)_j}}{\bra{\big( \widetilde{\rho_*} \big)_{j^{\prime}}}}             }   \bigg\}       \bigg\}            \\   +   \underset{\sigma_1 \in \Sigma_1}{\underset{0 < j^{\prime} \leq q}{\underset{0 \leq j \leq q}{\sum}}}         \mathrm{Tr}  \bigg\{    \bigg\{     {\tiny   \ket{1}  \bigotimes   \ket{\frac{\sigma_1}{\sigma_1 + \textit{noise}}}            }                 \bigg\}   \bigg\{   {\tiny \bra{\frac{\sigma_1}{\sigma_1 + \textit{noise}}}   \bigotimes    \bra{1}              }   \bigg\}                                 \bigg\{  {\tiny \frac{\ket{\big( \rho^* \big)_j}}{\ket{\big( \widetilde{\rho_*} \big)_{j^{\prime}}}}  \frac{\bra{\big( \rho^* \big)_j}}{\bra{\big( \widetilde{\rho_*} \big)_{j^{\prime}}}}         }        \bigg\}     \bigg\}   \\ \\              \\    \overset{\mathrm{(B.2.5)}}{\lesssim }          \underset{\sigma_0 \in \Sigma_0}{\underset{0 < j^{\prime} \leq q}{\underset{0 \leq j \leq q}{\sum}}}      \bigg\{ \bigg[    \mathrm{Tr}  \big\{  \ket{0} \big\}                \bigotimes  \mathrm{Tr}  \bigg\{   {\tiny \ket{\frac{\sigma_0}{\sigma_0 + \textit{noise}}} } \bigg\}  \bigg]   \bigg[ \mathrm{Tr}  \bigg\{   {\tiny \bra{\frac{\sigma_0}{\sigma_0 + \textit{noise}}} }  \bigg\}              \bigotimes  \mathrm{Tr}  \big\{  \bra{0} \big\}              \bigg]           \bigg\}  \\ \\ + \underset{\sigma_1 \in \Sigma_1}{\underset{0 < j^{\prime} \leq q}{\underset{0 \leq j \leq q}{\sum}}}      \bigg\{   \bigg[   \mathrm{Tr}  \big\{  \ket{1} \big\}                 \bigotimes        \mathrm{Tr}  \bigg\{    {\tiny \ket{\frac{\sigma_2}{\sigma_1 + \textit{noise}}} } \bigg\}                 \bigg]    \bigg[  \mathrm{Tr}  \bigg\{    \bra{\frac{\sigma_1}{\sigma_1 + \textit{noise}}} \bigg\}              \bigotimes  \mathrm{Tr}  \big\{  \bra{1} \big\}          \bigg]        \bigg\}  \\ \end{align*}

      \begin{align*} \Longrightarrow      \mathrm{Tr} \big[        \Pi \rho^*   \big] /     \mathrm{Tr} \big[  \widetilde{\Pi } \widetilde{\rho^{*}}    \big]              \overset{\mathrm{(B.2)}}{\lesssim}         \underset{\sigma_0 \in \Sigma_0}{\underset{0 < j^{\prime} \leq q}{\underset{0 \leq j \leq q}{\sum}}}      \bigg\{ \bigg[ {\tiny   \mathrm{Tr}  \big\{  \ket{0} \big\}                \bigotimes  \mathrm{Tr}  \bigg\{    \ket{\frac{\sigma_0}{\sigma_0 + \textit{noise}}} \bigg\}  \bigg]   \bigg[ \mathrm{Tr}  \bigg\{    \bra{\frac{\sigma_0}{\sigma_0 + \textit{noise}}} } \bigg\}      \\ \end{align*}

      \begin{align*} {\tiny     \bigotimes  \mathrm{Tr}  \big\{  \bra{0} \big\}   }            \bigg]           \bigg\}  \\ \\ + \underset{\sigma_1 \in \Sigma_1}{\underset{0 < j^{\prime} \leq q}{\underset{0 \leq j \leq q}{\sum}}}      \bigg\{   \bigg[  {\tiny \mathrm{Tr}  \big\{  \ket{1} \big\}                 \bigotimes        \mathrm{Tr}  \bigg\{    \ket{\frac{\sigma_2}{\sigma_1 + \textit{noise}}} \bigg\}                 \bigg]    \bigg[  \mathrm{Tr}  \bigg\{    \bra{\frac{\sigma_1}{\sigma_1 + \textit{noise}}} }  \bigg\}          \\ \\     \bigotimes {\tiny \mathrm{Tr}  \big\{  \bra{1} \big\}      }     \bigg]        \bigg\}        ,       \\              \end{align*} }

\noindent corresponding to \textbf{(2)},

      {\small \begin{align*} \frac{\ket{c, \sigma_c}}{\ket{\widetilde{c}, \widetilde{\sigma_c}}}       \overset{\mathrm{(B.3.1)}}{\lesssim}             \underset{0 < j^{\prime} \leq q}{\underset{0 \leq j \leq q}{\sum}}         \bigg\{    {\tiny \ket{\frac{c}{c^{\prime}}}     \bigotimes   \ket{\frac{\sigma_c}{\widetilde{\sigma_c}}}      }  \bigg\}     \overset{\mathrm{(B.3.2)}}{\lesssim}  \underset{0 < j^{\prime} \leq q}{\underset{0 \leq j \leq q}{\sum}}           \bigg\{   {\tiny \frac{ \ket{c} \bigotimes  \ket{\sigma_c} }{ \ket{\widetilde{c}} \bigotimes \ket{\widetilde{\sigma_c }}  } }  \bigg\}   \overset{\mathrm{(B.3.3)}}{\lesssim}  \underset{0 < j^{\prime} \leq q}{\underset{0 \leq j \leq q}{\sum}}           \bigg\{  {\tiny \frac{\ket{c}}{\ket{\widetilde{c}}} }  \bigg\} \\ \\  \bigotimes     \underset{0 < j^{\prime} \leq q}{\underset{0 \leq j \leq q}{\sum}}           \bigg\{   {\tiny \frac{\ket{\sigma_c }}{\ket{\widetilde{\sigma_c}}} }    \bigg\}  \\ \end{align*}

      \begin{align*} \Longrightarrow   \frac{\ket{c, \sigma_c}}{\ket{\widetilde{c}, \widetilde{\sigma_c}}}       \overset{\mathrm{(B.3)}}{\lesssim}  \underset{0 < j^{\prime} \leq q}{\underset{0 \leq j \leq q}{\sum}}           \bigg\{   \frac{\ket{c}}{\ket{\widetilde{c}}}  \bigg\}  \bigotimes     \underset{0 < j^{\prime} \leq q}{\underset{0 \leq j \leq q}{\sum}}           \bigg\{   \frac{\ket{\sigma_c }}{\ket{\widetilde{\sigma_c}}}    \bigg\}    ,  \\ \end{align*}  }

\noindent corresponding to \textbf{(3)},

{\small

       \begin{align*} \frac{\bra{c, \sigma_c}}{\bra{\widetilde{c}, \widetilde{\sigma_c}}}        \overset{\mathrm{(B.4.1)}}{\lesssim}  \underset{0 < j^{\prime} \leq q}{\underset{0 \leq j \leq q}{\sum}}           \bigg\{    \bra{\frac{c}{\widetilde{c}}}   \bigotimes    \bra{\frac{\sigma_c}{\widetilde{\sigma_c }}}     \bigg\}        \overset{\mathrm{(B.4.2)}}{\lesssim}  \underset{0 < j^{\prime} \leq q}{\underset{0 \leq j \leq q}{\sum}}           \bigg\{    \frac{ \bra{c} \bigotimes  \bra{\sigma_c} }{ \bra{\widetilde{c}} \bigotimes \bra{\widetilde{\sigma_c }}  }   \bigg\}     \overset{\mathrm{(B.4.3)}}{\lesssim}  \underset{0 < j^{\prime} \leq q}{\underset{0 \leq j \leq q}{\sum}}           \bigg\{   \frac{\bra{c}}{\bra{\widetilde{c}}}  \bigg\} \\ \\  \bigotimes     \underset{0 < j^{\prime} \leq q}{\underset{0 \leq j \leq q}{\sum}}           \bigg\{   \frac{\bra{\sigma_c }}{\bra{\widetilde{\sigma_c}}}    \bigg\}       \\ \end{align*}

       \begin{align*} \Longrightarrow     \frac{\bra{c, \sigma_c}}{\bra{\widetilde{c}, \widetilde{\sigma_c}}}        \overset{\mathrm{(B.4)}}{\lesssim}   \underset{0 < j^{\prime} \leq q}{\underset{0 \leq j \leq q}{\sum}}           \bigg\{   \frac{\bra{c}}{\bra{\widetilde{c}}}  \bigg\}  \bigotimes     \underset{0 < j^{\prime} \leq q}{\underset{0 \leq j \leq q}{\sum}}           \bigg\{   \frac{\bra{\sigma_c }}{\bra{\widetilde{\sigma_c}}}    \bigg\}            , \\ \\ 
    \end{align*}

}

\noindent corresponding to \textbf{(4)}, respectively. Altogether to obtain the desired up to constants lower bound for the trace distance, hence implying the desired estimate on the trace distance, it suffices to perform a computation of the form for which,

{\small

\begin{align*}
  \mathrm{Tr} \bigg\{  \frac{    \sqrt{ {\tiny    \big( \widetilde{\rho - \tau} \big)^{\dagger} \times  \big( \widetilde{\rho - \tau } \big)  } }   }{     \sqrt{  {\tiny \big( \rho - \tau \big)^{\dagger }  \times  \big( \rho - \tau \big)   }  }  }      \bigg\}    \lesssim    \mathrm{Tr} \bigg\{   \sqrt{ \frac{ {\tiny \big( \widetilde{\rho - \tau} \big)^{\dagger}}}{{\tiny \big( \rho - \tau \big)^{\dagger } }} \frac{{\tiny \big( \widetilde{\rho - \tau } \big)}}{{\tiny \big( \rho - \tau \big) } }}          \bigg\}       . \\ 
\end{align*}

}

\noindent To argue that (B),

{\small

\begin{align*}
 \mathrm{Td} \bigg[    \frac{ \big\{ \Pi \rho^* \Pi / \widetilde{\Pi} \widetilde{\rho^*} \widetilde{\Pi}  \big\}            }{  \big\{  \mathrm{Tr} \big[        \Pi \rho^*   \big] /     \mathrm{Tr} \big[  \widetilde{\Pi } \widetilde{\rho^{*}}    \big]    \big\}                           }  ,          \frac{ \big\{ \ket{c , \sigma_c}  \bra{c , \sigma_c }   \big\}     }{ \big\{ \ket{c , \widetilde{\sigma_c}}   \bra{c , \widetilde{\sigma_c }} \big\}   }             \bigg]       \lesssim   \big[  \textit{$(\mathrm{B.1})$} +  \textit{ $(\mathrm{B.2})$,} +   \textit{$(\mathrm{B.3})$, $(\mathrm{B.4})$}   \big]    . \\ 
 \end{align*}
 }

\noindent holds from the series of steps provided in estimates \textbf{(1)}, \textbf{(2)}, \textbf{(3)} and \textbf{(4)} it suffices to demonstrate that there exists the sequence of steps such that,

{\small

\begin{align*}
 \mathrm{Td} \bigg[    \frac{ \big\{ \Pi \rho^* \Pi / \widetilde{\Pi} \widetilde{\rho^*} \widetilde{\Pi}  \big\}            }{  \big\{  \mathrm{Tr} \big[        \Pi \rho^*   \big] /     \mathrm{Tr} \big[  \widetilde{\Pi } \widetilde{\rho^{*}}    \big]    \big\}                           }  ,          \frac{ \big\{ \ket{c , \sigma_c}  \bra{c , \sigma_c }   \big\}     }{ \big\{ \ket{c , \widetilde{\sigma_c}}   \bra{c , \widetilde{\sigma_c }} \big\}   }             \bigg]      \lesssim    ( \textit{Estimates obtained from $(\mathrm{B.1.1})$-$(\mathrm{B.1.6})$,} \\ \\  \textit{ $(\mathrm{B.2.1})$-$(\mathrm{B.2.5})$, $(\mathrm{B.3.1})$-$(\mathrm{B.3.3})$, $(\mathrm{B.4.1})$-$(\mathrm{B.4.3})$}   \big)   , \\ 
 \end{align*}
 }

\noindent corresponding to ($\mathrm{B.1}$), ($\mathrm{B.2}$), ($\mathrm{B.3}$) and ($\mathrm{B.4}$), respectively. To demonstrate that there exists a constant for which the above up to constants estimate can be sharpened to an inequality,

{\small

\begin{align*}
 \mathrm{Td} \bigg[    \frac{ \big\{ \Pi \rho^* \Pi / \widetilde{\Pi} \widetilde{\rho^*} \widetilde{\Pi}  \big\}            }{  \big\{  \mathrm{Tr} \big[        \Pi \rho^*   \big] /     \mathrm{Tr} \big[  \widetilde{\Pi } \widetilde{\rho^{*}}    \big]    \big\}                           }  ,          \frac{ \big\{ \ket{c , \sigma_c}  \bra{c , \sigma_c }   \big\}     }{ \big\{ \ket{c , \widetilde{\sigma_c}}   \bra{c , \widetilde{\sigma_c }} \big\}   }             \bigg]     \leq C_1  \times  ( \textit{Estimates obtained from $(\mathrm{B.1.1})$-$(\mathrm{B.1.6})$,} \\ \\  \textit{ $(\mathrm{B.2.1})$-$(\mathrm{B.2.5})$, $(\mathrm{B.3.1})$-$(\mathrm{B.3.3})$, $(\mathrm{B.4.1})$-$(\mathrm{B.4.3})$}   \big)     , 
\end{align*}

}

\noindent for some $C_1 >0$ in light of the estimates obtained above we further manipulate,

{\small

\begin{align*}
  \mathrm{Tr} \bigg\{   \sqrt{ \frac{ {\tiny \big( \widetilde{\rho - \tau} \big)^{\dagger}}}{{\tiny \big( \rho - \tau \big)^{\dagger } }} \frac{{\tiny \big( \widetilde{\rho - \tau } \big)}}{{\tiny \big( \rho - \tau \big) } }}          \bigg\}    , \\        
\end{align*}

}

\noindent for quantifying the constant for which,

{\small

\begin{align*}
 1 \lesssim   \mathrm{Tr} \bigg\{   \sqrt{ \frac{ {\tiny \big( \widetilde{\rho - \tau} \big)^{\dagger}}}{{\tiny \big( \rho - \tau \big)^{\dagger } }} \frac{{\tiny \big( \widetilde{\rho - \tau } \big)}}{{\tiny \big( \rho - \tau \big) } }}          \bigg\}  \bigg/        \mathrm{Tr} \bigg\{  \frac{    \sqrt{ {\tiny    \big( \widetilde{\rho - \tau} \big)^{\dagger} \times  \big( \widetilde{\rho - \tau } \big)  } }   }{     \sqrt{  {\tiny \big( \rho - \tau \big)^{\dagger }  \times  \big( \rho - \tau \big)   }  }  }      \bigg\}      , \\        
\end{align*}

}

\noindent satisfies,

{\small

\begin{align*}
  1 \leq    \mathrm{Tr} \bigg\{   \sqrt{ \frac{ {\tiny \big( \widetilde{\rho - \tau} \big)^{\dagger}}}{{\tiny \big( \rho - \tau \big)^{\dagger } }} \frac{{\tiny \big( \widetilde{\rho - \tau } \big)}}{{\tiny \big( \rho - \tau \big) } }}          \bigg\}  \bigg/       \mathrm{Tr} \bigg\{  \frac{    \sqrt{ {\tiny    \big( \widetilde{\rho - \tau} \big)^{\dagger} \times  \big( \widetilde{\rho - \tau } \big)  } }   }{     \sqrt{  {\tiny \big( \rho - \tau \big)^{\dagger }  \times  \big( \rho - \tau \big)   }  }  }      \bigg\}     . \\        
\end{align*}

}

\noindent As a result, to demonstrate for real $C,C^{\prime}, C^{\prime\prime}$,

{\small

\begin{align*}
        \mathrm{Tr} \bigg\{   \sqrt{ \frac{ {\tiny \big( \widetilde{\rho - \tau} \big)^{\dagger}}}{{\tiny \big( \rho - \tau \big)^{\dagger } }} \frac{{\tiny \big( \widetilde{\rho - \tau } \big)}}{{\tiny \big( \rho - \tau \big) } }}          \bigg\}   \geq  \big( C C^{\prime} C^{\prime\prime} \big)^{-1}   \cdot \underset{\rho,\tau,\widetilde{\rho},\widetilde{\tau}}{\mathrm{inf}} \bigg\{        \sqrt{\mathrm{Tr} \bigg\{   {\tiny \frac{\widetilde{\rho}^{\dagger}}{\big( \rho - \tau \big)^{\dagger} }  }  } \bigg\}   ,        \sqrt{\mathrm{Tr} \bigg\{  {\tiny   \frac{\widetilde{\rho}}{\big( \rho - \tau \big)}      }   \bigg\}}  ,    \sqrt{\mathrm{Tr} \bigg\{               {\tiny \frac{\widetilde{\rho}^{\dagger}}{\big( \rho - \tau \big)^{\dagger} } }          \bigg\}}    ,  \\   \\  \\    \sqrt{\mathrm{Tr} \bigg\{  {\tiny \frac{\widetilde{\tau}}{\big( \rho - \tau \big) } }   \bigg\}}  ,        \sqrt{\mathrm{Tr} \bigg\{              {\tiny   \frac{\widetilde{\tau}^{\dagger}}{\big( \rho - \tau \big)^{\dagger}}       }              \bigg\}}   ,     \sqrt{\mathrm{Tr} \bigg\{            {\tiny    \frac{\widetilde{\rho}}{\big( \rho - \tau \big) }     }         \bigg\}}   ,     \sqrt{\mathrm{Tr} \bigg\{     {\tiny   \frac{\widetilde{\tau}^{\dagger}}{\big( \rho - \tau \big)^{\dagger}}      }           \bigg\}} , \\ \\ \\       \sqrt{\mathrm{Tr} \bigg\{     {\tiny   \frac{\widetilde{\tau}}{\big( \rho - \tau \big)}      }           \bigg\}}               \bigg\}^8 , \\ 
\end{align*}

}

\noindent observe, from the fact that there exists $C^{-1}, \big\{ C^{\prime} \big\}^{-1}, \big\{ C^{\prime\prime} \big\}^{-1}$ taken sufficiently small one has,

{\small

\begin{align*}
  \mathrm{Tr} \bigg\{   \sqrt{ \frac{ {\tiny \big( \widetilde{\rho - \tau} \big)^{\dagger}}}{{\tiny \big( \rho - \tau \big)^{\dagger } }} \frac{{\tiny \big( \widetilde{\rho - \tau } \big)}}{{\tiny \big( \rho - \tau \big) } }}                    \bigg\}  =   \mathrm{Tr} \bigg\{   \sqrt{ \bigg[ \frac{ \widetilde{\rho}^{\dagger} }{\big( \rho - \tau \big)^{\dagger } } -  \frac{ \widetilde{\tau}^{\dagger} }{{\big( \rho - \tau \big)^{\dagger } }} \bigg]  \bigg[ \frac{ \widetilde{\rho  } }{\big( \rho - \tau \big) } -  \frac{ \widetilde{\tau}}{\big( \rho - \tau \big) } \bigg]   }               \bigg\}  \\ \\ \\  =                \mathrm{Tr} \bigg\{ \sqrt{  \bigg[   {\tiny \frac{\widetilde{\rho}^{\dagger} \widetilde{\rho}}{     \big( \rho - \tau \big)^{\dagger} \big( \rho - \tau \big)     }   - \frac{\widetilde{\rho}^{\dagger}    \widetilde{\tau}  }{\big( \rho - \tau \big)^{\dagger} \big( \rho - \tau \big)   }       - \frac{\widetilde{\tau}^{\dagger}\widetilde{\rho} }{\big( \rho - \tau \big)^{\dagger} \big( \rho - \tau \big) }  + \frac{\widetilde{\tau}^{\dagger} \widetilde{\tau}}{\big( \rho - \tau \big)^{\dagger} \big( \rho - \tau \big) }   }       \bigg]   }            \bigg\}                     \\ \end{align*}

  \begin{align*} \geq C^{-1} \cdot \mathrm{Tr} \bigg\{  \sqrt{ {\tiny \frac{\widetilde{\rho}^{\dagger} \widetilde{\rho}}{     \big( \rho - \tau \big)^{\dagger} \big( \rho - \tau \big)     }}   }  -        \sqrt{{\tiny \frac{\widetilde{\rho}^{\dagger}    \widetilde{\tau}  }{\big( \rho - \tau \big)^{\dagger} \big( \rho - \tau \big)   }   }} -  \sqrt{ {\tiny \frac{\widetilde{\tau}^{\dagger}\widetilde{\rho} }{\big( \rho - \tau \big)^{\dagger} \big( \rho - \tau \big) }}}  +        \sqrt{ {\tiny \frac{\widetilde{\tau}^{\dagger} \widetilde{\tau}}{\big( \rho - \tau \big)^{\dagger} \big( \rho - \tau \big) }}  }             \bigg\}         \\ \end{align*}

  \begin{align*}     \geq \big( C C^{\prime} \big)^{-1}   \cdot        \mathrm{Tr} \bigg\{  \sqrt{ {\tiny \frac{\widetilde{\rho}^{\dagger} \widetilde{\rho}}{     \big( \rho - \tau \big)^{\dagger} \big( \rho - \tau \big)     }}   }  \bigg\} \times  \mathrm{Tr} \bigg\{       \sqrt{{\tiny \frac{\widetilde{\rho}^{\dagger}    \widetilde{\tau}  }{\big( \rho - \tau \big)^{\dagger} \big( \rho - \tau \big)   }   }} \bigg\} \times  \mathrm{Tr} \bigg\{   \sqrt{ {\tiny \frac{\widetilde{\tau}^{\dagger}\widetilde{\rho} }{\big( \rho - \tau \big)^{\dagger} \big( \rho - \tau \big) }}}  \bigg\} \\ \times   \mathrm{Tr} \bigg\{      \sqrt{ {\tiny \frac{\widetilde{\tau}^{\dagger} \widetilde{\tau}}{\big( \rho - \tau \big)^{\dagger} \big( \rho - \tau \big) }}  }             \bigg\}          \\       \end{align*}

  \begin{align*} \overset{(\mathrm{JI})}{\geq} \big( C C^{\prime} \big)^{-1}   \cdot   \sqrt{\mathrm{Tr} \bigg\{  { {\tiny \frac{\widetilde{\rho}^{\dagger} \widetilde{\rho}}{     \big( \rho - \tau \big)^{\dagger} \big( \rho - \tau \big)     }}   }  \bigg\}} \times  \sqrt{\mathrm{Tr} \bigg\{       {{\tiny \frac{\widetilde{\rho}^{\dagger}    \widetilde{\tau}  }{\big( \rho - \tau \big)^{\dagger} \big( \rho - \tau \big)   }   }} \bigg\}} \times \sqrt{\mathrm{Tr} \bigg\{   { {\tiny \frac{\widetilde{\tau}^{\dagger}\widetilde{\rho} }{\big( \rho - \tau \big)^{\dagger} \big( \rho - \tau \big) }}}  \bigg\} } \\ \times \sqrt{ \mathrm{Tr} \bigg\{      { {\tiny \frac{\widetilde{\tau}^{\dagger} \widetilde{\tau}}{\big( \rho - \tau \big)^{\dagger} \big( \rho - \tau \big) }}  }             \bigg\} }   \\   \end{align*}

  \begin{align*}
  \geq  \big( C C^{\prime} C^{\prime\prime} \big)^{-1} \cdot   \underset{\rho,\tau,\widetilde{\rho},\widetilde{\tau}}{\mathrm{inf}} \bigg\{        \sqrt{\mathrm{Tr} \bigg\{   {\tiny \frac{\widetilde{\rho}^{\dagger}}{\big( \rho - \tau \big)^{\dagger} }  }  } \bigg\}   ,        \sqrt{\mathrm{Tr} \bigg\{  {\tiny   \frac{\widetilde{\rho}}{\big( \rho - \tau \big)}      }   \bigg\}}  ,    \sqrt{\mathrm{Tr} \bigg\{               {\tiny \frac{\widetilde{\rho}^{\dagger}}{\big( \rho - \tau \big)^{\dagger} } }          \bigg\}}    ,  \\   \\  \\    \sqrt{\mathrm{Tr} \bigg\{  {\tiny \frac{\widetilde{\tau}}{\big( \rho - \tau \big) } }   \bigg\}}  ,        \sqrt{\mathrm{Tr} \bigg\{              {\tiny   \frac{\widetilde{\tau}^{\dagger}}{\big( \rho - \tau \big)^{\dagger}}       }              \bigg\}}   ,     \sqrt{\mathrm{Tr} \bigg\{            {\tiny    \frac{\widetilde{\rho}}{\big( \rho - \tau \big) }     }         \bigg\}}   ,     \sqrt{\mathrm{Tr} \bigg\{     {\tiny   \frac{\widetilde{\tau}^{\dagger}}{\big( \rho - \tau \big)^{\dagger}}      }           \bigg\}} , \\ \\ \\       \sqrt{\mathrm{Tr} \bigg\{     {\tiny   \frac{\widetilde{\tau}}{\big( \rho - \tau \big)}      }           \bigg\}}               \bigg\}^8 , \\          
\end{align*}

}

\noindent where in the second to last inequality above Jensen's inequality is applied four times, each time with the square root function. Along similar lines, to argue that an estimate of the form, 

{\small

 \begin{align*} \mathrm{Tr} \bigg\{  \frac{    \sqrt{ {\tiny    \big( \widetilde{\rho - \tau} \big)^{\dagger} \times  \big( \widetilde{\rho - \tau } \big)  } }   }{     \sqrt{  {\tiny \big( \rho - \tau \big)^{\dagger }  \times  \big( \rho - \tau \big)   }  }  }      \bigg\}   \geq   \big( C C^{\prime} C^{\prime\prime} \big)^{-1}  \cdot  \underset{\rho,\tau,\widetilde{\rho},\widetilde{\tau}}{\mathrm{inf}} \bigg\{      {\mathrm{Tr} \bigg\{   {\tiny \frac{\sqrt{\widetilde{\rho}^{\dagger}}}{\sqrt{\big( \rho - \tau \big)^{\dagger} }}  }  }  \bigg\}  ,        {\mathrm{Tr} \bigg\{  {\tiny   \frac{\sqrt{\widetilde{\rho}}}{\sqrt{\big( \rho - \tau \big)}}      }   } \bigg\}  ,    {\mathrm{Tr}          \bigg\{       {\tiny \frac{\sqrt{\widetilde{\rho}^{\dagger}}}{\sqrt{\big( \rho - \tau \big)^{\dagger} }} }         }  \bigg\}   ,  \\   \\  \\    {\mathrm{Tr} \bigg\{   {\tiny \frac{\sqrt{\widetilde{\tau}}}{\sqrt{\big( \rho - \tau \big) } }}   } \bigg\}  ,       {\mathrm{Tr}              \bigg\{  {\tiny   \frac{\sqrt{\widetilde{\tau}^{\dagger}}}{\sqrt{\big( \rho - \tau \big)^{\dagger}} }      }              }  \bigg\}  ,    {\mathrm{Tr}         \bigg\{   {\tiny    \frac{\sqrt{\widetilde{\rho}}}{\sqrt{\big( \rho - \tau \big) }}     }      } \bigg\}   ,  {\mathrm{Tr}     \bigg\{  {\tiny   \frac{\sqrt{\widetilde{\tau}^{\dagger}}}{\sqrt{\big( \rho - \tau \big)^{\dagger}}}      }           } \bigg\} ,    {\mathrm{Tr}     \bigg\{  {\tiny   \frac{\sqrt{\widetilde{\tau}}}{\sqrt{\big( \rho - \tau \big)}}      }         }   \bigg\}             \bigg\}^8 , \\ \end{align*}

}

\noindent holds for the ratio of two square root functions under the trace,

{\small

\begin{align*}
    \frac{1}{2} \mathrm{Tr} \bigg\{  \frac{    \sqrt{ {\tiny    \big( \widetilde{\rho - \tau} \big)^{\dagger} \times  \big( \widetilde{\rho - \tau } \big)  } }   }{     \sqrt{  {\tiny \big( \rho - \tau \big)^{\dagger }  \times  \big( \rho - \tau \big)   }  }  }      \bigg\}    , \\ 
\end{align*}

}

\noindent as opposed to a single square root function under the trace, write,

  {\small  \begin{align*}  \frac{1}{2} \mathrm{Tr} \bigg\{  \frac{    \sqrt{ {\tiny    \big( \widetilde{\rho - \tau} \big)^{\dagger} \times  \big( \widetilde{\rho - \tau } \big)  } }   }{     \sqrt{  {\tiny \big( \rho - \tau \big)^{\dagger }  \times  \big( \rho - \tau \big)   }  }  }      \bigg\}    =  \frac{1}{2} \mathrm{Tr} \bigg\{   \sqrt{ \frac{  {\tiny   \big( \widetilde{\rho - \tau} \big)^{\dagger}}}{ {\tiny \big( \rho - \tau \big)^{\dagger }} } \frac{ {\tiny \big( \widetilde{\rho - \tau } \big) } }{ {\tiny \big( \rho - \tau \big) } }}      \times  \bigg\{      \frac{    \sqrt{ {\tiny  \big( \widetilde{\rho - \tau} \big)^{\dagger} \times  \big( \widetilde{\rho - \tau } \big)  }  } }{     \sqrt{  {\tiny \big( \rho - \tau \big)^{\dagger }  \times  \big( \rho - \tau \big)     } }  }    \bigg/ {\sqrt{ {\tiny \frac{\big( \widetilde{\rho - \tau} \big)^{\dagger}}{\big( \rho - \tau \big)^{\dagger } } \frac{\big( \widetilde{\rho - \tau } \big)}{\big( \rho - \tau \big) } } }  }           \bigg\}                              \bigg\} \\  \end{align*}
  
  \begin{align*} \geq \frac{1}{2 \mathcal{C}} \cdot \mathrm{Tr} \bigg\{  \sqrt{ \frac{  {\tiny   \big( \widetilde{\rho - \tau} \big)^{\dagger}}}{ {\tiny \big( \rho - \tau \big)^{\dagger }} } \frac{ {\tiny \big( \widetilde{\rho - \tau } \big) } }{ {\tiny \big( \rho - \tau \big) } }}      \bigg\} \times     \mathrm{Tr} \bigg\{  \frac{    \sqrt{ {\tiny  \big( \widetilde{\rho - \tau} \big)^{\dagger} \times  \big( \widetilde{\rho - \tau } \big)  }  } }{     \sqrt{ {\tiny  \big( \rho - \tau \big)^{\dagger }  \times  \big( \rho - \tau \big)     }  } } \bigg/   \sqrt{          {\tiny \frac{\big( \widetilde{\rho - \tau} \big)^{\dagger}}{\big( \rho - \tau \big)^{\dagger } } \frac{\big( \widetilde{\rho - \tau } \big)}{\big( \rho - \tau \big) } }   }          \bigg\}                            \\ \\      \\              \geq    \frac{1}{2 \mathcal{C} C C^{\prime} C^{\prime\prime} }  \cdot    \underset{\rho,\tau,\widetilde{\rho},\widetilde{\tau}}{\mathrm{inf}} \bigg\{        \sqrt{\mathrm{Tr} \bigg\{   {\tiny \frac{\widetilde{\rho}^{\dagger}}{\big( \rho - \tau \big)^{\dagger} }  }  } \bigg\}   ,        \sqrt{\mathrm{Tr} \bigg\{  {\tiny   \frac{\widetilde{\rho}}{\big( \rho - \tau \big)}      }   \bigg\}}  ,    \sqrt{\mathrm{Tr} \bigg\{               {\tiny \frac{\widetilde{\rho}^{\dagger}}{\big( \rho - \tau \big)^{\dagger} } }          \bigg\}}    , \sqrt{\mathrm{Tr} \bigg\{  {\tiny \frac{\widetilde{\tau}}{\big( \rho - \tau \big) } }   \bigg\}}  ,  \\   \\  \\             \sqrt{\mathrm{Tr} \bigg\{              {\tiny   \frac{\widetilde{\tau}^{\dagger}}{\big( \rho - \tau \big)^{\dagger}}       }              \bigg\}}   ,     \sqrt{\mathrm{Tr} \bigg\{            {\tiny    \frac{\widetilde{\rho}}{\big( \rho - \tau \big) }     }         \bigg\}}   ,     \sqrt{\mathrm{Tr} \bigg\{     {\tiny   \frac{\widetilde{\tau}^{\dagger}}{\big( \rho - \tau \big)^{\dagger}}      }           \bigg\}} ,       \sqrt{\mathrm{Tr} \bigg\{     {\tiny   \frac{\widetilde{\tau}}{\big( \rho - \tau \big)}      }           \bigg\}}               \bigg\}^8  \\ \\  \\     \times     \mathrm{Tr} \bigg\{  \frac{    \sqrt{  \big( \widetilde{\rho - \tau} \big)^{\dagger} \times  \big( \widetilde{\rho - \tau } \big)  }   }{     \sqrt{   \big( \rho - \tau \big)^{\dagger }  \times  \big( \rho - \tau \big)     }  }  \bigg/   \sqrt{ \frac{\big( \widetilde{\rho - \tau} \big)^{\dagger}}{\big( \rho - \tau \big)^{\dagger } } \frac{\big( \widetilde{\rho - \tau } \big)}{\big( \rho - \tau \big) } }             \bigg\}            
 , \\  \\ \\ \\ \geq     \frac{1}{2 \mathcal{C} \mathcal{C}^{\prime} C C^{\prime} C^{\prime\prime} }   \cdot   \underset{\rho,\tau,\widetilde{\rho},\widetilde{\tau}}{\mathrm{inf}} \bigg\{        \sqrt{\mathrm{Tr} \bigg\{   {\tiny \frac{\widetilde{\rho}^{\dagger}}{\big( \rho - \tau \big)^{\dagger} }  }  } \bigg\}   ,        \sqrt{\mathrm{Tr} \bigg\{  {\tiny   \frac{\widetilde{\rho}}{\big( \rho - \tau \big)}      }   \bigg\}}  ,    \sqrt{\mathrm{Tr} \bigg\{               {\tiny \frac{\widetilde{\rho}^{\dagger}}{\big( \rho - \tau \big)^{\dagger} } }          \bigg\}}    , \sqrt{\mathrm{Tr} \bigg\{  {\tiny \frac{\widetilde{\tau}}{\big( \rho - \tau \big) } }   \bigg\}} ,  \\   \\  \\           \sqrt{\mathrm{Tr} \bigg\{              {\tiny   \frac{\widetilde{\tau}^{\dagger}}{\big( \rho - \tau \big)^{\dagger}}       }              \bigg\}}   ,     \sqrt{\mathrm{Tr} \bigg\{            {\tiny    \frac{\widetilde{\rho}}{\big( \rho - \tau \big) }     }         \bigg\}}   ,     \sqrt{\mathrm{Tr} \bigg\{     {\tiny   \frac{\widetilde{\tau}^{\dagger}}{\big( \rho - \tau \big)^{\dagger}}      }           \bigg\}} ,        \sqrt{\mathrm{Tr} \bigg\{     {\tiny   \frac{\widetilde{\tau}}{\big( \rho - \tau \big)}      }           \bigg\}}               \bigg\}^8  \\ \\ \times  \mathrm{Tr} \bigg\{                \sqrt{\frac{{\tiny                  \big( \widetilde{\rho - \tau} \big)^{\dagger} \times \bigg(   \frac{\big( \widetilde{\rho - \tau} \big) }{ \rho - \tau }    \bigg)^{\dagger}     }}{{\tiny        \frac{\big( \rho - \tau \big)}{\big( \widetilde{\rho - \tau} \big)^2 }   }}}    \bigg/  \sqrt{  {\tiny \big( \rho - \tau \big)^{\dagger }  \times  \big( \rho - \tau \big)   }  }     \bigg\} .   \\ \\  \\  \tag{\textit{*}}               \end{align*}
 }

\noindent To demonstrate that,

{\small

\begin{align*}
 \mathrm{Tr} \bigg\{                \sqrt{\frac{{\tiny                  \big( \widetilde{\rho - \tau} \big)^{\dagger} \times \bigg(   \frac{\big( \widetilde{\rho - \tau} \big) }{ \rho - \tau }    \bigg)^{\dagger}     }}{{\tiny        \frac{\big( \rho - \tau \big)}{\big( \widetilde{\rho - \tau} \big)^2 }   }}}    \bigg/  \sqrt{  {\tiny \big( \rho - \tau \big)^{\dagger }  \times  \big( \rho - \tau \big)   }  }     \bigg\}    \geq   \frac{1}{\mathcal{C}^{\prime\prime}}  \cdot    \mathrm{Tr} \bigg\{              \bigg(    {\tiny              \frac{  \big( \widetilde{\rho - \tau } \big)^{\dagger} }{\big( \rho - \tau \big)^{\dagger}}  }     \bigg/    {\tiny               \frac{ \rho - \tau }{ \widetilde{\rho - \tau}   }                     }    \bigg)                                \bigg\}  ,  \\ 
\end{align*}

}

\noindent we perform the following computations. In particular the desired lower bound follows from the fact that further manipulations of the last trace in the lower bound above satisfy,

{\small

\begin{align*}
 \mathrm{Tr} \bigg\{                \sqrt{\frac{{\tiny                  \big( \widetilde{\rho - \tau} \big)^{\dagger} \times \bigg(   \frac{\big( \widetilde{\rho - \tau} \big) }{ \rho - \tau }    \bigg)^{\dagger}     }}{{\tiny        \frac{\big( \rho - \tau \big)}{\big( \widetilde{\rho - \tau} \big)^2 }   }}}    \bigg/  \sqrt{  {\tiny \big( \rho - \tau \big)^{\dagger }  \times  \big( \rho - \tau \big)   }  }     \bigg\}    \geq \frac{1}{\mathcal{C}^{\prime\prime}}  \cdot     \mathrm{Tr} \bigg\{           \sqrt{\frac{{\tiny                  \frac{\big( \widetilde{\rho - \tau} \big)^{\dagger}}{\big( \rho - \tau \big)^{\dagger}}  \times \bigg(   \frac{\big( \widetilde{\rho - \tau} \big) }{ \rho - \tau }    \bigg)^{\dagger}     }}{{\tiny       \bigg(  \frac{ \rho - \tau }{ \widetilde{\rho - \tau}   }     \bigg)^2        }}}                    \bigg\} \\ \\ \\    =         \frac{1}{\mathcal{C}^{\prime\prime}}   \cdot   \mathrm{Tr} \bigg\{             \sqrt{  \frac{  {\tiny          \bigg(    \frac{  \big( \widetilde{\rho - \tau } \big)^{\dagger} }{\big( \rho - \tau \big)^{\dagger}}   \bigg)^{2}  }    }{     {\tiny             \bigg(  \frac{ \rho - \tau }{ \widetilde{\rho - \tau}   }     \bigg)^2                }        }       }                \bigg\}   =      \frac{1}{\mathcal{C}^{\prime\prime}} \cdot      \mathrm{Tr} \bigg\{             \sqrt{  \bigg(   {\tiny              \frac{  \big( \widetilde{\rho - \tau } \big)^{\dagger} }{\big( \rho - \tau \big)^{\dagger}}  }     \bigg/    {\tiny               \frac{ \rho - \tau }{ \widetilde{\rho - \tau}   }                    }     \bigg)^2            }                   \bigg\}  \\ \\ \\  =   \frac{1}{\mathcal{C}^{\prime\prime}}  \cdot    \mathrm{Tr} \bigg\{              \bigg(    {\tiny              \frac{  \big( \widetilde{\rho - \tau } \big)^{\dagger} }{\big( \rho - \tau \big)^{\dagger}}  }     \bigg/    {\tiny               \frac{ \rho - \tau }{ \widetilde{\rho - \tau}   }                     }    \bigg)                                \bigg\}  .  \\ 
\end{align*}

}

\noindent Therefore,

{\small

\begin{align*}
   (\textit{*}) \geq       \frac{1}{2 \mathcal{C} \mathcal{C}^{\prime}  \mathcal{C}^{\prime\prime} C C^{\prime} C^{\prime\prime} }    \cdot \underset{\rho,\tau,\widetilde{\rho},\widetilde{\tau}}{\mathrm{inf}} \bigg\{        \sqrt{\mathrm{Tr} \bigg\{   {\tiny \frac{\widetilde{\rho}^{\dagger}}{\big( \rho - \tau \big)^{\dagger} }  }  } \bigg\}   ,        \sqrt{\mathrm{Tr} \bigg\{  {\tiny   \frac{\widetilde{\rho}}{\big( \rho - \tau \big)}      }   \bigg\}}  ,    \sqrt{\mathrm{Tr} \bigg\{               {\tiny \frac{\widetilde{\rho}^{\dagger}}{\big( \rho - \tau \big)^{\dagger} } }          \bigg\}}    , \sqrt{\mathrm{Tr} \bigg\{  {\tiny \frac{\widetilde{\tau}}{\big( \rho - \tau \big) } }   \bigg\}} ,  \\   \\  \\           \sqrt{\mathrm{Tr} \bigg\{              {\tiny   \frac{\widetilde{\tau}^{\dagger}}{\big( \rho - \tau \big)^{\dagger}}       }              \bigg\}}   ,     \sqrt{\mathrm{Tr} \bigg\{            {\tiny    \frac{\widetilde{\rho}}{\big( \rho - \tau \big) }     }         \bigg\}}   ,     \sqrt{\mathrm{Tr} \bigg\{     {\tiny   \frac{\widetilde{\tau}^{\dagger}}{\big( \rho - \tau \big)^{\dagger}}      }           \bigg\}} ,        \sqrt{\mathrm{Tr} \bigg\{     {\tiny   \frac{\widetilde{\tau}}{\big( \rho - \tau \big)}      }           \bigg\}}               \bigg\}^8  \\ \\ \times     \frac{1}{\mathcal{C}^{\prime\prime}} \cdot     \mathrm{Tr} \bigg\{              \bigg(    {\tiny              \frac{  \big( \widetilde{\rho - \tau } \big)^{\dagger} }{\big( \rho - \tau \big)^{\dagger}}  }     \bigg/    {\tiny               \frac{ \rho - \tau }{ \widetilde{\rho - \tau}   }                     }    \bigg)                                \bigg\}  \\ \\ \\  \\ \geq      \frac{1}{2 \mathcal{C} \mathcal{C}^{\prime}  \mathcal{C}^{\prime\prime}  \big( C C^{\prime} C^{\prime\prime} \big)^2 }  \cdot \underset{\rho,\tau,\widetilde{\rho},\widetilde{\tau}}{\mathrm{inf}} \bigg\{      {\mathrm{Tr} \bigg\{   {\tiny \frac{\sqrt{\widetilde{\rho}^{\dagger}}}{\sqrt{\big( \rho - \tau \big)^{\dagger} }}  }  }  \bigg\}  ,        {\mathrm{Tr} \bigg\{  {\tiny   \frac{\sqrt{\widetilde{\rho}}}{\sqrt{\big( \rho - \tau \big)}}      }   } \bigg\}  ,    {\mathrm{Tr}          \bigg\{       {\tiny \frac{\sqrt{\widetilde{\rho}^{\dagger}}}{\sqrt{\big( \rho - \tau \big)^{\dagger} }} }         }  \bigg\}   ,     {\mathrm{Tr} \bigg\{   {\tiny \frac{\sqrt{\widetilde{\tau}}}{\sqrt{\big( \rho - \tau \big) } }}   } \bigg\}  ,  \\   \\  \\       {\mathrm{Tr}              \bigg\{  {\tiny   \frac{\sqrt{\widetilde{\tau}^{\dagger}}}{\sqrt{\big( \rho - \tau \big)^{\dagger}} }      }              }  \bigg\}   ,    {\mathrm{Tr}         \bigg\{   {\tiny    \frac{\sqrt{\widetilde{\rho}}}{\sqrt{\big( \rho - \tau \big) }}     }      } \bigg\}   ,  {\mathrm{Tr}     \bigg\{  {\tiny   \frac{\sqrt{\widetilde{\tau}^{\dagger}}}{\sqrt{\big( \rho - \tau \big)^{\dagger}}}      }           } \bigg\} ,    {\mathrm{Tr}     \bigg\{  {\tiny   \frac{\sqrt{\widetilde{\tau}}}{\sqrt{\big( \rho - \tau \big)}}      }         }   \bigg\}             \bigg\}^8 \\ \\  \times          \frac{1}{\mathcal{C}^{\prime\prime}} \cdot      \mathrm{Tr} \bigg\{              \bigg(    {\tiny              \frac{  \big( \widetilde{\rho - \tau } \big)^{\dagger} }{\big( \rho - \tau \big)^{\dagger}}  }     \bigg/    {\tiny               \frac{ \rho - \tau }{ \widetilde{\rho - \tau}   }                     }    \bigg)                                \bigg\} \\ \\ \\  \end{align*}

   \begin{align*} \geq     \frac{1}{2 \mathcal{C} \mathcal{C}^{\prime}  \mathcal{C}^{\prime\prime}  \big( C C^{\prime} C^{\prime\prime} \big)^2 }  \cdot \underset{\rho,\tau,\widetilde{\rho},\widetilde{\tau}}{\mathrm{inf}} \bigg\{      {\mathrm{Tr} \bigg\{   {\tiny \frac{\sqrt{\widetilde{\rho}^{\dagger}}}{\sqrt{\big( \rho - \tau \big)^{\dagger} }}  }  }  \bigg\}  ,        {\mathrm{Tr} \bigg\{  {\tiny   \frac{\sqrt{\widetilde{\rho}}}{\sqrt{\big( \rho - \tau \big)}}      }   } \bigg\}  ,    {\mathrm{Tr}          \bigg\{       {\tiny \frac{\sqrt{\widetilde{\rho}^{\dagger}}}{\sqrt{\big( \rho - \tau \big)^{\dagger} }} }         }  \bigg\}   ,     {\mathrm{Tr} \bigg\{   {\tiny \frac{\sqrt{\widetilde{\tau}}}{\sqrt{\big( \rho - \tau \big) } }}   } \bigg\}  ,  \\   \\  \\       {\mathrm{Tr}              \bigg\{  {\tiny   \frac{\sqrt{\widetilde{\tau}^{\dagger}}}{\sqrt{\big( \rho - \tau \big)^{\dagger}} }      }              }  \bigg\}   ,    {\mathrm{Tr}         \bigg\{   {\tiny    \frac{\sqrt{\widetilde{\rho}}}{\sqrt{\big( \rho - \tau \big) }}     }      } \bigg\}   ,  {\mathrm{Tr}     \bigg\{  {\tiny   \frac{\sqrt{\widetilde{\tau}^{\dagger}}}{\sqrt{\big( \rho - \tau \big)^{\dagger}}}      }           } \bigg\} ,    {\mathrm{Tr}     \bigg\{  {\tiny   \frac{\sqrt{\widetilde{\tau}}}{\sqrt{\big( \rho - \tau \big)}}      }         }   \bigg\}             \bigg\}^8 \\ \\  \times          \frac{1}{\big( \mathcal{C}^{\prime\prime} \big)^2 }  \cdot    \mathrm{Tr} \bigg\{               {\tiny              \frac{  \big( \widetilde{\rho - \tau } \big)^{\dagger} }{\big( \rho - \tau \big)^{\dagger}}  }     \bigg\} \bigg/  \mathrm{Tr} \bigg\{       {\tiny               \frac{ \rho - \tau }{ \widetilde{\rho - \tau}   }                     }                    \bigg\}  \\ \\ \\  =   \frac{1}{2 \mathcal{C} \mathcal{C}^{\prime}  \big( \mathcal{C}^{\prime\prime} \big)^3  \big( C C^{\prime} C^{\prime\prime} \big)^2 }  \cdot \underset{\rho,\tau,\widetilde{\rho},\widetilde{\tau}}{\mathrm{inf}} \bigg\{         {\mathrm{Tr} \bigg\{   {\tiny \frac{\sqrt{\widetilde{\rho}^{\dagger}}}{\sqrt{\big( \rho - \tau \big)^{\dagger} }}  }  }  \bigg\}    \times  \bigg[ \mathrm{Tr} \bigg\{               {\tiny              \frac{  \big( \widetilde{\rho - \tau } \big)^{\dagger} }{\big( \rho - \tau \big)^{\dagger}}  }     \bigg\} \bigg/  \mathrm{Tr} \bigg\{       {\tiny               \frac{ \rho - \tau }{ \widetilde{\rho - \tau}   }                     }                    \bigg\}      \bigg]              ,        {\mathrm{Tr} \bigg\{  {\tiny   \frac{\sqrt{\widetilde{\rho}}}{\sqrt{\big( \rho - \tau \big)}}      }   } \bigg\}   \\   \\  \\       \times  \bigg[ \mathrm{Tr} \bigg\{               {\tiny              \frac{  \big( \widetilde{\rho - \tau } \big)^{\dagger} }{\big( \rho - \tau \big)^{\dagger}}  }     \bigg\} \bigg/  \mathrm{Tr} \bigg\{       {\tiny               \frac{ \rho - \tau }{ \widetilde{\rho - \tau}   }                     }                    \bigg\}      \bigg]   ,  {\mathrm{Tr}          \bigg\{       {\tiny \frac{\sqrt{\widetilde{\rho}^{\dagger}}}{\sqrt{\big( \rho - \tau \big)^{\dagger} }} }         }  \bigg\}    \times  \bigg[ \mathrm{Tr} \bigg\{               {\tiny              \frac{  \big( \widetilde{\rho - \tau } \big)^{\dagger} }{\big( \rho - \tau \big)^{\dagger}}  }     \bigg\} \bigg/  \mathrm{Tr} \bigg\{       {\tiny               \frac{ \rho - \tau }{ \widetilde{\rho - \tau}   }                     }                    \bigg\}      \bigg]  ,  \\ \\ \\       {\mathrm{Tr} \bigg\{   {\tiny \frac{\sqrt{\widetilde{\tau}}}{\sqrt{\big( \rho - \tau \big) } }}   } \bigg\}    \times  \bigg[ \mathrm{Tr} \bigg\{               {\tiny              \frac{  \big( \widetilde{\rho - \tau } \big)^{\dagger} }{\big( \rho - \tau \big)^{\dagger}}  }     \bigg\} \bigg/  \mathrm{Tr} \bigg\{       {\tiny               \frac{ \rho - \tau }{ \widetilde{\rho - \tau}   }                     }                    \bigg\}      \bigg]   ,     {\mathrm{Tr}              \bigg\{  {\tiny   \frac{\sqrt{\widetilde{\tau}^{\dagger}}}{\sqrt{\big( \rho - \tau \big)^{\dagger}} }      }              }  \bigg\}    \\ \end{align*}

   \begin{align*}      \times  \bigg[ \mathrm{Tr} \bigg\{               {\tiny              \frac{  \big( \widetilde{\rho - \tau } \big)^{\dagger} }{\big( \rho - \tau \big)^{\dagger}}  }     \bigg\} \bigg/  \mathrm{Tr} \bigg\{       {\tiny               \frac{ \rho - \tau }{ \widetilde{\rho - \tau}   }                     }                    \bigg\}      \bigg]    ,    {\mathrm{Tr}         \bigg\{   {\tiny    \frac{\sqrt{\widetilde{\rho}}}{\sqrt{\big( \rho - \tau \big) }}     }      } \bigg\} \\ \\ \\     \times  \bigg[ \mathrm{Tr} \bigg\{               {\tiny              \frac{  \big( \widetilde{\rho - \tau } \big)^{\dagger} }{\big( \rho - \tau \big)^{\dagger}}  }     \bigg\} \bigg/  \mathrm{Tr} \bigg\{       {\tiny               \frac{ \rho - \tau }{ \widetilde{\rho - \tau}   }                     }                    \bigg\}      \bigg]    ,   {\mathrm{Tr}     \bigg\{  {\tiny   \frac{\sqrt{\widetilde{\tau}^{\dagger}}}{\sqrt{\big( \rho - \tau \big)^{\dagger}}}      }           } \bigg\}  \\ \\ \\   \times  \bigg[ \mathrm{Tr} \bigg\{               {\tiny              \frac{  \big( \widetilde{\rho - \tau } \big)^{\dagger} }{\big( \rho - \tau \big)^{\dagger}}  }     \bigg\} \bigg/  \mathrm{Tr} \bigg\{       {\tiny               \frac{ \rho - \tau }{ \widetilde{\rho - \tau}   }                     }                    \bigg\}      \bigg]   ,    {\mathrm{Tr}     \bigg\{  {\tiny   \frac{\sqrt{\widetilde{\tau}}}{\sqrt{\big( \rho - \tau \big)}}      }         }   \bigg\}  \\ \\ \\   \times  \bigg[ \mathrm{Tr} \bigg\{               {\tiny              \frac{  \big( \widetilde{\rho - \tau } \big)^{\dagger} }{\big( \rho - \tau \big)^{\dagger}}  }     \bigg\} \bigg/  \mathrm{Tr} \bigg\{       {\tiny               \frac{ \rho - \tau }{ \widetilde{\rho - \tau}   }                     }                    \bigg\}      \bigg]             \bigg\}^8    \\ \\ \\ \end{align*}

   \begin{align*} \geq        1 -  \bigg(  \frac{1}{2 \mathcal{C} \mathcal{C}^{\prime}  \big( \mathcal{C}^{\prime\prime} \big)^3  \big( C C^{\prime} C^{\prime\prime} \big)^2 } \bigg)^2  \cdot         \underset{\rho,\tau,\widetilde{\rho},\widetilde{\tau}}{\mathrm{inf}} \bigg\{         {\mathrm{Tr} \bigg\{   {\tiny \frac{\sqrt{\widetilde{\rho}^{\dagger}}}{\sqrt{\big( \rho - \tau \big)^{\dagger} }}  }  }  \bigg\}    \times  \bigg[ \mathrm{Tr} \bigg\{               {\tiny              \frac{  \big( \widetilde{\rho - \tau } \big)^{\dagger} }{\big( \rho - \tau \big)^{\dagger}}  }     \bigg\} \bigg/  \mathrm{Tr} \bigg\{       {\tiny               \frac{ \rho - \tau }{ \widetilde{\rho - \tau}   }                     }                    \bigg\}      \bigg]              ,        {\mathrm{Tr} \bigg\{  {\tiny   \frac{\sqrt{\widetilde{\rho}}}{\sqrt{\big( \rho - \tau \big)}}      }   } \bigg\}   \\   \\  \\       \times  \bigg[ \mathrm{Tr} \bigg\{               {\tiny              \frac{  \big( \widetilde{\rho - \tau } \big)^{\dagger} }{\big( \rho - \tau \big)^{\dagger}}  }     \bigg\} \bigg/  \mathrm{Tr} \bigg\{       {\tiny               \frac{ \rho - \tau }{ \widetilde{\rho - \tau}   }                     }                    \bigg\}      \bigg]   ,  {\mathrm{Tr}          \bigg\{       {\tiny \frac{\sqrt{\widetilde{\rho}^{\dagger}}}{\sqrt{\big( \rho - \tau \big)^{\dagger} }} }         }  \bigg\}    \times  \bigg[ \mathrm{Tr} \bigg\{               {\tiny              \frac{  \big( \widetilde{\rho - \tau } \big)^{\dagger} }{\big( \rho - \tau \big)^{\dagger}}  }     \bigg\} \bigg/  \mathrm{Tr} \bigg\{       {\tiny               \frac{ \rho - \tau }{ \widetilde{\rho - \tau}   }                     }                    \bigg\}      \bigg]  ,  \\ \\ \\       {\mathrm{Tr} \bigg\{   {\tiny \frac{\sqrt{\widetilde{\tau}}}{\sqrt{\big( \rho - \tau \big) } }}   } \bigg\}    \times  \bigg[ \mathrm{Tr} \bigg\{               {\tiny              \frac{  \big( \widetilde{\rho - \tau } \big)^{\dagger} }{\big( \rho - \tau \big)^{\dagger}}  }     \bigg\} \bigg/  \mathrm{Tr} \bigg\{       {\tiny               \frac{ \rho - \tau }{ \widetilde{\rho - \tau}   }                     }                    \bigg\}      \bigg]   ,     {\mathrm{Tr}              \bigg\{  {\tiny   \frac{\sqrt{\widetilde{\tau}^{\dagger}}}{\sqrt{\big( \rho - \tau \big)^{\dagger}} }      }              }  \bigg\}    \\ \\ \\      \times  \bigg[ \mathrm{Tr} \bigg\{               {\tiny              \frac{  \big( \widetilde{\rho - \tau } \big)^{\dagger} }{\big( \rho - \tau \big)^{\dagger}}  }     \bigg\} \bigg/  \mathrm{Tr} \bigg\{       {\tiny               \frac{ \rho - \tau }{ \widetilde{\rho - \tau}   }                     }                    \bigg\}      \bigg]    ,    {\mathrm{Tr}         \bigg\{   {\tiny    \frac{\sqrt{\widetilde{\rho}}}{\sqrt{\big( \rho - \tau \big) }}     }      } \bigg\} \\ \\ \\     \times  \bigg[ \mathrm{Tr} \bigg\{               {\tiny              \frac{  \big( \widetilde{\rho - \tau } \big)^{\dagger} }{\big( \rho - \tau \big)^{\dagger}}  }     \bigg\} \bigg/  \mathrm{Tr} \bigg\{       {\tiny               \frac{ \rho - \tau }{ \widetilde{\rho - \tau}   }                     }                    \bigg\}      \bigg]    ,   {\mathrm{Tr}     \bigg\{  {\tiny   \frac{\sqrt{\widetilde{\tau}^{\dagger}}}{\sqrt{\big( \rho - \tau \big)^{\dagger}}}      }           } \bigg\}  \\ \\ \\   \times  \bigg[ \mathrm{Tr} \bigg\{               {\tiny              \frac{  \big( \widetilde{\rho - \tau } \big)^{\dagger} }{\big( \rho - \tau \big)^{\dagger}}  }     \bigg\} \bigg/  \mathrm{Tr} \bigg\{       {\tiny               \frac{ \rho - \tau }{ \widetilde{\rho - \tau}   }                     }                    \bigg\}      \bigg]   ,    {\mathrm{Tr}     \bigg\{  {\tiny   \frac{\sqrt{\widetilde{\tau}}}{\sqrt{\big( \rho - \tau \big)}}      }         }   \bigg\}  \\ \\ \\   \times  \bigg[ \mathrm{Tr} \bigg\{               {\tiny              \frac{  \big( \widetilde{\rho - \tau } \big)^{\dagger} }{\big( \rho - \tau \big)^{\dagger}}  }     \bigg\} \bigg/  \mathrm{Tr} \bigg\{       {\tiny               \frac{ \rho - \tau }{ \widetilde{\rho - \tau}   }                     }                    \bigg\}      \bigg]             \bigg\}^8  \\ \\ \\   \equiv 1 - \epsilon                             \\ \\ \\ \gtrsim  {\tiny \mathrm{NEGL} \big(      \lambda^{\prime}  - \lambda     \big)}  \\ \\ \\ \gtrsim 1 -   {\tiny \mathrm{NEGL} \big(      \lambda^{\prime}  - \lambda     \big)}          , \end{align*}

   \begin{align*} \Longrightarrow                \frac{1}{2} \mathrm{Tr} \bigg\{  \frac{    \sqrt{ {\tiny    \big( \widetilde{\rho - \tau} \big)^{\dagger} \times  \big( \widetilde{\rho - \tau } \big)  } }   }{     \sqrt{  {\tiny \big( \rho - \tau \big)^{\dagger }  \times  \big( \rho - \tau \big)   }  }  }      \bigg\}    \geq 1 - \epsilon                         , \\ 
\end{align*}

   \begin{align*} \Longrightarrow                \frac{1}{2} \mathrm{Tr} \bigg\{  \frac{    \sqrt{ {\tiny    \big( \widetilde{\rho - \tau} \big)^{\dagger} \times  \big( \widetilde{\rho - \tau } \big)  } }   }{     \sqrt{  {\tiny \big( \rho - \tau \big)^{\dagger }  \times  \big( \rho - \tau \big)   }  }  }      \bigg\}         \gtrsim   {\tiny \mathrm{NEGL} \big(      \lambda^{\prime}  - \lambda     \big) }                               \gtrsim 1 -   {\tiny \mathrm{NEGL} \big(      \lambda^{\prime}  - \lambda     \big) }         , \\ 
\end{align*}

}

\noindent where $\epsilon$ above is taken as the parameter in the lower bound on the trace appearing in the statement of GML (a parameter such that $0 < \epsilon \leq 1$). The above constant,

{\small

\begin{align*}
    \bigg(  \frac{1}{2 \mathcal{C} \mathcal{C}^{\prime}  \big( \mathcal{C}^{\prime\prime} \big)^3  \big( C C^{\prime} C^{\prime\prime} \big)^2 } \bigg)^2         , 
\end{align*}

}

\noindent can be taken sufficiently small so that,

{\small

\begin{align*}
   \bigg(  \frac{1}{2 \mathcal{C} \mathcal{C}^{\prime}  \big( \mathcal{C}^{\prime\prime} \big)^3  \big( C C^{\prime} C^{\prime\prime} \big)^2 } \bigg)^2        \geq     \underset{\rho,\tau,\widetilde{\rho},\widetilde{\tau}}{\mathrm{inf}} \bigg\{         {\mathrm{Tr} \bigg\{   {\tiny \frac{\sqrt{\widetilde{\rho}^{\dagger}}}{\sqrt{\big( \rho - \tau \big)^{\dagger} }}  }  }  \bigg\}    \times  \bigg[ \mathrm{Tr} \bigg\{               {\tiny              \frac{  \big( \widetilde{\rho - \tau } \big)^{\dagger} }{\big( \rho - \tau \big)^{\dagger}}  }     \bigg\} \bigg/  \mathrm{Tr} \bigg\{       {\tiny               \frac{ \rho - \tau }{ \widetilde{\rho - \tau}   }                     }                    \bigg\}      \bigg]              ,        {\mathrm{Tr} \bigg\{  {\tiny   \frac{\sqrt{\widetilde{\rho}}}{\sqrt{\big( \rho - \tau \big)}}      }   } \bigg\}   \\   \\  \\       \times  \bigg[ \mathrm{Tr} \bigg\{               {\tiny              \frac{  \big( \widetilde{\rho - \tau } \big)^{\dagger} }{\big( \rho - \tau \big)^{\dagger}}  }     \bigg\} \bigg/  \mathrm{Tr} \bigg\{       {\tiny               \frac{ \rho - \tau }{ \widetilde{\rho - \tau}   }                     }                    \bigg\}      \bigg]   ,  {\mathrm{Tr}          \bigg\{       {\tiny \frac{\sqrt{\widetilde{\rho}^{\dagger}}}{\sqrt{\big( \rho - \tau \big)^{\dagger} }} }         }  \bigg\}    \times  \bigg[ \mathrm{Tr} \bigg\{               {\tiny              \frac{  \big( \widetilde{\rho - \tau } \big)^{\dagger} }{\big( \rho - \tau \big)^{\dagger}}  }     \bigg\} \bigg/  \mathrm{Tr} \bigg\{       {\tiny               \frac{ \rho - \tau }{ \widetilde{\rho - \tau}   }                     }                    \bigg\}      \bigg]  ,  \\ \\ \\       {\mathrm{Tr} \bigg\{   {\tiny \frac{\sqrt{\widetilde{\tau}}}{\sqrt{\big( \rho - \tau \big) } }}   } \bigg\}    \times  \bigg[ \mathrm{Tr} \bigg\{               {\tiny              \frac{  \big( \widetilde{\rho - \tau } \big)^{\dagger} }{\big( \rho - \tau \big)^{\dagger}}  }     \bigg\} \bigg/  \mathrm{Tr} \bigg\{       {\tiny               \frac{ \rho - \tau }{ \widetilde{\rho - \tau}   }                     }                    \bigg\}      \bigg]   ,     {\mathrm{Tr}              \bigg\{  {\tiny   \frac{\sqrt{\widetilde{\tau}^{\dagger}}}{\sqrt{\big( \rho - \tau \big)^{\dagger}} }      }              }  \bigg\}    \\ \\ \\      \times  \bigg[ \mathrm{Tr} \bigg\{               {\tiny              \frac{  \big( \widetilde{\rho - \tau } \big)^{\dagger} }{\big( \rho - \tau \big)^{\dagger}}  }     \bigg\} \bigg/  \mathrm{Tr} \bigg\{       {\tiny               \frac{ \rho - \tau }{ \widetilde{\rho - \tau}   }                     }                    \bigg\}      \bigg]    ,    {\mathrm{Tr}         \bigg\{   {\tiny    \frac{\sqrt{\widetilde{\rho}}}{\sqrt{\big( \rho - \tau \big) }}     }      } \bigg\} \\ \\ \\     \times  \bigg[ \mathrm{Tr} \bigg\{               {\tiny              \frac{  \big( \widetilde{\rho - \tau } \big)^{\dagger} }{\big( \rho - \tau \big)^{\dagger}}  }     \bigg\} \bigg/  \mathrm{Tr} \bigg\{       {\tiny               \frac{ \rho - \tau }{ \widetilde{\rho - \tau}   }                     }                    \bigg\}      \bigg]    ,   {\mathrm{Tr}     \bigg\{  {\tiny   \frac{\sqrt{\widetilde{\tau}^{\dagger}}}{\sqrt{\big( \rho - \tau \big)^{\dagger}}}      }           } \bigg\}  \\ \\ \\   \times  \bigg[ \mathrm{Tr} \bigg\{               {\tiny              \frac{  \big( \widetilde{\rho - \tau } \big)^{\dagger} }{\big( \rho - \tau \big)^{\dagger}}  }     \bigg\} \bigg/  \mathrm{Tr} \bigg\{       {\tiny               \frac{ \rho - \tau }{ \widetilde{\rho - \tau}   }                     }                    \bigg\}      \bigg]   ,    {\mathrm{Tr}     \bigg\{  {\tiny   \frac{\sqrt{\widetilde{\tau}}}{\sqrt{\big( \rho - \tau \big)}}      }         }   \bigg\}  \\ \\ \\   \times  \bigg[ \mathrm{Tr} \bigg\{               {\tiny              \frac{  \big( \widetilde{\rho - \tau } \big)^{\dagger} }{\big( \rho - \tau \big)^{\dagger}}  }     \bigg\} \bigg/  \mathrm{Tr} \bigg\{       {\tiny               \frac{ \rho - \tau }{ \widetilde{\rho - \tau}   }                     }                    \bigg\}      \bigg]             \bigg\}^8       > 0   .
\end{align*}

}

\noindent Hence, from the fact that,

{\small

\begin{align*}
   \mathrm{Td} \bigg[    \frac{ \big\{ \Pi \rho^* \Pi / \widetilde{\Pi} \widetilde{\rho^*} \widetilde{\Pi}  \big\}            }{  \big\{  \mathrm{Tr} \big[        \Pi \rho^*   \big] /     \mathrm{Tr} \big[  \widetilde{\Pi } \widetilde{\rho^{*}}    \big]    \big\}                           }  ,          \frac{ \big\{ \ket{c , \sigma_c}  \bra{c , \sigma_c }   \big\}     }{ \big\{ \ket{\widetilde{c} , \widetilde{\sigma_c}}   \bra{\widetilde{c} , \widetilde{\sigma_c }} \big\}   }         \bigg] \leq \mathrm{Td} \bigg[     \frac{ \big\{ \Pi \rho^* \Pi / \widetilde{\Pi} \widetilde{\rho^*} \widetilde{\Pi}  \big\}            }{  \big\{  \mathrm{Tr} \big[        \Pi \rho^*   \big] /     \mathrm{Tr} \big[  \widetilde{\Pi } \widetilde{\rho^{*}}    \big]    \big\}                           }          ,        \rho^* \big/ \widetilde{\rho^*}               \bigg]  , 
\end{align*}

}

\noindent by the GML,

{\small

\begin{align*}
 \mathrm{Td} \bigg[    \frac{ \big\{ \Pi \rho^* \Pi / \widetilde{\Pi} \widetilde{\rho^*} \widetilde{\Pi}  \big\}            }{  \big\{  \mathrm{Tr} \big[        \Pi \rho^*   \big] /     \mathrm{Tr} \big[  \widetilde{\Pi } \widetilde{\rho^{*}}    \big]    \big\}                           }  ,          \frac{ \big\{ \ket{c , \sigma_c}  \bra{c , \sigma_c }   \big\}     }{ \big\{ \ket{\widetilde{c} , \widetilde{\sigma_c}}   \bra{\widetilde{c} , \widetilde{\sigma_c }} \big\}   }         \bigg] \leq 2 \sqrt{\epsilon}    ,  
\end{align*}

}

\noindent because,

{\small

\begin{align*}
   \mathrm{Td} \bigg[     \frac{ \big\{ \Pi \rho^* \Pi / \widetilde{\Pi} \widetilde{\rho^*} \widetilde{\Pi}  \big\}            }{  \big\{  \mathrm{Tr} \big[        \Pi \rho^*   \big] /     \mathrm{Tr} \big[  \widetilde{\Pi } \widetilde{\rho^{*}}    \big]    \big\}                           }          ,        \rho^* \big/ \widetilde{\rho^*}               \bigg]  \leq 2 \sqrt{\epsilon}  , 
\end{align*}

}

\noindent from the choice of the parameter $\epsilon$,

{\small

\begin{align*}
  {\tiny \epsilon \equiv \epsilon \big( \rho , \tau , \widetilde{\rho} , \widetilde{\tau} , C , C^{\prime} , C^{\prime\prime} , \mathcal{C} , \mathcal{C}^{\prime} , \mathcal{C}^{\prime\prime} \big) }   , 
\end{align*}

}

\noindent derived above. Similarly another application of the GML allows one to conclude,

{\small

\begin{align*}
  \mathrm{Td} \bigg[    \frac{ \big\{ \Pi \rho^* \Pi / \widetilde{\Pi} \widetilde{\rho^*} \widetilde{\Pi}  \big\}            }{  \big\{  \mathrm{Tr} \big[        \Pi \rho^*   \big] /     \mathrm{Tr} \big[  \widetilde{\Pi } \widetilde{\rho^{*}}    \big]    \big\}                           }  ,   \rho^* \big/ \widetilde{\rho^*}                                \bigg] \leq 2 \sqrt{   {\tiny \mathrm{NEGL} \big(      \lambda^{\prime}  - \lambda     \big)    }          }     , 
\end{align*}

}

\noindent and therefore,

{\small

\begin{align*}
  \mathrm{Td} \bigg[    \frac{ \big\{ \Pi \rho^* \Pi / \widetilde{\Pi} \widetilde{\rho^*} \widetilde{\Pi}  \big\}            }{  \big\{  \mathrm{Tr} \big[        \Pi \rho^*   \big] /     \mathrm{Tr} \big[  \widetilde{\Pi } \widetilde{\rho^{*}}    \big]    \big\}                           }  ,          \frac{ \big\{ \ket{c , \sigma_c}  \bra{c , \sigma_c }   \big\}     }{ \big\{ \ket{\widetilde{c} , \widetilde{\sigma_c}}   \bra{\widetilde{c} , \widetilde{\sigma_c }} \big\}   }         \bigg] \leq 2 \sqrt{    {\tiny \mathrm{NEGL} \big(      \lambda^{\prime}  - \lambda     \big) }              }     ,    \\ \end{align*}

  \begin{align*} 2 \sqrt{     {\tiny \mathrm{NEGL} \big(      \lambda^{\prime}  - \lambda     \big) }            }  \leq 2 \cdot  {\tiny \mathrm{exp} \big[ \lambda^{\prime} - \lambda  \big] }   \lesssim {\tiny \mathrm{exp} \big[ \lambda^{\prime} - \lambda  \big]}  , 
\end{align*}

}

\noindent because, as in the previous application of the GML,

{\small

\begin{align*}
   \mathrm{Td} \bigg[    \frac{ \big\{ \Pi \rho^* \Pi / \widetilde{\Pi} \widetilde{\rho^*} \widetilde{\Pi}  \big\}            }{  \big\{  \mathrm{Tr} \big[        \Pi \rho^*   \big] /     \mathrm{Tr} \big[  \widetilde{\Pi } \widetilde{\rho^{*}}    \big]    \big\}                           }  ,          \frac{ \big\{ \ket{c , \sigma_c}  \bra{c , \sigma_c }   \big\}     }{ \big\{ \ket{\widetilde{c} , \widetilde{\sigma_c}}   \bra{\widetilde{c} , \widetilde{\sigma_c }} \big\}   }         \bigg] \leq \mathrm{Td} \bigg[     \frac{ \big\{ \Pi \rho^* \Pi / \widetilde{\Pi} \widetilde{\rho^*} \widetilde{\Pi}  \big\}            }{  \big\{  \mathrm{Tr} \big[        \Pi \rho^*   \big] /     \mathrm{Tr} \big[  \widetilde{\Pi } \widetilde{\rho^{*}}    \big]    \big\}                           }          ,        \rho^* \big/ \widetilde{\rho^*}               \bigg]  . \\ 
\end{align*}

}

\noindent from which it has been demonstrated that the desired upper bound on the trace distance holds. We conclude the argument as it has been demonstrated that items \textit{(A)}-\textit{(C)} hold. \boxed{}

\subsection{Proof of Theorem}

\noindent \textit{Proof of Theorem}. To demonstrate that the everlasting security notion,

{\small

\begin{align*}
  \mathrm{Td} \big(    {\tiny \widetilde{\mathrm{Exp}}^{A_{\lambda^{\prime}}}  \big( 1^{\lambda^{\prime}} , 1 \big)     }      ,   {\tiny \widetilde{\mathrm{Exp}}^{A_{\lambda^{\prime}}} \big( 1^{\lambda^{\prime}} , 0 \big)     }    \big)  \lesssim {\tiny \mathrm{NEGL} \big( \lambda^{\prime} \big) }   , \\ 
\end{align*}

}

\noindent holds recall,

{\small

\begin{align*}
  \mathcal{P} \bigg[ {\tiny  1 = \widetilde{\mathrm{Ver}} \big( \widetilde{\mathrm{vk}} , \widetilde{m^{*}} , \widetilde{\sigma^{*}} \big)}  \textit{ with } {\tiny \big( \widetilde{m} , \widetilde{\sigma} \big) \neq \big( \widetilde{m^{*}} , \widetilde{\sigma^{*}} \big) :                     \big\{ \big( \widetilde{\mathrm{vk}} , \widetilde{\mathrm{sk}} \big) \leftarrow \widetilde{\mathrm{SGen}} \big( 1^{\lambda^{\prime}} \big) \big\} , \big\{ \sigma \leftarrow \widetilde{\mathrm{Sign}} \big( \widetilde{\mathrm{sk}} , \widetilde{m} \big) \big\} ,}  \\ {\tiny \big\{        \big( \widetilde{m^{*}} , \widetilde{\sigma^{*}} \big) \leftarrow A_{\lambda^{\prime}} \big( \mathrm{vk} , \widetilde{m} , \widetilde{\sigma} \big)       \big\}  }     \bigg] = {\tiny \mathrm{NEGL} \big( \lambda^{\prime} \big) }  , \\ 
\end{align*}

}

\noindent from \textbf{Definition} \textit{40} is an alternative formulation of the neglibility function $\mathrm{NEGL} {\tiny \big( \cdot \big)}$ and,

{\small

\begin{align*}
 \mathcal{P} \big[  {\tiny  1 = \widetilde{\mathrm{Ver}} \big( \widetilde{\mathrm{vk}} , m^{\prime} ,  \widetilde{\mathrm{Sign}} \big(       \widetilde{\mathrm{sk}} , m^{\prime}    \big)  \big)  :                \big( \widetilde{\mathrm{vk}} , \widetilde{\mathrm{sk}} \big) \leftarrow  \widetilde{\mathrm{SGen}} \big( 1^{\lambda^{\prime}} \big)    }        \big] = 1    , \\ 
\end{align*}

}

\noindent from \textbf{Definition} \textit{35} is the correctness property of the signature generation scheme. Under the correctness conditions involving the special abort symbol '$\bot$' in the NQPKE-QKD setting manipulations in the previous result demonstrated that,

{\small

\begin{align*}
 \mathrm{Td} \bigg[    \frac{ \big\{ \Pi \rho^* \Pi / \widetilde{\Pi} \widetilde{\rho^*} \widetilde{\Pi}  \big\}            }{  \big\{  \mathrm{Tr} \big[        \Pi \rho^*   \big] /     \mathrm{Tr} \big[  \widetilde{\Pi } \widetilde{\rho^{*}}    \big]    \big\}                           }  ,          \frac{ \big\{ \ket{c , \sigma_c}  \bra{c , \sigma_c }   \big\}     }{ \big\{ \ket{\widetilde{c} , \widetilde{\sigma_c}}   \bra{\widetilde{c} , \widetilde{\sigma_c }} \big\}   }             \bigg]      \lesssim  \sqrt{ \bigg\{  {\tiny \frac{\mathrm{negl} \big(    \lambda    \big)}{\mathrm{NEGL} \big(      \lambda^{\prime}      \big)}} \bigg\}^{-1} }    \overset{(\mathrm{GM})}{\leq}    2 \sqrt{ {\tiny \mathrm{NEGL} \big(      \lambda^{\prime}      \big)}}     \\ \lesssim                   {\tiny \mathrm{exp} \big[   \lambda - \lambda^{\prime}        \big] }   ,  \\ 
\end{align*}

}

\noindent namely that it suffices to argue that the desired statement holds from the conditions,

{\small

\begin{align*} 
     \mathrm{Td} \big(    {\tiny \widetilde{\mathrm{Exp}}^{A_{\lambda^{\prime}}}  \big( 1^{\lambda^{\prime}} , 1 \big)     }      ,   {\tiny \widetilde{\mathrm{Exp}}^{A_{\lambda^{\prime}}} \big( 1^{\lambda^{\prime}} , 0 \big)     }    \big)  \lesssim  {\tiny \mathrm{NEGL} \big( \lambda^{\prime} \big) }        , \\ \\    \mathrm{Td} \big(    {\tiny {\mathrm{Exp}}^{A_{\lambda}}  \big( 1^{\lambda} , 1 \big)     }      ,   {\tiny {\mathrm{Exp}}^{A_{\lambda}} \big( 1^{\lambda} , 0 \big)     }    \big) =  {\tiny \mathrm{negl} \big( \lambda \big) }   ,  
\end{align*} 

}

\noindent which encapsulate everlasting security in the noisy and noiseless settings, respectively. To upper bound,

{\small

\begin{align*} 
    \mathrm{Td} \bigg[       \bigg\{   {\tiny \widetilde{\mathrm{Exp}}^{A_{\lambda^{\prime}}}  \big( 1^{\lambda^{\prime}} , 1 \big)}  \big/   {\tiny {\mathrm{Exp}}^{A_{\lambda}}  \big( 1^{\lambda} , 1 \big)   }  \bigg\}     ,  \bigg\{  {\tiny \widetilde{\mathrm{Exp}}^{A_{\lambda^{\prime}}} \big( 1^{\lambda^{\prime}} , 0 \big) }  \big/   {\tiny {\mathrm{Exp}}^{A_{\lambda}} \big( 1^{\lambda} , 0 \big)            }     \bigg\}   \bigg]  \\ \\  =        \bigg| \bigg|           \bigg\{   {\tiny \widetilde{\mathrm{Exp}}^{A_{\lambda^{\prime}}}  \big( 1^{\lambda^{\prime}} , 1 \big)}  \big/   {\tiny {\mathrm{Exp}}^{A_{\lambda}}  \big( 1^{\lambda} , 1 \big)   }  \bigg\}        -  \bigg\{  {\tiny \widetilde{\mathrm{Exp}}^{A_{\lambda^{\prime}}} \big( 1^{\lambda^{\prime}} , 0 \big) }  \big/   {\tiny {\mathrm{Exp}}^{A_{\lambda}} \big( 1^{\lambda} , 0 \big)            }     \bigg\}  \bigg| \bigg|_1        \end{align*}

    \begin{align*} =     \bigg| \bigg|           \bigg\{   {\tiny \widetilde{\mathrm{Exp}}^{A_{\lambda^{\prime}}}  \big( 1^{\lambda^{\prime}} , 1 \big)} \times  {\tiny {\mathrm{Exp}}^{A_{\lambda}} \big( 1^{\lambda} , 0 \big)            }    -        {\tiny \widetilde{\mathrm{Exp}}^{A_{\lambda^{\prime}}} \big( 1^{\lambda^{\prime}} , 0 \big) }        \times      {\tiny {\mathrm{Exp}}^{A_{\lambda}}  \big( 1^{\lambda} , 1 \big)   }    \bigg\} \bigg/   \bigg\{    {\tiny {\mathrm{Exp}}^{A_{\lambda}}  \big( 1^{\lambda} , 1 \big)   }          \\ \\ \times    {\tiny {\mathrm{Exp}}^{A_{\lambda}} \big( 1^{\lambda} , 0 \big)            }        \bigg\}        \bigg| \bigg|_1                , \\ 
\end{align*} 

}

\noindent with the negligibility function $\mathrm{NEGL} {\tiny \big( \cdot \big)}$, directly apply the computations from the previous above result, particularly through the GML, to conclude that an upper bound on,

{\small

\begin{align*}
   \mathrm{Tr} \bigg\{   \sqrt{ \frac{ {\tiny \big( \widetilde{\rho - \tau} \big)^{\dagger}}}{{\tiny \big( \rho - \tau \big)^{\dagger } }} \frac{{\tiny \big( \widetilde{\rho - \tau } \big)}}{{\tiny \big( \rho - \tau \big) } }}                    \bigg\}  , 
\end{align*}

}

\noindent readily implies the last desired result for everlasting security under noise stated in $\textbf{Theorem}$, as,

{\small

\begin{align*}
   \bigg\{   \mathrm{Tr} \bigg\{   \sqrt{ \frac{ {\tiny \big( \widetilde{\rho - \tau} \big)^{\dagger}}}{{\tiny \big( \rho - \tau \big)^{\dagger } }} \frac{{\tiny \big( \widetilde{\rho - \tau } \big)}}{{\tiny \big( \rho - \tau \big) } }}                    \bigg\}            \geq      1 -  \bigg(  \frac{1}{2 \mathcal{C} \mathcal{C}^{\prime}  \big( \mathcal{C}^{\prime\prime} \big)^3  \big( C C^{\prime} C^{\prime\prime} \big)^2 } \bigg)^2  \cdot         \underset{\rho,\tau,\widetilde{\rho},\widetilde{\tau}}{\mathrm{inf}} \bigg\{         {\mathrm{Tr} \bigg\{   {\tiny \frac{\sqrt{\widetilde{\rho}^{\dagger}}}{\sqrt{\big( \rho - \tau \big)^{\dagger} }}  }  }  \bigg\}     \times  \bigg[ \mathrm{Tr} \bigg\{               {\tiny              \frac{  \big( \widetilde{\rho - \tau } \big)^{\dagger} }{\big( \rho - \tau \big)^{\dagger}}  }     \bigg\}   \\   \\ \bigg/  \mathrm{Tr} \bigg\{       {\tiny               \frac{ \rho - \tau }{ \widetilde{\rho - \tau}   }                     }                    \bigg\}      \bigg]              ,        {\mathrm{Tr} \bigg\{  {\tiny   \frac{\sqrt{\widetilde{\rho}}}{\sqrt{\big( \rho - \tau \big)}}      }   } \bigg\}          \bigg[ \mathrm{Tr} \bigg\{               {\tiny              \frac{  \big( \widetilde{\rho - \tau } \big)^{\dagger} }{\big( \rho - \tau \big)^{\dagger}}  }     \bigg\} \bigg/  \mathrm{Tr} \bigg\{       {\tiny               \frac{ \rho - \tau }{ \widetilde{\rho - \tau}   }                     }                    \bigg\}      \bigg]   ,  {\mathrm{Tr}          \bigg\{       {\tiny \frac{\sqrt{\widetilde{\rho}^{\dagger}}}{\sqrt{\big( \rho - \tau \big)^{\dagger} }} }         }  \bigg\}    \times  \bigg[ \mathrm{Tr} \bigg\{               {\tiny              \frac{  \big( \widetilde{\rho - \tau } \big)^{\dagger} }{\big( \rho - \tau \big)^{\dagger}}  }     \bigg\} \\  \\ \bigg/  \mathrm{Tr} \bigg\{       {\tiny               \frac{ \rho - \tau }{ \widetilde{\rho - \tau}   }                     }                    \bigg\}      \bigg]  ,    {\mathrm{Tr} \bigg\{   {\tiny \frac{\sqrt{\widetilde{\tau}}}{\sqrt{\big( \rho - \tau \big) } }}   } \bigg\}    \times  \bigg[ \mathrm{Tr} \bigg\{               {\tiny              \frac{  \big( \widetilde{\rho - \tau } \big)^{\dagger} }{\big( \rho - \tau \big)^{\dagger}}  }     \bigg\} \bigg/  \mathrm{Tr} \bigg\{       {\tiny               \frac{ \rho - \tau }{ \widetilde{\rho - \tau}   }                     }                    \bigg\}      \bigg]   ,     {\mathrm{Tr}              \bigg\{  {\tiny   \frac{\sqrt{\widetilde{\tau}^{\dagger}}}{\sqrt{\big( \rho - \tau \big)^{\dagger}} }      }              }  \bigg\}    \\ \\ \\      \times  \bigg[ \mathrm{Tr} \bigg\{               {\tiny              \frac{  \big( \widetilde{\rho - \tau } \big)^{\dagger} }{\big( \rho - \tau \big)^{\dagger}}  }     \bigg\} \bigg/  \mathrm{Tr} \bigg\{       {\tiny               \frac{ \rho - \tau }{ \widetilde{\rho - \tau}   }                     }                    \bigg\}      \bigg]    ,    {\mathrm{Tr}         \bigg\{   {\tiny    \frac{\sqrt{\widetilde{\rho}}}{\sqrt{\big( \rho - \tau \big) }}     }      } \bigg\} \\ \\ \\     \times  \bigg[ \mathrm{Tr} \bigg\{               {\tiny              \frac{  \big( \widetilde{\rho - \tau } \big)^{\dagger} }{\big( \rho - \tau \big)^{\dagger}}  }     \bigg\} \bigg/  \mathrm{Tr} \bigg\{       {\tiny               \frac{ \rho - \tau }{ \widetilde{\rho - \tau}   }                     }                    \bigg\}      \bigg]    ,   {\mathrm{Tr}     \bigg\{  {\tiny   \frac{\sqrt{\widetilde{\tau}^{\dagger}}}{\sqrt{\big( \rho - \tau \big)^{\dagger}}}      }           } \bigg\}  \\ \\ \\   \times  \bigg[ \mathrm{Tr} \bigg\{               {\tiny              \frac{  \big( \widetilde{\rho - \tau } \big)^{\dagger} }{\big( \rho - \tau \big)^{\dagger}}  }     \bigg\} \bigg/  \mathrm{Tr} \bigg\{       {\tiny               \frac{ \rho - \tau }{ \widetilde{\rho - \tau}   }                     }                    \bigg\}      \bigg]   ,    {\mathrm{Tr}     \bigg\{  {\tiny   \frac{\sqrt{\widetilde{\tau}}}{\sqrt{\big( \rho - \tau \big)}}      }         }   \bigg\}  \\ \\ \\  \times  \bigg[ \mathrm{Tr} \bigg\{               {\tiny              \frac{  \big( \widetilde{\rho - \tau } \big)^{\dagger} }{\big( \rho - \tau \big)^{\dagger}}  }     \bigg\} \bigg/  \mathrm{Tr} \bigg\{       {\tiny               \frac{ \rho - \tau }{ \widetilde{\rho - \tau}   }                     }                    \bigg\}      \bigg]             \bigg\}^8                       \bigg\} \\ \end{align*}

   \begin{align*}   \overset{(\mathrm{GML})}{\Longrightarrow} \bigg\{                                    \mathrm{Td} \bigg[    \frac{ \big\{ \Pi \rho^* \Pi / \widetilde{\Pi} \widetilde{\rho^*} \widetilde{\Pi}  \big\}            }{  \big\{  \mathrm{Tr} \big[        \Pi \rho^*   \big] /     \mathrm{Tr} \big[  \widetilde{\Pi } \widetilde{\rho^{*}}    \big]    \big\}                           }  ,          \frac{ \big\{ \ket{c , \sigma_c}  \bra{c , \sigma_c }   \big\}     }{ \big\{ \ket{\widetilde{c} , \widetilde{\sigma_c}}   \bra{\widetilde{c} , \widetilde{\sigma_c }} \big\}   }         \bigg] \leq 2 \sqrt{    {\tiny \mathrm{NEGL} \big(      \lambda^{\prime}  - \lambda     \big)  }              }                                         \bigg\}  \\ \\ \\     \overset{(\textbf{Lemma})}{\Longrightarrow} \bigg\{                    \mathrm{Td} \bigg[    \frac{ \big\{ \Pi \rho^* \Pi / \widetilde{\Pi} \widetilde{\rho^*} \widetilde{\Pi}  \big\}            }{  \big\{  \mathrm{Tr} \big[        \Pi \rho^*   \big] /     \mathrm{Tr} \big[  \widetilde{\Pi } \widetilde{\rho^{*}}    \big]    \big\}                           }  ,          \frac{ \big\{ \ket{c , \sigma_c}  \bra{c , \sigma_c }   \big\}     }{ \big\{ \ket{\widetilde{c} , \widetilde{\sigma_c}}   \bra{\widetilde{c} , \widetilde{\sigma_c }} \big\}   }         \bigg]                                                                     \lesssim {\tiny \mathrm{exp} \big[ \lambda^{\prime} - \lambda  \big]}          \bigg\} \\ \\ \\  {\Longrightarrow} \bigg\{                                       \mathrm{Td} \bigg[    \frac{ \big\{ \Pi \rho^* \Pi / \widetilde{\Pi} \widetilde{\rho^*} \widetilde{\Pi}  \big\}            }{  \big\{  \mathrm{Tr} \big[        \Pi \rho^*   \big] /     \mathrm{Tr} \big[  \widetilde{\Pi } \widetilde{\rho^{*}}    \big]    \big\}                           }  ,          \frac{ \big\{ \ket{c , \sigma_c}  \bra{c , \sigma_c }   \big\}     }{ \big\{ \ket{\widetilde{c} , \widetilde{\sigma_c}}   \bra{\widetilde{c} , \widetilde{\sigma_c }} \big\}   }         \bigg]      \lesssim  {\tiny  \mathrm{NEGL}  \big( \lambda^{\prime} - \lambda  \big)       }                                                         \bigg\}   \\ \\ \\   {\Longrightarrow} \bigg\{                                       \mathrm{Td} \bigg[    \frac{ \big\{ \Pi \rho^* \Pi / \widetilde{\Pi} \widetilde{\rho^*} \widetilde{\Pi}  \big\}            }{  \big\{  \mathrm{Tr} \big[        \Pi \rho^*   \big] /     \mathrm{Tr} \big[  \widetilde{\Pi } \widetilde{\rho^{*}}    \big]    \big\}                           }  ,          \frac{ \big\{ \ket{c , \sigma_c}  \bra{c , \sigma_c }   \big\}     }{ \big\{ \ket{\widetilde{c} , \widetilde{\sigma_c}}   \bra{\widetilde{c} , \widetilde{\sigma_c }} \big\}   }         \bigg]      \lesssim  {\tiny  \mathrm{NEGL}  \big( \lambda^{\prime}   \big)       }                                                         \bigg\} \\ \end{align*}

   \begin{align*}   {\Longrightarrow}      \bigg\{           \mathrm{Td} \bigg(       \bigg\{   {\tiny \widetilde{\mathrm{Exp}}^{A_{\lambda^{\prime}}}  \big( 1^{\lambda^{\prime}} , 1 \big)}  \big/   {\tiny {\mathrm{Exp}}^{A_{\lambda}}  \big( 1^{\lambda} , 1 \big)   }  \bigg\}     ,  \bigg\{  {\tiny \widetilde{\mathrm{Exp}}^{A_{\lambda^{\prime}}} \big( 1^{\lambda^{\prime}} , 0 \big) }  \big/   {\tiny {\mathrm{Exp}}^{A_{\lambda}} \big( 1^{\lambda} , 0 \big)            }     \bigg\}   \bigg)      \leq  {\tiny  \mathrm{NEGL}  \big( \lambda^{\prime}  - \lambda  \big)       }                                                               \bigg\}  \\ \end{align*}

   \begin{align*} {\Longrightarrow}      \bigg\{           \mathrm{Td} \bigg(       \bigg\{   {\tiny \widetilde{\mathrm{Exp}}^{A_{\lambda^{\prime}}}  \big( 1^{\lambda^{\prime}} , 1 \big)}  \big/   {\tiny {\mathrm{Exp}}^{A_{\lambda}}  \big( 1^{\lambda} , 1 \big)   }  \bigg\}     ,  \bigg\{  {\tiny \widetilde{\mathrm{Exp}}^{A_{\lambda^{\prime}}} \big( 1^{\lambda^{\prime}} , 0 \big) }  \big/   {\tiny {\mathrm{Exp}}^{A_{\lambda}} \big( 1^{\lambda} , 0 \big)            }     \bigg\}   \bigg)      \leq  {\tiny  \mathrm{NEGL}  \big( \lambda^{\prime}   \big)       }                                                               \bigg\}      \\ \end{align*}

   \begin{align*}   {\Longrightarrow}      \bigg\{           \mathrm{Td} \bigg(       \bigg\{   {\tiny \widetilde{\mathrm{Exp}}^{A_{\lambda^{\prime}}}  \big( 1^{\lambda^{\prime}} , 1 \big)}  \big/   {\tiny {\mathrm{Exp}}^{A_{\lambda}}  \big( 1^{\lambda} , 1 \big)   }  \bigg\}     ,  \bigg\{  {\tiny \widetilde{\mathrm{Exp}}^{A_{\lambda^{\prime}}} \big( 1^{\lambda^{\prime}} , 0 \big) }  \big/   {\tiny {\mathrm{Exp}}^{A_{\lambda}} \big( 1^{\lambda} , 0 \big)            }     \bigg\}   \bigg)      \lesssim  {\tiny  \mathrm{NEGL}  \big( \lambda^{\prime} - \lambda   \big)       }                                                               \bigg\} \\ \end{align*}

   \begin{align*}  {\Longrightarrow}      \bigg\{           \mathrm{Td} \bigg(       \bigg\{   {\tiny \widetilde{\mathrm{Exp}}^{A_{\lambda^{\prime}}}  \big( 1^{\lambda^{\prime}} , 1 \big)}  \big/   {\tiny {\mathrm{Exp}}^{A_{\lambda}}  \big( 1^{\lambda} , 1 \big)   }  \bigg\}     ,  \bigg\{  {\tiny \widetilde{\mathrm{Exp}}^{A_{\lambda^{\prime}}} \big( 1^{\lambda^{\prime}} , 0 \big) }  \big/   {\tiny {\mathrm{Exp}}^{A_{\lambda}} \big( 1^{\lambda} , 0 \big)            }     \bigg\}   \bigg)      \lesssim  {\tiny  \mathrm{NEGL}  \big( \lambda^{\prime}   \big)       }                                                               \bigg\}   \\ \\ \\ \Longleftrightarrow    \bigg\{              \bigg| \bigg|           \bigg\{   {\tiny \widetilde{\mathrm{Exp}}^{A_{\lambda^{\prime}}}  \big( 1^{\lambda^{\prime}} , 1 \big)} \times  {\tiny {\mathrm{Exp}}^{A_{\lambda}} \big( 1^{\lambda} , 0 \big)            }    -        {\tiny \widetilde{\mathrm{Exp}}^{A_{\lambda^{\prime}}} \big( 1^{\lambda^{\prime}} , 0 \big) }        \times      {\tiny {\mathrm{Exp}}^{A_{\lambda}}  \big( 1^{\lambda} , 1 \big)   }    \bigg\} \bigg/   \bigg\{    {\tiny {\mathrm{Exp}}^{A_{\lambda}}  \big( 1^{\lambda} , 1 \big)   }          \\ \\ \times    {\tiny {\mathrm{Exp}}^{A_{\lambda}} \big( 1^{\lambda} , 0 \big)            }        \bigg\}        \bigg| \bigg|_1    \lesssim  {\tiny  \mathrm{NEGL}  \big( \lambda^{\prime}   \big)       }                                                               \bigg\} , \\ 
\end{align*}

}

\noindent from which we conclude the argument. \boxed{}

\subsection{Proof of Corollary}

\noindent \textit{Proof of Corollary}. It suffices to show,

{\small

\begin{align*}
  \widetilde{\mathrm{Adv}} \big( 0 \big)    \big/ \mathrm{NEGL} \big( \lambda^{\prime} \big) \lesssim 1          , \\
\end{align*}

}

 \noindent which observe, by direct computation,

{\small

\begin{align*}
      \widetilde{\mathrm{Adv}} \big( 0 \big)    \big/ \mathrm{NEGL} \big( \lambda^{\prime} \big)  \overset{(\textbf{Theorem})}{\lesssim}             \widetilde{\mathrm{Adv}} \big( 0 \big)    \bigg/      \mathrm{Td} \bigg[     \frac{ \big\{ \Pi \rho^* \Pi / \widetilde{\Pi} \widetilde{\rho^*} \widetilde{\Pi}  \big\}            }{  \big\{  \mathrm{Tr} \big[        \Pi \rho^*   \big] /     \mathrm{Tr} \big[  \widetilde{\Pi } \widetilde{\rho^{*}}    \big]    \big\}                           }          ,        \rho^* \big/ \widetilde{\rho^*}               \bigg] \lesssim    \widetilde{\mathrm{Adv}} \big( 0 \big)   \\ \\  \bigg/      \mathrm{Td} \bigg[ \frac{ \big\{ \Pi \rho^* \Pi / \widetilde{\Pi} \widetilde{\rho^*} \widetilde{\Pi}  \big\}            }{  \big\{  \mathrm{Tr} \big[        \Pi \rho^*   \big] /     \mathrm{Tr} \big[  \widetilde{\Pi } \widetilde{\rho^{*}}    \big]    \big\}                           }   ,          \frac{ \big\{ \ket{c , \sigma_c}  \bra{c , \sigma_c }   \big\}     }{ \big\{ \ket{\widetilde{c} , \widetilde{\sigma_c}}   \bra{\widetilde{c} , \widetilde{\sigma_c }} \big\}   }   \bigg]        \end{align*}

      \begin{align*} =            \mathrm{Td} \big[   {\tiny \widetilde{\mathrm{Hyb}^{\mathcal{A}_{\lambda^{\prime}}}_{0,0}}     ,   \widetilde{\mathrm{Hyb}^{\mathcal{A}_{\lambda^{\prime}}}_{1,0} }  }  \big]        \bigg/      \mathrm{Td} \bigg[ \frac{ \big\{ \Pi \rho^* \Pi / \widetilde{\Pi} \widetilde{\rho^*} \widetilde{\Pi}  \big\}            }{  \big\{  \mathrm{Tr} \big[        \Pi \rho^*   \big] /     \mathrm{Tr} \big[  \widetilde{\Pi } \widetilde{\rho^{*}}    \big]    \big\}                           }   ,          \frac{ \big\{ \ket{c , \sigma_c}  \bra{c , \sigma_c }   \big\}     }{ \big\{ \ket{\widetilde{c} , \widetilde{\sigma_c}}   \bra{\widetilde{c} , \widetilde{\sigma_c }} \big\}   }   \bigg]     \\ \end{align*}

      \begin{align*} <   (\mathrm{Const} \big(      \epsilon , \rho , \tau , \widetilde{\rho} , \widetilde{\tau} , C , C^{\prime} , C^{\prime\prime} , \mathcal{C} , \mathcal{C}^{\prime} , \mathcal{C}^{\prime\prime} \big)  \big)  \times      \mathrm{Td} \bigg[ {\tiny \widetilde{\mathrm{Hyb}^{\mathcal{A}_{\lambda^{\prime}}}_{0,0}} } \bigg/  \bigg\{ \frac{ \big\{ \Pi \rho^* \Pi / \widetilde{\Pi} \widetilde{\rho^*} \widetilde{\Pi}  \big\}            }{  \big\{  \mathrm{Tr} \big[        \Pi \rho^*   \big] /     \mathrm{Tr} \big[  \widetilde{\Pi } \widetilde{\rho^{*}}    \big]    \big\}                           }   \bigg\}     ,   {\tiny \widetilde{\mathrm{Hyb}^{\mathcal{A}_{\lambda^{\prime}}}_{1,0} }  } \\ \\ \bigg/  \bigg\{    \frac{ \big\{ \ket{c , \sigma_c}  \bra{c , \sigma_c }   \big\}     }{ \big\{ \ket{\widetilde{c} , \widetilde{\sigma_c}}   \bra{\widetilde{c} , \widetilde{\sigma_c }} \big\}   }              \bigg\}             \bigg]   \end{align*}

\begin{align*} 
   \\  =      (\mathrm{Const} \big(      \epsilon , \rho , \tau , \widetilde{\rho} , \widetilde{\tau} , C , C^{\prime} , C^{\prime\prime} , \mathcal{C} , \mathcal{C}^{\prime} , \mathcal{C}^{\prime\prime} \big)  \big)  \times           \bigg| \bigg|          {\tiny \widetilde{\mathrm{Hyb}^{\mathcal{A}_{\lambda^{\prime}}}_{0,0}} } \bigg/  \bigg\{ \frac{ \big\{ \Pi \rho^* \Pi / \widetilde{\Pi} \widetilde{\rho^*} \widetilde{\Pi}  \big\}            }{  \big\{  \mathrm{Tr} \big[        \Pi \rho^*   \big] /     \mathrm{Tr} \big[  \widetilde{\Pi } \widetilde{\rho^{*}}    \big]    \big\}                           }   \bigg\}    -    {\tiny \widetilde{\mathrm{Hyb}^{\mathcal{A}_{\lambda^{\prime}}}_{1,0} }  }  \\ \\ \bigg/  \bigg\{    \frac{ \big\{ \ket{c , \sigma_c}  \bra{c , \sigma_c }   \big\}     }{ \big\{ \ket{\widetilde{c} , \widetilde{\sigma_c}}   \bra{\widetilde{c} , \widetilde{\sigma_c }} \big\}   }              \bigg\}     \bigg| \bigg|_1      \end{align*}

\begin{align*}
  \\   \leq    (\mathrm{Const} \big(      \epsilon , \rho , \tau , \widetilde{\rho} , \widetilde{\tau} , C , C^{\prime} , C^{\prime\prime} , \mathcal{C} , \mathcal{C}^{\prime} , \mathcal{C}^{\prime\prime} \big)  \big)  \times          \bigg| \bigg| \bigg[            {\tiny \widetilde{\mathrm{Hyb}^{\mathcal{A}_{\lambda^{\prime}}}_{0,0}} } \times   \bigg\{    \frac{ \big\{ \ket{c , \sigma_c}  \bra{c , \sigma_c }   \big\}     }{ \big\{ \ket{\widetilde{c} , \widetilde{\sigma_c}}   \bra{\widetilde{c} , \widetilde{\sigma_c }} \big\}   }              \bigg\}   \bigg]    -         \bigg[      {\tiny \widetilde{\mathrm{Hyb}^{\mathcal{A}_{\lambda^{\prime}}}_{1,0} }  } \\  \times  \bigg\{ \frac{ \big\{ \Pi \rho^* \Pi / \widetilde{\Pi} \widetilde{\rho^*} \widetilde{\Pi}  \big\}            }{  \big\{  \mathrm{Tr} \big[        \Pi \rho^*   \big] /     \mathrm{Tr} \big[  \widetilde{\Pi } \widetilde{\rho^{*}}    \big]    \big\}                           }   \bigg\}     \bigg]          \bigg| \bigg|_1      \\     \\ \times \bigg| \bigg|              \bigg\{    \frac{ \big\{ \ket{c , \sigma_c}  \bra{c , \sigma_c }   \big\}     }{ \big\{ \ket{\widetilde{c} , \widetilde{\sigma_c}}   \bra{\widetilde{c} , \widetilde{\sigma_c }} \big\}   }              \bigg\}    \times   \bigg\{ \frac{ \big\{ \Pi \rho^* \Pi / \widetilde{\Pi} \widetilde{\rho^*} \widetilde{\Pi}  \big\}            }{  \big\{  \mathrm{Tr} \big[        \Pi \rho^*   \big] /     \mathrm{Tr} \big[  \widetilde{\Pi } \widetilde{\rho^{*}}    \big]    \big\}                           }   \bigg\}    \bigg| \bigg|^{-1}_1   
\end{align*}

      \begin{align*} \lesssim 1       . \\
\end{align*}

}

\noindent Identical results, through the statements,

{\small

\begin{align*}
  \widetilde{\mathrm{Adv}} {\tiny \big( 1 \big)  }   \big/ \mathrm{NEGL} {\tiny \big( \lambda^{\prime} \big) }  \lesssim 1          , \\ \\   \widetilde{\mathrm{Adv}} {\tiny \big( 2 \big) }    \big/ \mathrm{NEGL} {\tiny \big( \lambda^{\prime} \big) } \lesssim 1        , 
\end{align*}

}

\noindent can be immediately obtained by directly applying the above rearrangements with $  \widetilde{\mathrm{Adv}} \big( 1 \big)$ or $  \widetilde{\mathrm{Adv}} \big( 2 \big)$ in place of $  \widetilde{\mathrm{Adv}} \big( 0 \big)$, hence yielding the desired estimate,

{\small

\begin{align*}
    {\tiny \widetilde{\mathrm{Adv}} \big( 0 \big) \lesssim  \widetilde{\mathrm{Adv}} \big( 1 \big) \lesssim  \widetilde{\mathrm{Adv}} \big( 2 \big) \lesssim \mathrm{NEGL} \big( \lambda^{\prime} \big)     }         , 
\end{align*}

}

\noindent from which we conclude the argument. \boxed{}

\section{Conclusion}

\noindent In this paper we demonstrated how noise can be injected into a protocol previously examined for Quantum public key encryption in {\color{blue}[3]}. To replace an equality obtained by the authors between the trace distance and the negligibility function we describe (as stated in the $\textbf{Lemma}$ from a previous section), how the Gentle Measurement Lemma from Quantum information theory can be used to infer that a desired upper bound for the trace distance holds from a lower bound on the trace. More specifically under the presence of noise in the Quantum public key encryption protocol we relate computations for obtaining the lower bound on the trace to the correctness of the protocol under noise. However nevertheless limitations to implementing the NQPKE-QKD protocol in the presence of noise emerge as the security parameter $\lambda^{\prime}$ was shown to be related to the upper bound on the trace distance from the fact that,

{\small

\begin{align*}
  \bigg\{ {\tiny \frac{\lambda_i}{\lambda_j}  } \bigg\}_{\{ i \neq j, 1 \leq i \leq j \leq N \} }  \equiv  \frac{ {\tiny   \big\{ \lambda_{i,\mathrm{Encoding}} \big\}_{1 \leq i \leq C}   +  \big\{ \lambda_{i,\mathrm{Decoding}} \big\}_{1 \leq i \leq C}   +  \big\{ \lambda_{i,\text{Secret Key Generation}}  \big\}_{1 \leq i \leq C}         \cdots }}{ {\tiny \big\{ \lambda_{j,\mathrm{Encoding}} \big\}_{1 \leq j \leq C, j \neq i }   +  \big\{ \lambda_{j,\mathrm{Decoding}} \big\}_{1 \leq i \leq C, j \neq i }   +     \cdots }  } \\  \end{align*}
  
  \begin{align*} \frac{ {\tiny +  \big\{   \lambda_{i,\text{Public Key Generation}}      \big\}_{1 \leq i \leq C}  } }{ {\tiny +   \big\{ \lambda_{j,\text{Secret Key Generation}}  \big\}_{1 \leq i \leq C, j \neq i }      +   \big\{   \lambda_{j,\text{Public Key Generation}}   \big\}_{1 \leq j \leq C, j \neq i } }}. \\ 
\end{align*}

}

\noindent as stated in the subsequentiality assumption stated in \textit{3.2}. It remains of further interest to generalize the computations performed in this work to closely related cryptographic settings in which prospective Quantum advantage can be realized (specifically, as to whether any computations surrounding the statement,

  {\small
    
    \begin{align*}
  \mathrm{Td} \big[   \big\{              \widetilde{\mathrm{QKDSec}^{\mathcal{A}_{\lambda^{\prime}}}}       , \widetilde{k_0} , \widetilde{k_1} \big\}       ,     \big\{   \widetilde{\mathrm{QKDSec}^{\mathcal{A}_{\lambda^{\prime}}}}        , \widetilde{k^{\prime}_0} , \widetilde{k^{\prime}_1} \big\}      \big]     \lesssim   {\tiny \mathrm{NEGL} \big( \lambda^{\prime} \big)  }  . \\
    \end{align*}

    }

\noindent provided in \textbf{Definition} \textit{39} can be further examined).

\section{Declarations}

\subsection{Ethics approval and consent to participate}

The author consents to participate in the peer review process.

\subsection{Consent for publication}

The author consents to submit the following work for publication.

\subsection{Availability of data and materials}

Not applicable

\subsection{Competing interests}

Not applicable

\subsection{Funding}

Not applicable

\section{References}

\noindent [1] Camenisch, J., Enderlein, R.R., Maurer, U.
Security and Cryptography for Networks (SCN), Lecture Notes in Computer Science, Springer \textbf{9841}: pp. 104–125, Aug 2016. https://doi.org/10.1007/978-3-319-44618-9 6.


\bigskip






\noindent [2] Demay, G., Gaži, P., Hirt, M., Maurer, U. (2013). Resource-Restricted Indifferentiability. In: Johansson, T., Nguyen, P.Q. (eds) Advances in Cryptology – EUROCRYPT. Lecture Notes in Computer Science, vol 7881 (2013). Springer, Berlin, Heidelberg. https://doi.org/10.1007/978-3-642-38348-9 39.

\bigskip

\noindent [3] Malavolta, G., Walter, M. (2024). Robust Quantum Public-Key Encryption with Applications to Quantum Key Distribution. In: Reyzin, L., Stebila, D. (eds) Advances in Cryptology – CRYPTO 2024. CRYPTO 2024. Lecture Notes in Computer Science, vol 14926. Springer, Cham. https://doi.org/10.1007/978-3-031-68394-7.



\bigskip

\noindent [4] Konig, R., Renner, R., Bariska, A., Maurer, U. Small Accessible Quantum Information Does Not Imply Security.
Phys. Rev. Lett. 98(14): 140502 (2007). https://link.aps.org/doi/10.1103/PhysRevLett.98.140502.

\bigskip

\noindent [5] Lanzenberger, D., Maurer, U. Direct Product Hardness Amplification. In: Nissim, K., Waters, B. (eds) Theory of Cryptography. TCC 2021. Lecture Notes in Computer Science, vol 13043 (2021). Springer, Cham. https://doi.org/10.1007/978-3-030-90453-1 21.

\bigskip

\noindent [6] Maurer, U., Rüedlinger, A., Tackmann, B. Confidentiality and Integrity: A Constructive Perspective. In: Cramer, R. (eds) Theory of Cryptography. TCC. Lecture Notes in Computer Science, vol 7194 (2012). Springer, Berlin, Heidelberg. https://doi.org/10.1007/978-3-642-28914-9 12.

\bigskip

\noindent [7] Maurer, U., Pietrzak, K., Renner, R. Indistinguishability Amplification. In: Menezes, A. (eds) Advances in Cryptology - CRYPTO. Lecture Notes in Computer Science, vol 4622 (2007). Springer, Berlin, Heidelberg. https://doi.org/10.1007/978-3-540-74143-5 8.

\bigskip

\noindent  [8] Maurer, U., Renner, R., Holenstein, C. Indifferentiability, Impossibility Results on Reductions, and Applications to the Random Oracle Methodology. Cryptology {ePrint} Archive, Paper 2003/161 (2003). https://eprint.iacr.org/2003/161.

\bigskip

\noindent [9] Maurer, U., Tackmann, B. On the soundness of authenticate-then-encrypt: formalizing the malleability of symmetric encryption. In Proceedings of the 17th ACM conference on Computer and communications security (2010). Association for Computing Machinery, New York, NY, USA, 505–515. https://doi.org/10.1145/1866307.1866364.

\bigskip

\noindent [10] Ostrev, D. QKD parameter estimation by two-universal hashing. Quantum \textbf{7}, 894 (2023) https://doi.org/10.22331/q-2023-01-13-894.


\bigskip

\noindent [11] Portmann, C., Renner, R. Cryptographic security of quantum key distribution. arXiv:1409.3525 (2014). 
https://doi.org/10.48550/arXiv.1409.3525.

\bigskip





\noindent [12] Renner, R. Security of Quantum Key Distribution. International Journal of Quantum Information 06:01: 1-127 (2008). https://doi.org/10.1142/S0219749908003256.

\bigskip

\noindent [13] Renner, R., Wolf, S. The Exact Price for Unconditionally Secure Asymmetric Cryptography. In: Cachin, C., Camenisch, J.L. (eds) Advances in Cryptology - EUROCRYPT 2004. EUROCRYPT 2004. Lecture Notes in Computer Science, \textbf{3027}. Springer, Berlin, Heidelberg. $https://doi.org/10.1007/978-3-540-24676-3 7$.

\bigskip

\noindent [14] Rigas, P. Probability distributions over CSS codes: two-universality, QKD hashing, collision bounds, security. arXiv:2510.02402, submitted (2025). https://doi.org/10.48550/arXiv.2510.02402.

\bigskip

\noindent [15] Rigas, P. Eve's forgery probability from her false acceptance probability: interactive authentication, Holevo information and the min-entropy. arXiv: 2603.06645, submitted (2026). 
https://doi.org/10.48550/arXiv.2603.06645.

\bigskip

\noindent [16] Rigas, P. Multiplayer parallel repetition without dependency-breaking and anchoring variables: monotonic, concave amplification. arXiv: 2605.08259, submitted (2026). 
https://doi.org/10.48550/arXiv.2605.08259.

\bigskip

\noindent [17] Rigas, P. Composable, unconditional security without a Quantum secret key: public broadcast channels and their conceptualizations, adaptive bit transmission rates, fidelity pruning under wiretaps. arXiv:2512.19759, submitted (2026). 
https://doi.org/10.48550/arXiv.2512.19759.

\bigskip

\noindent [18] Wilde, M. Quantum information theory lecture notes (2015). https://markwilde.com/teaching/2015-fall-qit/lectures/lecture-16.pdf

\end{document}